\documentclass[fleqn,usenatbib]{mnras}
\usepackage{tikz}
\usetikzlibrary{shapes,arrows,chains}

\usepackage{newtxtext,newtxmath}

\tikzstyle{startstop} = [rectangle, rounded corners, minimum width=2.2cm, minimum height=0.9cm,text centered, draw=black]
\tikzstyle{arrow} = [thick,->,>=stealth]

\usepackage[T1]{fontenc}

\DeclareRobustCommand{\VAN}[3]{#2}
\let\VANthebibliography\thebibliography
\def\thebibliography{\DeclareRobustCommand{\VAN}[3]{##3}\VANthebibliography}

\usepackage{graphicx}	
\usepackage{amsmath}	

\title[Neutron star natal kicks]{Combined analysis of neutron star natal kicks using proper motions and parallax measurements for radio pulsars and Be X-ray binaries  }

\author[A.P. Igoshev et al.]{
Andrei P. Igoshev,$^{1}$\thanks{E-mail: ignotur@gmail.com}
Martyna Chruslinska,$^{2}$
Andris Dorozsmai, $^{3}$ 
and Silvia Toonen $^{3,4}$
\\
$^{1}$Department of Applied Mathematics, University of Leeds, LS2 9JT Leeds, UK\\
$^{2}$Institute of Mathematics, Astrophysics and Particle Physics, Radboud University Nijmegen, PO Box 9010, 6500 GL Nijmegen, The Netherlands\\
$^{3}$Institute of Gravitational Wave Astronomy, School of Physics and Astronomy, University of Birmingham, Birmingham, B15 2TT, United Kingdom\\
$^{4}$Anton Pannekoek Institute for Astronomy, University of Amsterdam, 1090 GE Amsterdam, The Netherlands
}

\date{Accepted XXX. Received YYY; in original form ZZZ}

\pubyear{2015}

\begin{document}
\label{firstpage}
\pagerange{\pageref{firstpage}--\pageref{lastpage}}
\maketitle

\begin{abstract}
Supernova explosion and the associated neutron star natal kicks are important events on a pathway of a binary to become a gravitational wave source, an X-ray binary or a millisecond radio pulsar. Weak natal kicks often lead to binary survival, while strong kicks frequently disrupt the binary.
In this article, we aim to further constrain neutron star natal kicks in binaries.  We explore binary population synthesis models by varying prescription for natal kick, remnant mass and mass accretion efficiency. 
We introduce a robust statistical technique to analyse combined observations of different nature.
Using this technique, we further test different models using parallax and proper motion measurements for young isolated radio pulsars and similar measurements for Galactic Be X-ray binaries.
Our best model for natal kicks is consistent with both measurements and contains a fraction of $w=0.2\pm 0.1$ weak natal kicks with $\sigma_1 = 45^{+25}_{-15}$~km/s, the remaining natal kicks are drawn from the high-velocity component, same as in previous works: $\sigma_2 = 336$~km/s. 
We found that currently used models for natal kicks of neutron stars produced by electron capture supernova (combination of maxwellian $\sigma=265$~km/s and $\sigma = 30$~km/s for electron capture) are inconsistent or marginally consistent with parallaxes and proper motions measured for isolated radio pulsars.
We suggest a new model for natal kicks of ecSN, which satisfy both observations of isolated radio pulsars and Be X-ray binaries. 
\end{abstract}




\section{Introduction}

Neutron stars (NSs) often receive a significant birth velocity (natal kick) at the moment of supernova explosion \citep{lyne1994}. On one hand, this kick is frequently strong with amplitude $|v| > 100$~km~s$^{-1}$ (in some cases reaching 800~km~s$^{-1}$) while typical velocity dispersion of NS progenitors are $<20$~km~s$^{-1}$. On the other hand, formation of certain Galactic double neutron stars (DNSs) systems and presence of pulsars in globular clusters (with escape velocities $\lesssim$50 km/s) clearly require small natal kicks \citep{tauris2017}. 
It is unclear whether the NS natal kick velocity distribution is broad and extends from $\approx 0$~--~$1000$~km~s$^{-1}$, or if NSs with some particular characteristics are formed with much weaker natal kicks ($ 0-30$~km~s$^{-1}$).

The problem of estimating the NS natal kick distribution has two parts, which were not sufficiently addressed in the previous research on the subject: (1) NSs receiving weak natal kick should be less common among isolated radio pulsars because they less frequently disrupt the binary and (2) different formation channels could lead to different velocity distributions. In studies of observed velocity distribution of radio pulsars e.g. \cite{hobbs2005, Verbunt2017,Igoshev2020} young pulsar velocities were assumed to be equal to the isolated NS natal kick velocities and the influence of binarity of the NS progenitors was never analysed.

Individual velocities of radio pulsars (mode of activity typical for young, magnetised NSs with fast rotation, see e.g. \citealt{deller2019}) are estimated based on VLBI measurements of parallaxes and proper motions. Individual velocities for high-mass X-ray binaries (HMXBs) and double neutron stars are estimated based on analysis of orbital properties (see e.g. \citealt{Dewi2005,beniamini2016} and \citealt{tauris2017}).

On the one hand, young radio pulsars with reliably measured parallaxes and proper motions seem to be isolated \citep{IgoshevPerets2019}. Also, recent analysis of all radio pulsars recorded in the ATNF pulsar catalogue\footnote{ATNF pulsar catalogue http://www.atnf.csiro.au/research/pulsar/psrcat} \citep{Manchester2005} found an upper limit of 5.3~per~cent on the fraction of radio pulsars with wide binary companions \citep{Antoniadis2020,Antoniadis20202}. 
Close binary companions with orbital periods shorter than $\sim 10$~years (timescales comparable with observation spans for radio pulsars) are easy to detect using the pulsar timing technique \citep{timing2006}. 
On the other hand, progenitors of NSs are massive ($\ge 7$~M$_\odot$) and the majority of such stars are formed in binaries. A fraction of truly isolated massive stars is less than 15~per~cent \citep{Moe2017}. This large discrepancy between the fraction of massive stars in binaries and the fraction of radio pulsars in binaries can be explained by a supernova explosion, which often disrupt a binary due to mass-loss and NS natal kick, see e.g. \cite{Renzo2019}. 

A stronger natal kick is more likely to disrupt a binary and leave behind an isolated radio pulsar,
while an NS formed with natal kick weaker or comparable to orbital velocity of the secondary (typically less than 100~km~s$^{-1}$) is more likely to remain bound within the binary.
Therefore, a universal natal kick distribution (if it exists) should be probed simultaneously using binaries with NS and isolated radio pulsars. 
A particular class of systems that can be used in such an analysis, and is explored in this study,
is the population of  Be X-ray binaries (BeX).
Those are high-mass X-ray binaries (HMXB) hosting NS, whose formation path involves a phase of (stable) mass transfer before the supernova explosion \citep[e.g.][]{Vinciguerra2020}.

In this article, we aim to model the observed velocities of isolated radio pulsars and the peculiar velocities of Galactic Be X-ray binaries using the binary population synthesis. Our approach is an extension to one suggested by \cite{andrews2015}. First, we identify the model which best describes the population of Be X-ray binaries and then we vary the parameters of the natal kick distribution only.
To allow for a qualitative comparison of different models, we introduce a new likelihood method for parallaxes and proper motion measurements, which allows us to directly use the results of binary stellar population synthesis as a model for peculiar velocities.

While we were preparing this article, \cite{Willcox2021} published a study on a similar topic. The author used a non-parametric description for the natal kicks of core-collapse NSs. They found that ecSN events should only occur in stars stripped of hydrogen envelope to satisfy observed transverse velocities of isolated radio pulsars. This research does not take into account a comparison with Be X-ray binaries and leaves us the opportunity to investigate low natal kicks. 

This article is structured as follows: in Section~\ref{s:background} we give a brief overview for Be X-ray binaries, electron capture supernova and modelling efforts; in Section~\ref{s:data} we introduce observational data sets for young isolated radio pulsars and Be X-ray binaries, and perform simple analysis of peculiar velocities for Be X-ray binaries. In Section~\ref{s:combined_analysis} we describe our binary population synthesis. In Section~\ref{s:all_results} we present results for these exploratory models. In Section~\ref{s:combined} we introduce our combined analysis and discuss electron capture supernova explosion in details, which is followed by Discussions and Conclusions Sections.

\section{Be X-ray binaries and electron capture supernova explosions}
\label{s:background}

Be stars are B spectral type stars with enhanced emission lines \citep[see e.g. review by][]{Reig2011}. Their emission spectral lines are formed in decretion disk. The secondary star rotates faster than 70~per~cent of its critical spinning rate \citep{Slettebak1982,Porter1996,Yudin2001,Townsend2004} and loses material from equatorial regions forming a large decretion disk.  When combined with an NS in a binary these stars produced the Be X-ray binary phenomenon. NS orbiting around the Be star enters the decretion disc close to the periastron position and accretes material. The accretion causes a strong heating of the polar regions, which is seen in X-rays.

The decretion disk around Be is thought to be formed due to fast rotation of B stars. \cite{Kriz1975, rappaport1982} suggested that the star is spun up due to accretion of large amount of material with angular momentum. It is important to note that masses of the isolated Be stars and Be X-ray stars differ. The most common masses of isolated Be stars are around 3$M_\odot$ while Be stars in Be X-ray systems are more massive with masses 8~$M_\odot$. \cite{Vinciguerra2020} suggested that the mass distribution of Be stars in binaries could be a good indicator for the mass accretion efficiency. A large sample of Be X-ray binaries is observed in the Small Magellanic Cloud where researchers discovered $\approx 70$ active systems.

\cite{kinematicsHMXB} recently analysed proper motions of Be X-ray binaries in the Galaxy using the Gaia DR2. He found that transverse systemic velocities of Be X-ray binaries are 29$\pm 11$~km~s$^{-1}$ for short period binaries and $16\pm 8$~km~s$^{-1}$ for long period binaries. He confirmed that this difference is statistically significant. In another recent analysis by \cite{Bodaghee2021}, researchers used cross correlation function between position of high-mass X-ray binaries in Small Magellanic Cloud and OB associations. 
They found that initial systemic velocities (called kick velocities) are in range 2-34~km~s$^{-1}$. In this work, we plan to analyse the parallaxes and proper motions of Be X-ray binaries using new Gaia EDR3.

Over the years, it was suggested that NS born in binaries could receive very different natal kicks due to supernova explosion mechanisms which prevail or only operate in binaries, such as: (1) the electron capture supernova explosion (ecSN; \citealt{Nomoto1984,Nomoto1987}) and (2) the explosion of ultra-stripped helium stars \citep[e.g.][]{Tauris13}. The latter mechanism relies on a reduction of ejecta mass \citep{tugboat}, while the former mechanism leads to weaker natal kicks because stellar cores of ecSN progenitors are easier to explode and less asymmetries (responsible for formation of the natal kick) are developed \citep{Podsiadlowski2004}. 
All of these mechanisms are related to mass transfer in close interacting binaries, which constitute up to 70\% of all massive binaries \citep{sana_mult}. Once again, NSs formed through these mechanisms will more frequently stay in a binary due to reduced kick and mass-loss.

EcSN progenitors are believed to originate from stars that
are too lightweight to ignite oxygen in their cores (stars that ignite oxygen eventually undergo CCSN) and too heavy to end up as oxygen-neon white dwarfs.
Such stars develop degenerate ONeMg cores after the carbon burning phase that exceed the Chandrasekhar limit $\sim1.37 M_\odot$ and they can collapse to NS in the process triggered by electron captures on magnesium and neon
\citep[e.g.][]{Miyaji80,Nomoto1984,NomotoKondo91,WoosleyHeger15,Jones16}.
The range of initial masses allowed for ecSN progenitors is expected to be narrow for single stars, but can be effectively broadened by binary interactions, allowing for massive degenerate ONeMg core formation from initially more/less massive progenitors \citep[e.g.][]{Podsiadlowski2004,Doherty17,Poelarends17,SiessLebreuilly18}.
EcSN are typically implemented in binary population synthesis codes
based on the helium core mass of a star at the beginning of the AGB stage $M_{\rm C; BAGB}$ (with the commonly adopted ecSN progenitor $M_{\rm C; BAGB}$ range [1.8, 2.25]~M$_\odot$, following \citealt{Hurley2000, EldridgeTout04} - as also used in this study).
It should be noted that the exact limits on this mass range are very sensitive to the adopted assumptions about the stellar and binary physics (e.g. the treatment of convection and overshooting/mixing in stellar interiors, wind mass loss rates, treatment of mass transfer, e.g. \citealt{Poelarends08,Jones16,Doherty17,SiessLebreuilly18}) and depend on metallicity (e.g. \citealt{Poelarends08}, but see \citealt{Doherty17} - where the metallicity dependence is shown to be very limited).
Furthermore, a fraction of the ecSN progenitors estimated that way may not collapse to NS, but undergo a partial theronuclear explosion and instead leave behind a white dwarf \citep[e.g.][]{Jones16}.

Overall, the evolution of the binary after supernova explosion was modelled multiple times in the literature, see e.g. \cite{Hills1983,Dewey1987,Brandt1995,Kalogera1996,Fryer1998}.
Multiple studies were focused on the influence of NS natal kicks on the rate of compact binary mergers, especially in relation to gravitational wave progenitors, e.g. \cite{Lipunov1997,tauris2017}. 
More recently, \cite{Kuranov2009} performed binary population synthesis assuming that ecSN are only produced in binaries. They also assumed that NSs born in such explosions receive a significantly smaller natal kick than seen in truly isolated NS population. They demonstrated that a contribution of ecSN, which disrupted their parent binary, is negligible in comparison to amount of isolated and binary origin NS with large velocities. The result of their population synthesis showed that the final velocity distribution of isolated NSs followed closely the distribution of natal kicks in isolated NSs besides a small region around 30~km$^{-1}$. In this region, they saw a slight increase (amplitude of a few percents) in number NSs. Such a difference is impossible to notice when sample size of parallaxes and proper motions of isolated radio pulsars is limited by 20-100 measurements. 

\section{Data}
\label{s:data}

In this article, we study parallaxes and proper motion measurements. For young (spin-down ages less than $3$~Myr) isolated radio pulsars we use the same catalogue as \cite{Igoshev2020} (see their table 1 and table A1 for details), which includes VLBI measurements by \cite{bbgt02,btgg03,ccl+01,ccv+04,cbv+09,kvw+15,dtbr09,deller2019}. For Be X-ray binaries we use Gaia EDR3 \citep{gaiamission,edr3,edr3astrometry}. Below we describe in detail our analysis of Be X-ray binaries. The result of similar analysis for isolated radio pulsars are available in article by \cite{Igoshev2020}.

\subsection{Gaia EDR3 parallaxes and proper motions of Galactic Be X-ray binaries} 

We use the Gaia EDR3 catalogue to identify Be X-ray binaries. The identification procedure is similar to \cite{IgoshevPerets2019}. Namely, we use the catalogue of HMXBs by \cite{hmxb_catalogue} and check the type of these systems in the Simbad database \citep{Wenger2000}. To proceed with identification, we request all stars from the Gaia EDR3 around the position of each HMXB in a radius of 15~arcsec. We select from these stars only ones with gaia G magnitude in range  $|G - V| < 1.5$ (V is visual magnitude available in Simbad) located at the smallest angular separation from the catalogue coordinate. We notice that the astrometry analysis procedure improved significantly since Gaia DR2 \citep{dr2astrometry,edr3astrometry}, and precision checks related to \texttt{astrometric\_chi2\_al} and \texttt{astrometric\_n\_good\_obs\_al} are already implemented. Therefore, we select only stars with five and six-parameter astrometric solutions  with values of re-normalised unite weight error close to 1, namely  $0.6<$\texttt{ruwe}$< 1.4$ \citep{edr3astrometry}. We additionally filter out all measurements with formal errors of parallax  $\varpi'/\sigma_\varpi < 3$. After this procedure we end up with 45 systems. We summarise the Be X-ray binaries identified in Gaia EDR3 in Table~\ref{t:xrb_gaia}.

At the next stage, we inspect the catalogue and remove all systems with unknown spectral type or spectral type different from Be or Oe. In particular we remove multiple wind-fed supergiant systems because the formation of these systems does not require an angular momentum transfer from primary to the secondary star, therefore these systems could have been formed via the dynamically unstable mass transfer or without any mass transfer at all. In Appendix~\ref{s:systems_BeX} we list some systems where we perform an additional literature search to decide if they are to be included in the velocity analysis or not. Our final list contains 31 objects.

\begin{figure*}
	\begin{minipage}{0.48\linewidth}
	\includegraphics[width=\columnwidth]{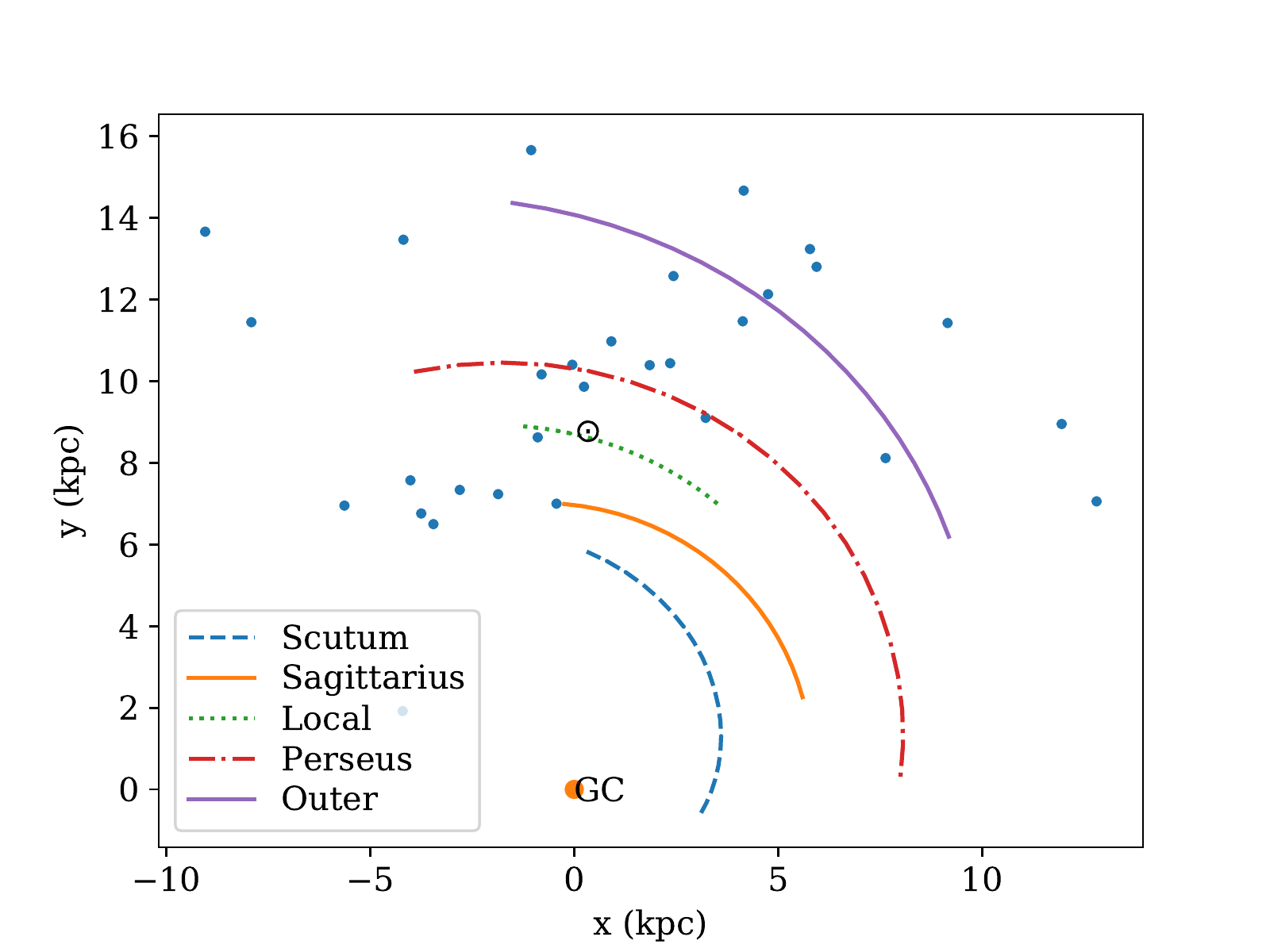}
	\end{minipage}
	\begin{minipage}{0.48\linewidth}
	\includegraphics[width=\columnwidth]{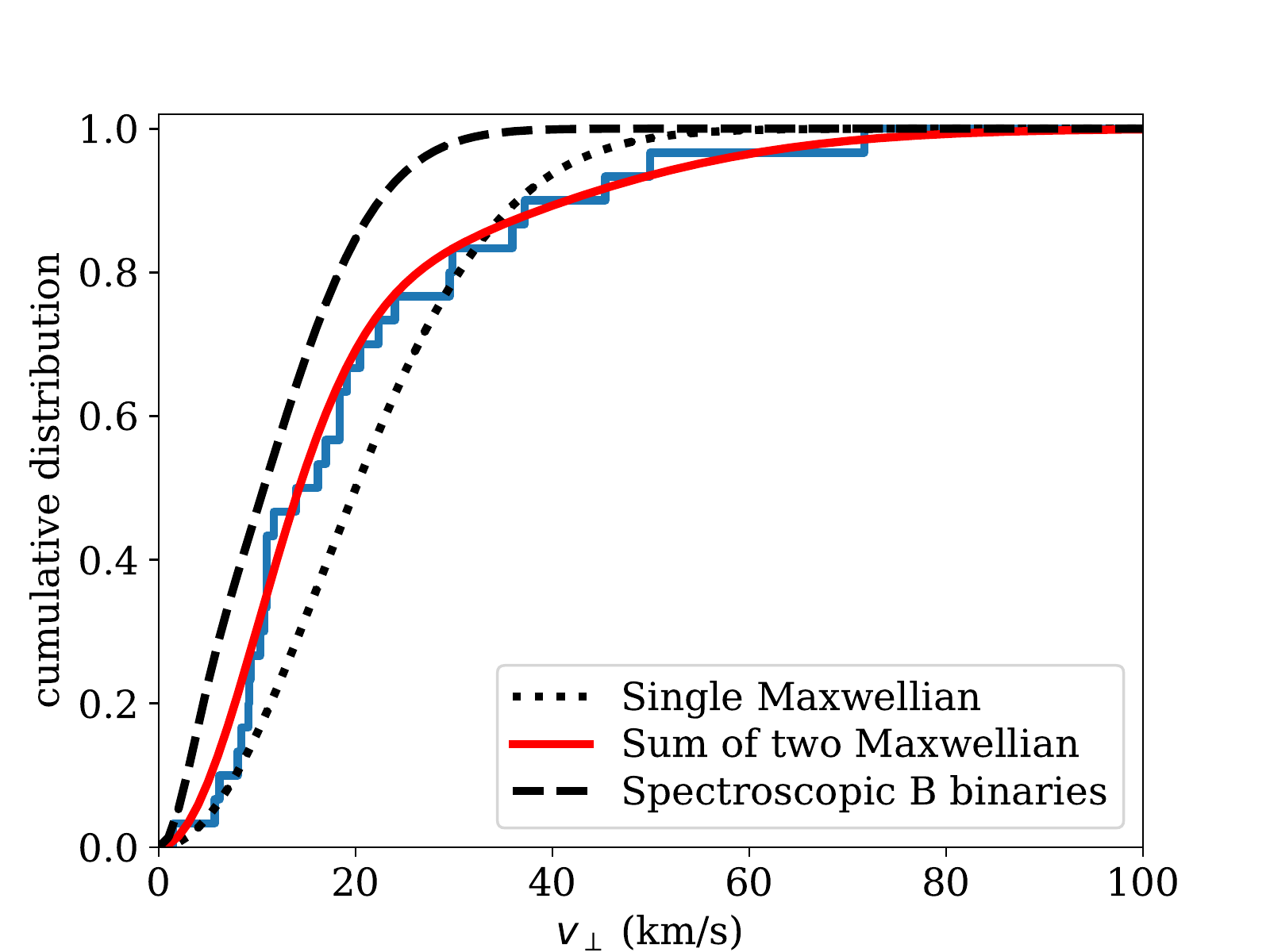}
	\end{minipage}
    \caption{The distribution of Be X-ray binaries identified in Gaia EDR3 in the Galaxy (left panel) and distribution of their and binary B spectroscopic stars nominal peculiar velocities corrected for the Galactic rotation (right panel). The parameters of spiral arms are estimated by \protect\cite{reid2014} and confirmed using the Gaia data \protect\citep{xu2018}. }
    \label{fig:galaxy_BeX}
\end{figure*}

We plot location of these Be X-ray binaries projected to the Galactic plane in Figure~\ref{fig:galaxy_BeX}. We also plot the cumulative distribution of nominal velocities in same Figure, where the nominal velocity in a particular direction (e.g. right ascension) is computed as: 
\begin{equation}
v_\alpha' = c \frac{\mu_\alpha}{\varpi'},    
\end{equation}
where $c = 4.74$ is a numerical coefficient to transform proper motion  measurements $\mu_\alpha$ [mas/year] and parallax measurements $\varpi$ [mas] to units [km/s].  
The transverse nominal velocities of Be X-ray binaries are computed as the following:
\begin{equation}
v'  = \sqrt{\left(c \frac{\mu_l}{\varpi'} - \Delta v_l (l, b, 1/\varpi')\right)^2 + \left(c \frac{\mu_b}{\varpi'} - \Delta v_b (l, b, 1/\varpi')\right)^2}.     
\end{equation}
Here $\Delta v_l (l, b, D)$ and $\Delta v_b (l,b,D)$ are the corrections for the rotation of the Galaxy at location with Galactic coordinates $l,b$ and distance $D$ from the Earth. Exact equations for these corrections can be found in appendix B by \cite{Verbunt2017}. It is easy to see in Figure~\ref{fig:galaxy_BeX} (right panel) that these transverse velocities are limited by 80~km/s and half of velocities are below $\approx 20$~km/s. Typical velocity dispersion inside regions where OB stars formed are below 10~km/s \citep{deBruijne1999,Kiminki2018}. These star formation regions could have a velocity dispersion of $\approx 5$~km/s in respect to regular Galactic rotation \citep{reid2014}. We also analyse velocities of spectroscopic binaries with primary of B type (possible progenitors of Be X-ray binaries). All details are summarised in Appendix~\ref{s:bbinary}. We find that these stars have velocities, which can be described as a sum of two Maxwellians with $w = 0.2$, $\sigma = 3$~km~s$^{-1}$ and $\sigma_2 = 11$~km~s$^{-1}$ (see black dashed line in right panel of Figure~\ref{fig:galaxy_BeX}). 
Therefore, some Be X-ray binaries have velocities which are clearly above ones typical for their binary progenitors.

\subsection{Peculiar velocities for Be X-ray binaries }

We analyse the parallaxes and proper motions of Be X-ray binaries from Table~\ref{t:xrb_gaia} using the same maximum likelihood technique as was developed and described in detail by \cite{verbuntCator2017} and \cite{Verbunt2017}. Brief summary of the method is available in Appendix~\ref{s:appendix_ml}. The results of this analysis are summarised in Table~\ref{tab:mxw_res}, and we plot these velocity distributions in Figure~\ref{fig:galaxy_BeX} (right panel). A visual inspection confirms that the sum of two Maxwellians describes the peculiar velocities of Be X-ray binaries significantly better than a single Maxwellian. The values for $\sigma_1$ and $\sigma_2$ which we found in our analysis are similar to one by \cite{kinematicsHMXB}.

For completeness, we perform the same analysis using newer measurements for solar rotation speed $v_\odot = 233$~km/s and distance toward the Galactic centre $R_\mathrm{G} = 8.122$~kpc, see e.g. \cite{Mroz2019}. Results are also available in Table~\ref{tab:mxw_res}. Different values for Galactic rotation parameters could change the distribution for peculiar velocities of Be X-ray binaries by a couple of km/s. These new values are within the confidence interval of values estimated using the old $v_\odot$ and $R_\odot$.

\begin{table}
    \centering
    \begin{tabular}{lllcc}
    \hline
    Fixed Parameters & Model & Parameters & $2\Delta\mathcal {L}$ \\
    \hline
    $v_\odot = 220$~km/s & Single Maxwellian    & $\sigma = 17\pm 3$~km~s$^{-1}$ & - \\ 
    $R_\odot = 8.5$~kpc \\
                         & Sum of two           & $w = 0.74\pm 0.15$ & 15.2 \\
                         &  Maxwellians         & $\sigma_1 = 10\pm 3$~km~s$^{-1}$ \\ 
                         &                         & $\sigma_2 = 29\pm 12$~km~s$^{-1}$ \\
    \hline
    $v_\odot = 233$~km/s  & Single Maxwellian   & $\sigma = 16\pm 3$~km~s$^{-1}$ & - \\
    $R_\odot = 8.122$~kpc \\
                          & Sum of two          & $w = 0.82\pm 0.13$ & 10.6 \\
                          & Maxwellians        & $\sigma_1 = 11\pm 2$~km~s$^{-1}$ \\ 
                          &                         & $\sigma_2 = 31\pm 15$~km~s$^{-1}$ \\    
    \hline
    \end{tabular}
    \caption{The parameters of the velocity distribution for Be X-ray binaries in the Milky Way.}
    \label{tab:mxw_res}
\end{table}

Binary surviving the supernova explosion receives a peculiar velocity comparable to orbital speed of the exploded star. Therefore, the peculiar velocity of Be X-ray binary is a complicated function of mass transfer, natal kick strength and orientation, as well as orbital parameters at the moment of supernova explosion. Therefore, it is necessary to use a realistic binary population synthesis to meaningfully analyse Be X-ray peculiar velocities. 

\cite{Renzo2019} computed peculiar velocities of binaries with NS. The authors did not model Be X-ray binaries specifically, but looked at the velocity distribution of all main sequence - NS binaries including ones which will never be seen as Be X-ray binary. Therefore a direct comparison of results by \cite{Renzo2019} with Gaia parallaxes and proper motions of Be X-ray binaries is difficult. Thus, we perform our own binary population synthesis. Details of this synthesis are described in the following Section.

\section{Model exploration}
\label{s:combined_analysis}

\subsection{Formation paths for Be X-ray binaries and isolated radio pulsars}

In this article we concentrate on isolated radio pulsar and Be X-ray binaries formed via binary interactions in Milky Way (MW or Galaxy). We also examine Be X-ray binaries population in Small Magellanic Cloud (SMC) to allow for comparison with work by \cite{Vinciguerra2020}. 
We do not consider objects formed via dynamical formation channel in crowded environments such as globular clusters. The main advantages of restricting analysis to isolated radio pulsars and Be X-ray binaries is that formation of these binaries could only weakly depend on poorly understood common envelope stage (CE; \citealt{Ivanova2013})\footnote{A small fraction of isolated radio pulsars could originate from massive stars which is a result of a merger after common envelope stage. However, that is not the dominant channel and we disregard such cases. } and stars in these systems are not expected to undergo ultra-stripped SN (which could lead to weaker natal kicks).

\subsubsection{Isolated radio pulsars}

A significant fraction (up to 85 per cent) of NS progenitors are born in binaries or stellar multiples \citep{Kobulnicky2007,sana_mult,Moe2017}. A strong natal kick and fast mass loss during the SN explosion often leads to binary disruption. A fraction of $86_{-9}^{+11}$~per~cent of NS born in binaries may end up as isolated free-floating NSs \citep{Renzo2019} for natal kicks drawn from Maxwellian distribution with $\sigma=265$~km/s \citep{hobbs2005} and treatment of binary disruption by \cite{TaurisT1998}. 
Thus, a high fraction of the observed isolated radio pulsars \citep{Antoniadis2020} does not contradict their expected binary origin.
\\
NS could operate as an isolated radio pulsar up to Gyr depending on their initial rotational period, magnetic field strength and evolution history. In practice, most NSs fade or stop emitting radio after 100~Myr, see e.g. \cite{Faucher2006}. The radio emission produced by pulsars is strongly beamed, so we miss up to $90$~per~cent of pulsar population simply due to geometry \citep{Tauris1998}. 

In our simulations, we assume that every NS is born as a radio pulsar which operates for 100~Myr. We do not integrate their motion in the Galactic gravitational potential and we do not take beaming and sensitivity of modern pulsars survey into account.  Those factors become much more important when the whole population of radio pulsars is considered, in particular old pulsars, which could travel a significant distance within the Galaxy. Therefore, our synthetic pulsars can only be compared with young radio pulsars (spin-down ages $\tau<3$~Myr). The velocities of our synthetic pulsars stay representative of the natal kicks, but get some correction due to the binary origin. A primary effect of binary origin is that low-natal kick NSs stay more frequently bound in a binary and therefore are excluded from isolated radio pulsar population.  
There is no indication that beaming or radio luminosity correlates with natal kick velocity or orientation, so these factors are probably irrelevant in natal kick studies. 

However, there is a bias which could affect the result of our research and is poorly understood at the moment. Pulsars with well measured parallaxes and proper motions typically have high Galactic latitudes \citep{deller2019}. The low-natal kick pulsars should be concentrated toward the galactic plane and might be less represented in VLBI measurements. Future pulsar population synthesis is necessary to quantify the role of this effect on analysis of NS natal kicks.

\subsubsection{Be X-ray binaries and binary stellar evolution before the NS formation}
\label{s:bex}


In this section we first briefly discuss the most common evolutionary channels described in the literature (and found in our SeBa simulations) and then concentrate on the one producing Be X-ray binaries.

If the binary is wide enough,
its components never interact and evolve as single stars.
In such cases, the secondary star is not spun up and does not form a decretion disk. Therefore, the NS has no material to accrete and the system does not appear as a Be X-ray binary. Such systems could still be visible at later stages of evolution as supergiant wind-fed X-ray binaries. We further refer to this type of evolution as effectively single binary evolution channel.

\begin{figure}
	\includegraphics[width=\columnwidth]{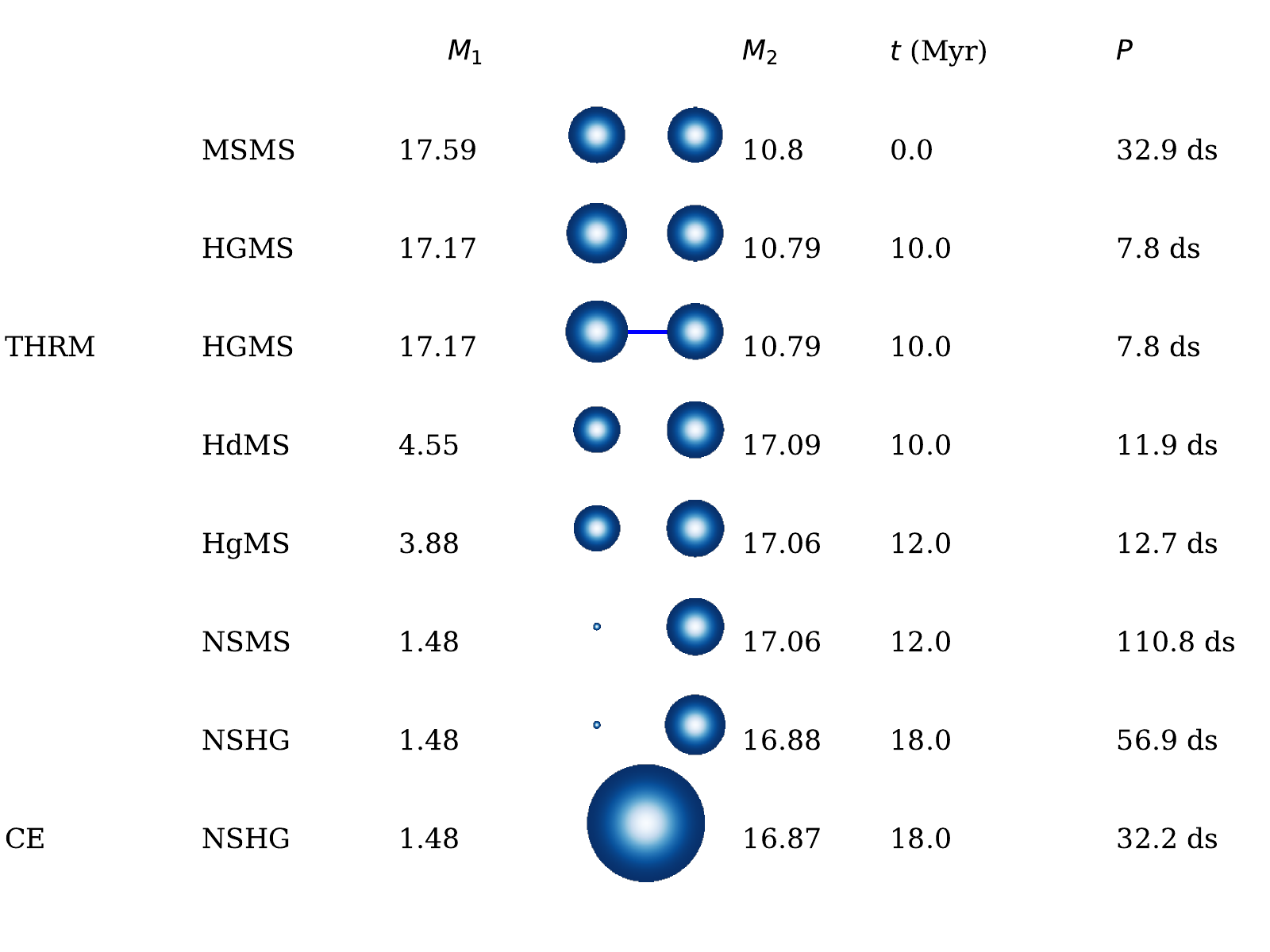}
    \caption{The formation channel for Be X stars for mass accretion efficiency of 0.5 accreted mass.  $M_1$ and $M_2$ are masses of primary and secondary star in solar masses respectively, $t$ is the age of the system and $P$ is the orbital period. The size of each circle indicates stellar radius. THRM stands for mass transfer on thermal timescale and CE stands for common envelope. MSMS stands for two main sequence stars, HG is Hertzsprung gap, Hd is a helium star, Hg is a helium giant, NS is neutron star. }
    \label{f:formationBeX_0.5}
\end{figure}

If stars are born sufficiently close, the primary fills its Roche-lobe and initiates a mass transfer at some point of its evolution. This could result both in stable and unstable mass transfer.
Dynamically unstable mass transfer can lead to a merger, resulting in formation of a solitary rejuvenated star which could explode as SN and form isolated NS  (see discussion of possible caveats in Section~\ref{s:caveats}). 
Stable mass transfer at this stage is a necessary condition for the formation of a Be X-ray binary (e.g. \citealt{Portegies_Zwart1995, ShaoLi2014} and \citealt{Vinciguerra2020}).
These previous studies found that the most common formation channel involves an episode of non-conservative, dynamically stable mass transfer initiated after the primary star depleted its core hydrogen. 
This channel is illustrated in Figure~\ref{f:formationBeX_0.5}. 
The post-mass transfer binary may get disrupted during the subsequent NS formation. If the binary is unbound due to the natal kick, we consider the NS as an isolated radio pulsar.

If instead of stable mass transfer the binary goes though a CE stage two things could happen: (1) CE is usually very short (see e.g. \citealt{IgoshevPerets2020}) and only a small fraction of mass is transferred to the secondary, so a transfer of  angular momentum is limited and no decretion disk is formed further in the evolution. (2) If the CE is initiated after one or multiple mass transfers, the CE could destroy the decretion disk.

In the interacting formation channel, the mass transfer is non-conservative, and the mass accretion efficiency controls how massive the secondary star (Be star) will be \citep{Vinciguerra2020}. A highly non-conservative mass transfer leads to the formation of Be X-ray binaries with Be star masses $3-5$~M$_\odot$. On the other hand, if 50-75 per cent of transferred mass is successfully accreted by the secondary, the peak of the Be star mass distribution shifts towards  8-10~M$_\odot$, as it is expected from observations. 

\subsection{Assumptions for the binary population synthesis}
\label{s:model_exploration}

We briefly summarise the details of fourteen evolutionary models which we consider in this paper in Table~\ref{tab:models}. We use the binary population synthesis code SeBa\footnote{The code is publicly available https://github.com/amusecode/SeBa} (\citealt{PortegiesZwart1996}; up-to-date description is available \citealt{toonen2012}). For our exploratory runs, we model the evolution of 100000 zero-age main sequence (ZAMS) binaries with primaries mass ranging from 6~M$_\odot$ to 40~M$_\odot$. The primary masses are sampled from a power-law distribution with an exponent of -2.35 \citep{Salpeter1955ApJ}. We restrict the mass of the secondary to the range $q\in [0.001, 1]$ where $q=M_2/M_1$. We draw  mass ratio from the uniform distribution \citep{duchene2013}. For the stellar evolution simulations in the Galaxy, we use the solar metallicity $0.02$; for the Small Magellanic Cloud, we use a metallicity of $3.5\times 10^{-3}$. The semi-major axis distribution is chosen in the form of power-law with exponent of $-1$ ranging from Roche lobe radius to $10^6\; R_\odot$ \citep{Abt1983}. The eccentricity distribution is thermal \citep{Ambartsumian1937,heggie1975}.  We assume 100~per~cent binary fraction, see Section~\ref{s:lower_binary_fraction} for test of a smaller binary fraction. 

\begin{table*}
    \centering
    \begin{tabular}{cllcccccccc}
    \hline
    Model & Natal kick        & Remnant mass  & Mass accretion efficiency  & ecSN & Mass range for ecSN  & $2\Delta \mathcal L$ & $N_\mathrm{SMC}$\\
    \hline
    A     & Hobbs   & Fryer  & default & No   &                        & 99.98   & 200\\
    B     & Hobbs   & Antoniadis  & default & No   &                   & 107.8   & 210\\
    C     & Igoshev & Fryer  & default & No   &                        & 50.44   & 320\\
    D     & Igoshev & Fryer  & semi-conservative & No &                & 1.3     & 180\\
    E     & No kick     & Fryer & semi-conservative & No &             & 20802.4 & 820\\
    F      & Hobbs  & Fryer  & semi-conservative & No &                & 11.9    & 60\\
    G0     & Hobbs  & Fryer  & semi-conservative & Yes & 1.83-2.25 (*) & 12.4    & 90\\ 
    G1     & Hobbs  & Fryer  & semi-conservative & Yes & 1.83-2.25     & 8.82    & 100\\
    H0     & Hobbs  & Fryer  & semi-conservative & Yes & 1.63-2.45 (*) & 10.9    & 170\\
    H1     & Hobbs  & Fryer  & semi-conservative & Yes & 1.63-2.45     & 7.84    & 180\\
    J0     & Hobbs  & Fryer  & semi-conservative & Yes & 1.6-2.25 (*)  & 10.66   & 90\\
    J1     & Hobbs  & Fryer  & semi-conservative & Yes & 1.6-2.25      & 6.22    & 100\\
    K      & Optimum  & Fryer & semi-conservative & No &               & -       & 140\\
    L      & New ecSN & Fryer & semi-conservative & Yes & 1.63-2.45    &  5.18   & 140 \\
    \hline
    \end{tabular}
    \caption{List of models with relevant assumptions; Fryer and Antoniadis stand for a relation between progenitor and remnant masses based on \protect\cite{Fryer2012} delayed explosion and \protect\cite{bimodal_msp_mass} respectively. Hobbs corresponds to single Maxwellian natal kick with parameter $\sigma = 265$~km~s$^{-1}$ \protect\citep{hobbs2005}; Igoshev corresponds to sum of two Maxwellians natal kick with $w=0.2$, and parameters $\sigma_1 = 56$~km~s$^{-1}$ and $\sigma_2 = 336$~km~s$^{-1}$ \protect\citep{Igoshev2020}; Optimum stands for sum of two Maxwellians natal kick with $w=0.2$, and parameters $\sigma_1 = 45$~km~s$^{-1}$ and $\sigma_2 = 336$~km~s$^{-1}$. In model L core collapse SN receive natal kick in form of single Maxwellian with $\sigma=336$~km/s and ecSN receive single Maxwellian natal kick with $\sigma = 45$~km/s. The asterisk (*) means that ecSN occurs only in stars stripped of their hydrogen envelope. $N_\mathrm{SMC}$ is the total number of Be X-ray binaries predicted by each model in the SMC.  }
    \label{tab:models}
\end{table*}

We consider two mass accretion efficiency schemes: following the default SeBa implementation and a fixed semi-conservative case. 
The default mass accretion efficiency implemented in SeBa is based on \cite{PolsMarinus1994}. The accretion rate is limited by:
\begin{equation}
\dot m_\mathrm{max} = \dot m_\mathrm{KH} \left(10^{-\mathcal{B}} \log \frac{R_L}{R_\mathrm{ms}}\right),    
\label{e:mass_transfer}
\end{equation}
where all parameters correspond to accretor: $R_L$ the Roche-lobe radius, $\dot m_\mathrm{KH}$ is the Kelvin-Helmholtz time scale accretion rate, $R_\mathrm{ms}$ is the main sequence radius. Numerical coefficient $\mathcal{B}$ is different for different types of stars and can be found in table 1 of \cite{PolsMarinus1994}. This $\dot m_\mathrm{max}$ corresponds to the condition that accretor stays within its Roche lobe. Eq.(\ref{e:mass_transfer}) based on detailed simulations of stellar evolution for stars with masses below 17~M$_\odot$.

\cite{Vinciguerra2020} analysed a range of fixed mass accretion efficiencies $\alpha$ and  found that values of $\alpha=0.5$ and $\alpha=0.75$ could explain the distribution of Be masses in Be X-ray binaries. They noticed that a fixed value of $\alpha=0.5$ describes the observed population of Be X-ray binaries in SMC best in terms of masses of Be stars and period distributions. There is no consensus on the unique mass accretion efficiency. For example, \cite{deMink2007} showed using detailed binary evolution simulations that there exists a spread in mass accretion efficiency. There are a few works where authors showed that the mass accretion efficiency should be less than 0.5, e.g. \cite{Petrovic2005,Kruckow2018MNRAS}.
Nevertheless, in our work we follow prescription by \cite{Vinciguerra2020}. We implement a semi-conservative mass transfer where the accretion rate is limited by:
\begin{equation}
\dot m_\mathrm{max} = \frac{1}{2} \dot m,    
\end{equation}
where $\dot m$ is the mass transfer rate set by donor expansion. 

Another factor which could affect the final orbital parameters of binary is the relation between NS mass and the mass of its progenitor. Thus, in one of the variations we use the fitting equations by \cite{Fryer2012} delayed explosion. We also modify the maximum mass of a new born NS from $1.8$~M$_\odot$ to $2.1$~M$_\odot$. In another variation we draw masses of NS from bimodal mass distribution found by \cite{bimodal_msp_mass} for millisecond radio pulsars,
implemented as follows: for 42.5~per~cent of all NSs we draw masses using normal distribution with mean $1.393$~M$_\odot$ and standard deviation of $\sigma_{m1} = 0.064$, for remaining NS we draw masses from another normal distribution with mean $1.807$~M$_\odot$ and standard deviation of $\sigma_{m2} = 0.178$. 
Note that in this procedure the NS mass is drawn randomly from the distribution, independent of its progenitor properties.

As for the natal kick, we consider seven different prescriptions. In our models A and B we assign to each newly born NS a natal kick drawn from Maxwellian velocity distribution (see e.g. eq. \ref{e:maxw}) with $\sigma=265$~km/s \citep{hobbs2005}. In model E we assume NS receives no natal kick. Some isolated radio pulsars still receive a small velocity due to Blaauw effect \citep{Blaauw1961}. 
In models C,D,F natal kick is chosen from two Maxwellians with $\sigma_1 = 56$~km/s for 20~per~cent of NS and with $\sigma_2 = 336$~km/s for remaining NSs \citep{Igoshev2020}. 

In our model G0, H0 and J0 we select helium giants (evolved stars stripped of hydrogen envelope) and check if their initial helium core mass is within the ecSN mass range. We assume that such stars undergo an ecSN, receiving weak natal kicks drawn from the Maxwellian velocity distribution with $\sigma = 30$~km/s. In these three models we assume that ecSN occurs exclusively in stars which lost their hydrogen envelope due to mass transfer in line with original idea by \cite{Podsiadlowski2004}, and more recently used by \cite{Willcox2021}.
In our models G1, H1 and J1 we also check if effectively isolated stars have helium core mass at the beginning of AGB in ranges as mentioned in Table~\ref{tab:models}. Such stars are also considered to undergo ecSN and receive a natal kick drawn from the Maxwellian velocity distribution with $\sigma = 30$~km/s.

In our optimal (statistical) model K, we vary the fraction of weak natal kicks $w$ and parameter of Maxwellian distribution $\sigma_1$ for weak natal kicks while keeping the parameter $\sigma_2$ for strong kicks fixed at value $\sigma_2 = 336$~km/s.
In our model L (physically motivated) we use the prescription of the ecSN from H1 (model producing enough Be X-ray binaries in the SMC) and assume that core-collapse SN receive single Maxwellian natal kick with parameter $\sigma = 336$~km/s while ecSN receive natal kick drawn from the Maxwellian velocity distribution with $\sigma = 45$~km/s.

To analyse the population synthesis simulation results, we proceed as follows. First, we keep track of the unbound NSs which formed in supernova explosion of the primary star. Primary star becomes isolated NS 17 times more frequently than the secondary star in our calculations, so we do not look at the velocities of secondaries which form NSs\footnote{We examine the outcome of models D and H0 and noticed that addition of secondaries NSs is hardly noticeable. We estimate that the histogram for isolated NS velocities changes at the level less than 0.5~per~cent.}. Second, we select binaries which, at some point of their evolution, have the NS primary and main sequence secondary more massive than $3$~M$_\odot$ (i.e. the potential Be X-ray binaries). 
We mark each of those binaries as:
\begin{itemize}
    \item Be X if it experienced a mass transfer from the primary to the secondary star on thermal timescale prior to SN explosion, see Figure~\ref{f:formationBeX_0.5} .
    \item CE if the binary experienced only CE stage prior to SN explosion. 
    \item Effectively single if the binary experienced no mass transfer prior to SN explosion.
    \item Mixed if the binary experienced multiple mass transfer episodes of different nature prior to SN explosion.
\end{itemize}
In order to find the fraction of currently observable systems with particular orbital parameters we follow the procedure by \cite{Vinciguerra2020}, see Appendix~\ref{s:orbitperiod}. We use the star formation history by \cite{rubele2015} to model the population of SMC and a constant star formation rate $1.65$~M$_\odot$/year for MW.
We also keep track of the systemic velocities for Be X-ray binaries. In our combined analysis, we use velocities of isolated NS and systemic velocities of Be X-ray binaries weighted with the star formation rate for the Milky Way.

\begin{figure*}
    \centering
    \begin{minipage}{0.40\linewidth}
    \includegraphics[width=0.99\columnwidth]{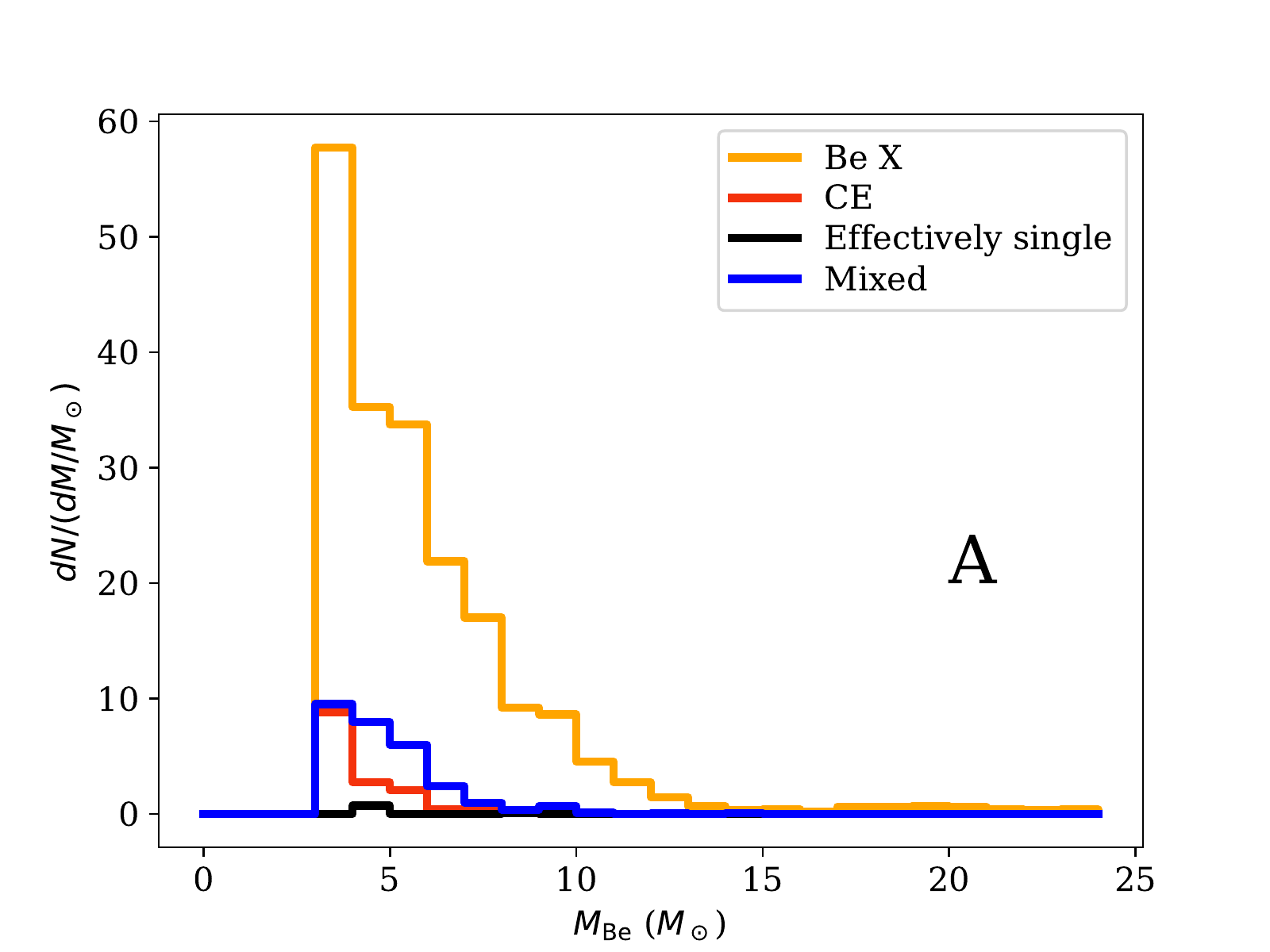}
    \end{minipage}
    \begin{minipage}{0.40\linewidth}
    \includegraphics[width=0.99\columnwidth]{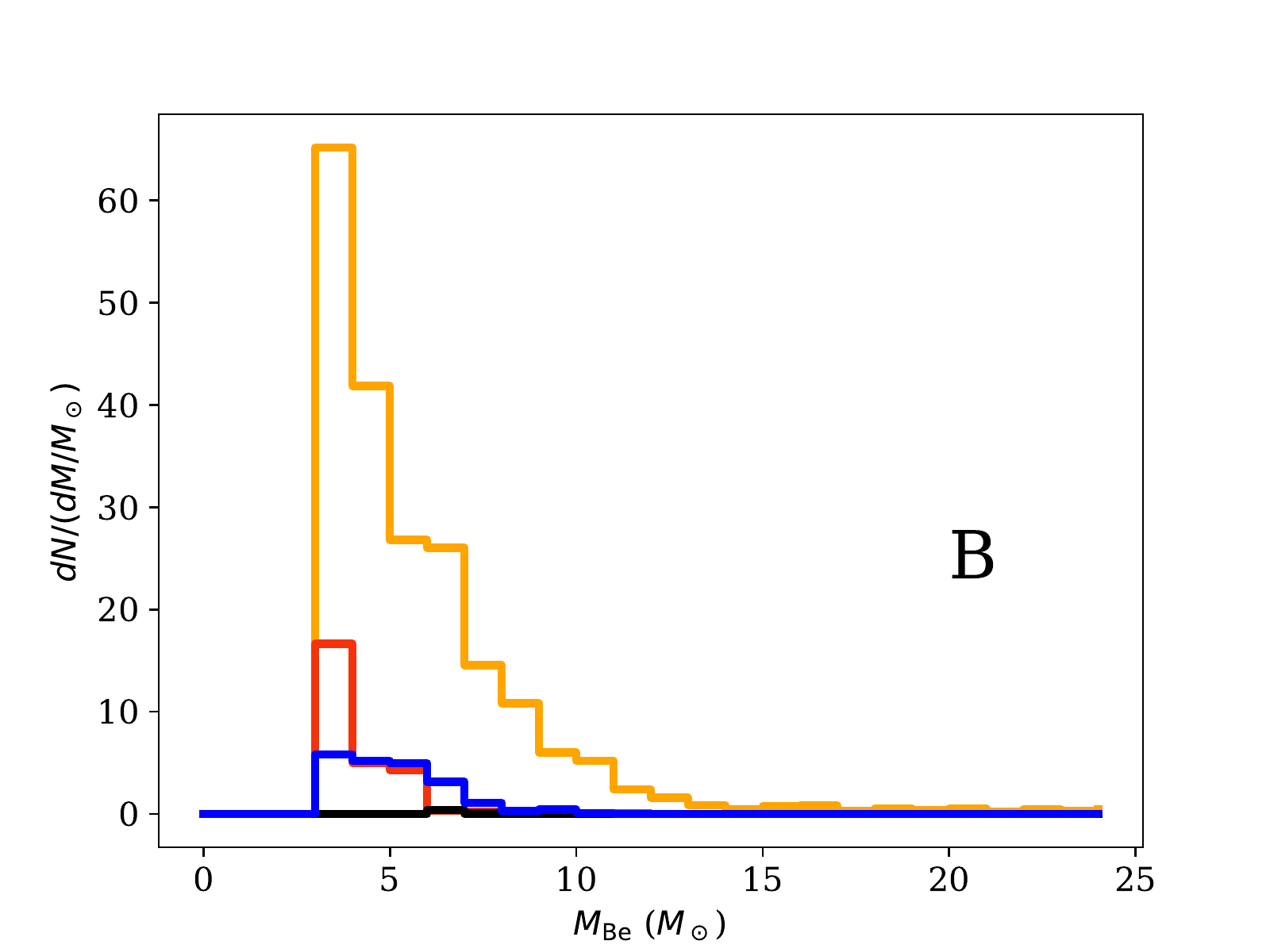}
    \end{minipage}
    \begin{minipage}{0.40\linewidth}
    \includegraphics[width=0.99\columnwidth]{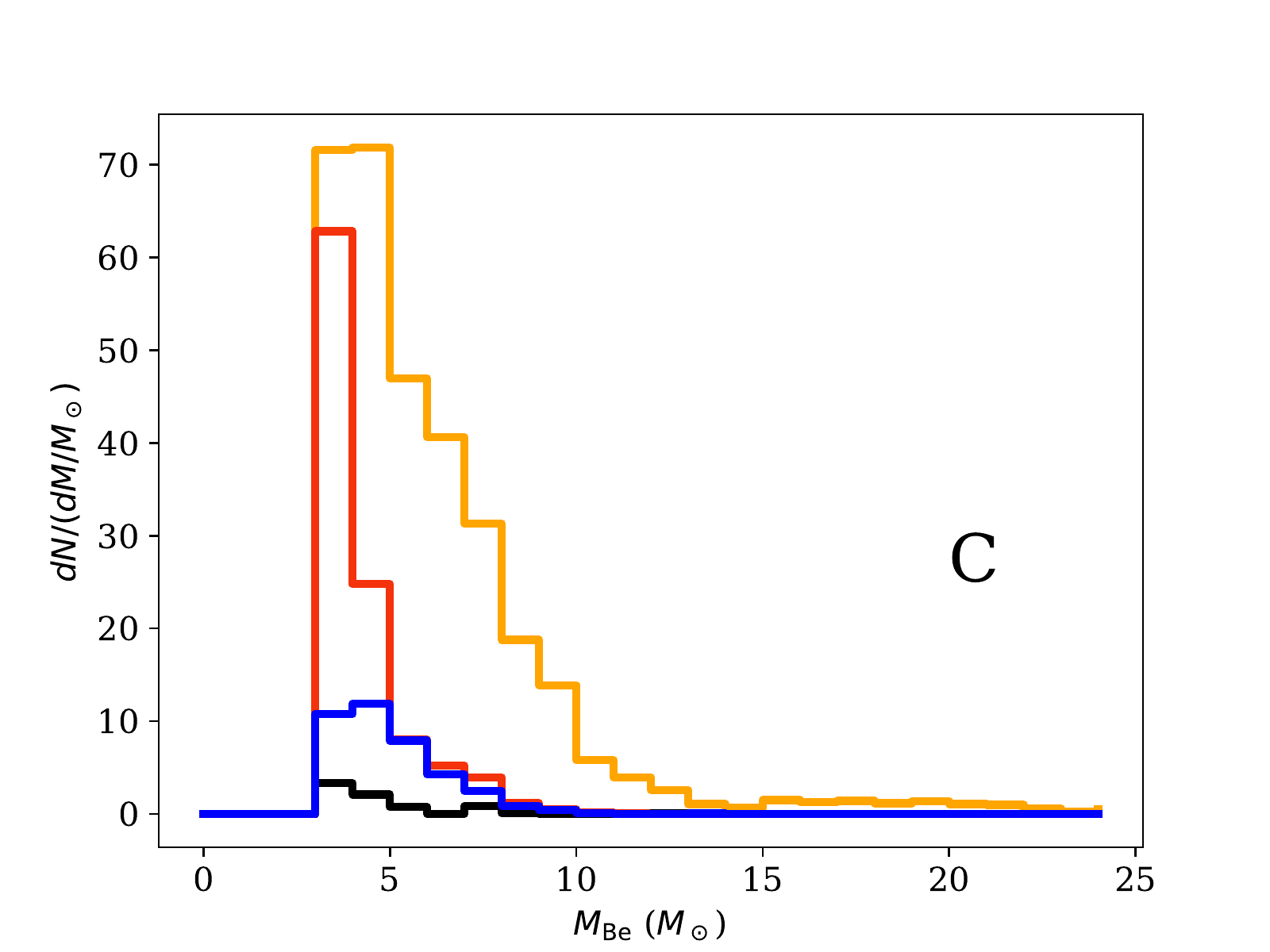}
    \end{minipage}
    \begin{minipage}{0.40\linewidth}
    \includegraphics[width=0.99\columnwidth]{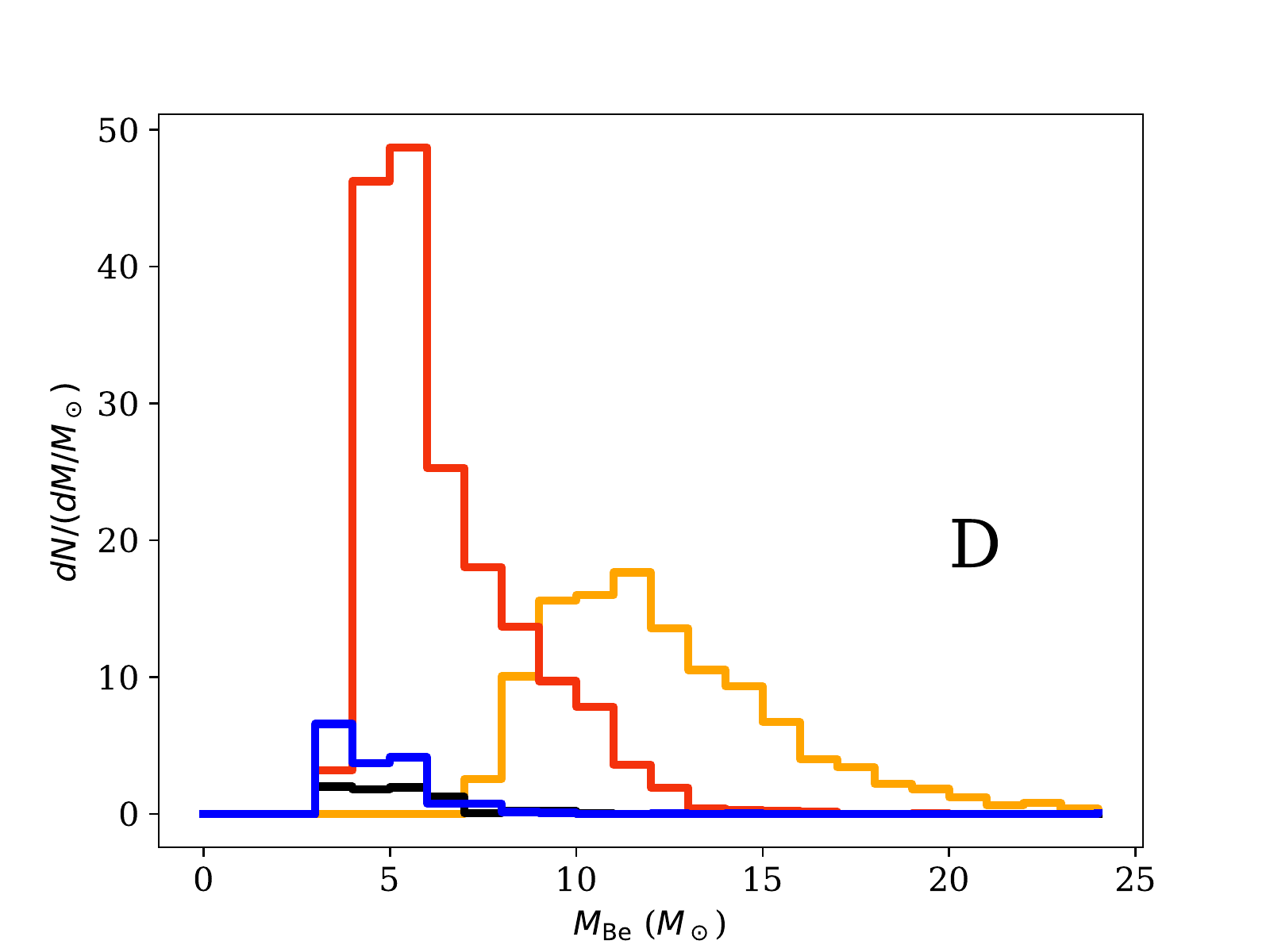}
    \end{minipage}
    \begin{minipage}{0.40\linewidth}
    \includegraphics[width=0.99\columnwidth]{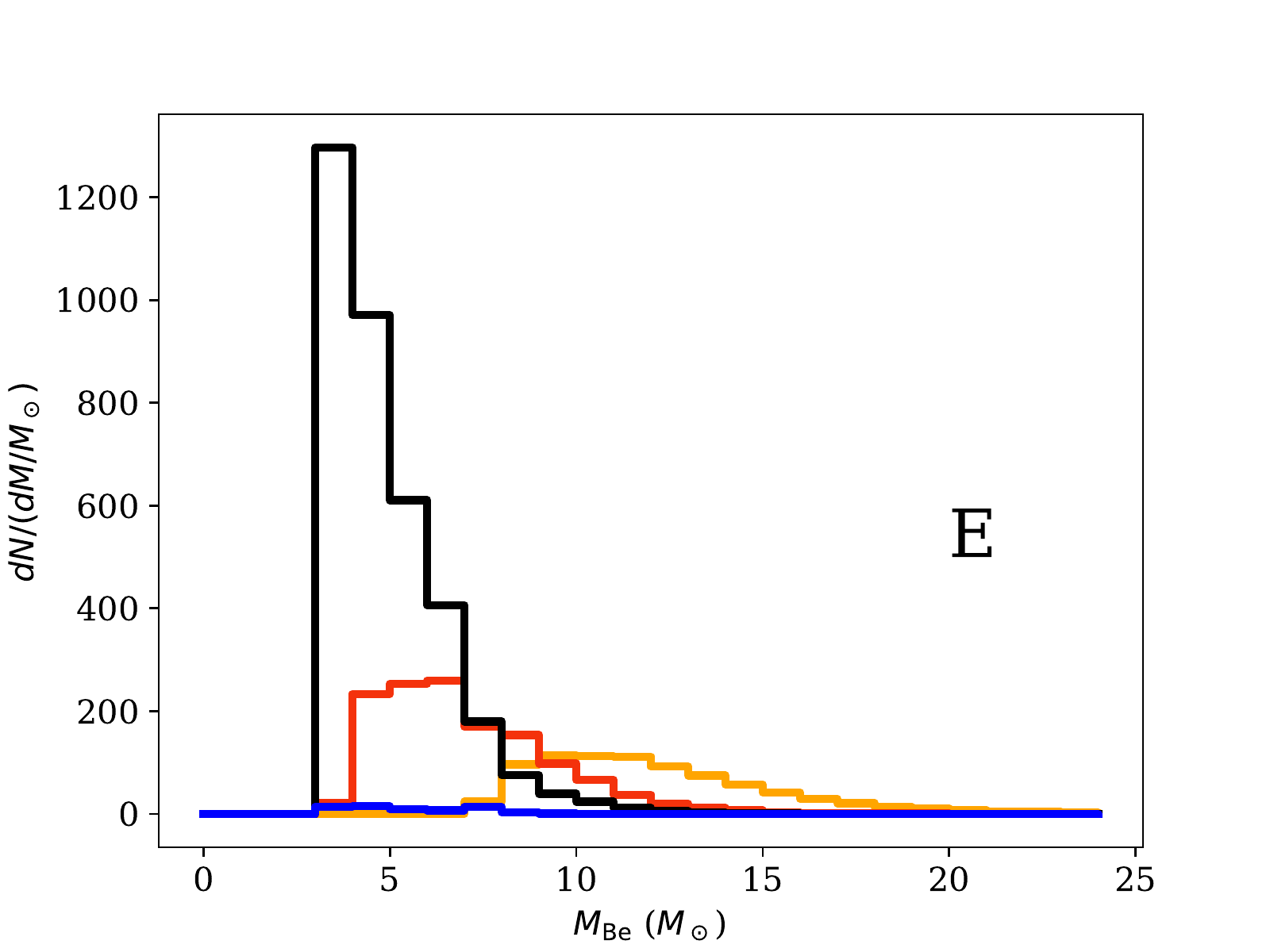}
    \end{minipage}
    \begin{minipage}{0.40\linewidth}
    \includegraphics[width=0.99\columnwidth]{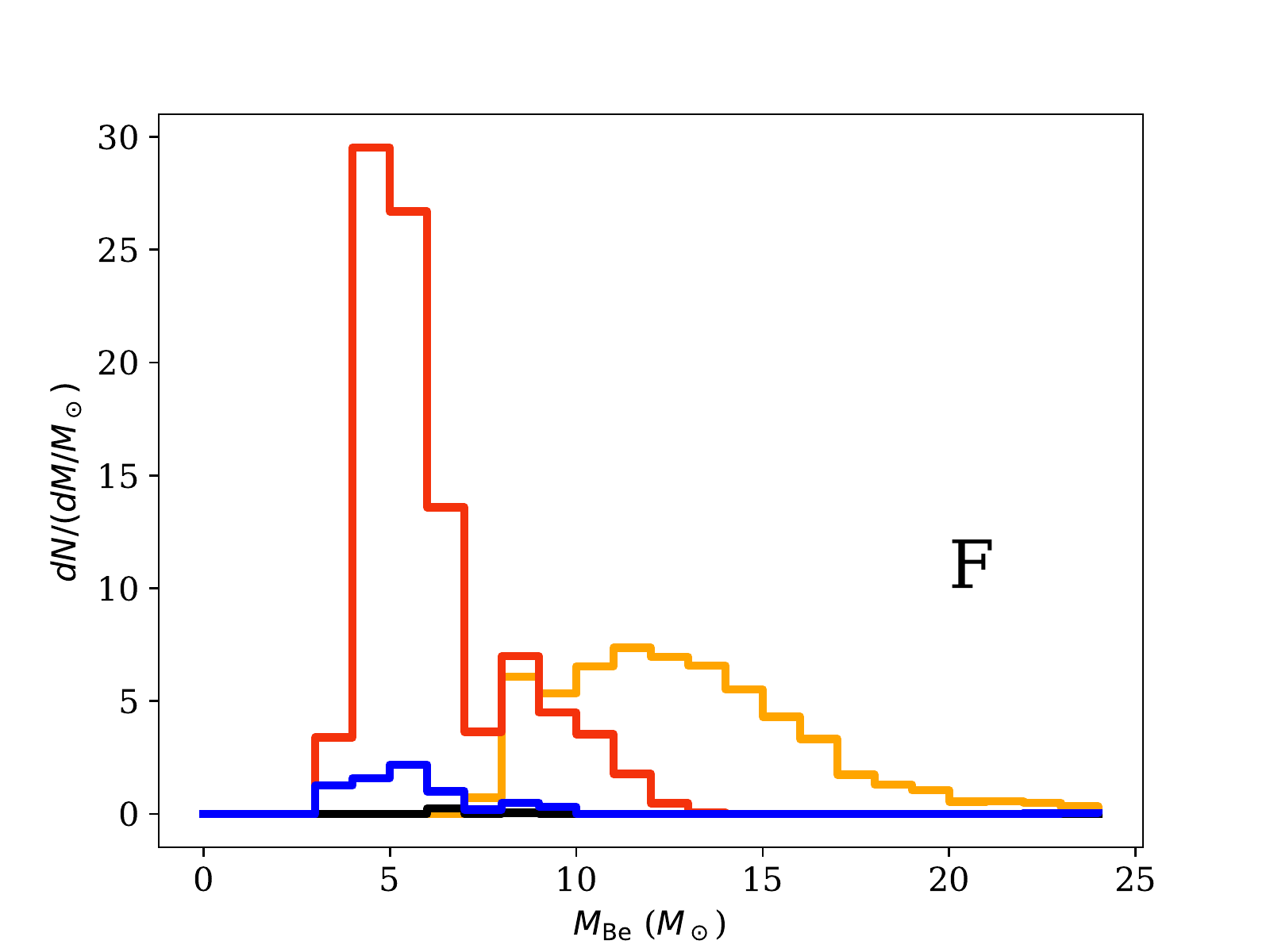}
    \end{minipage}
    \begin{minipage}{0.40\linewidth}
    \includegraphics[width=0.99\columnwidth]{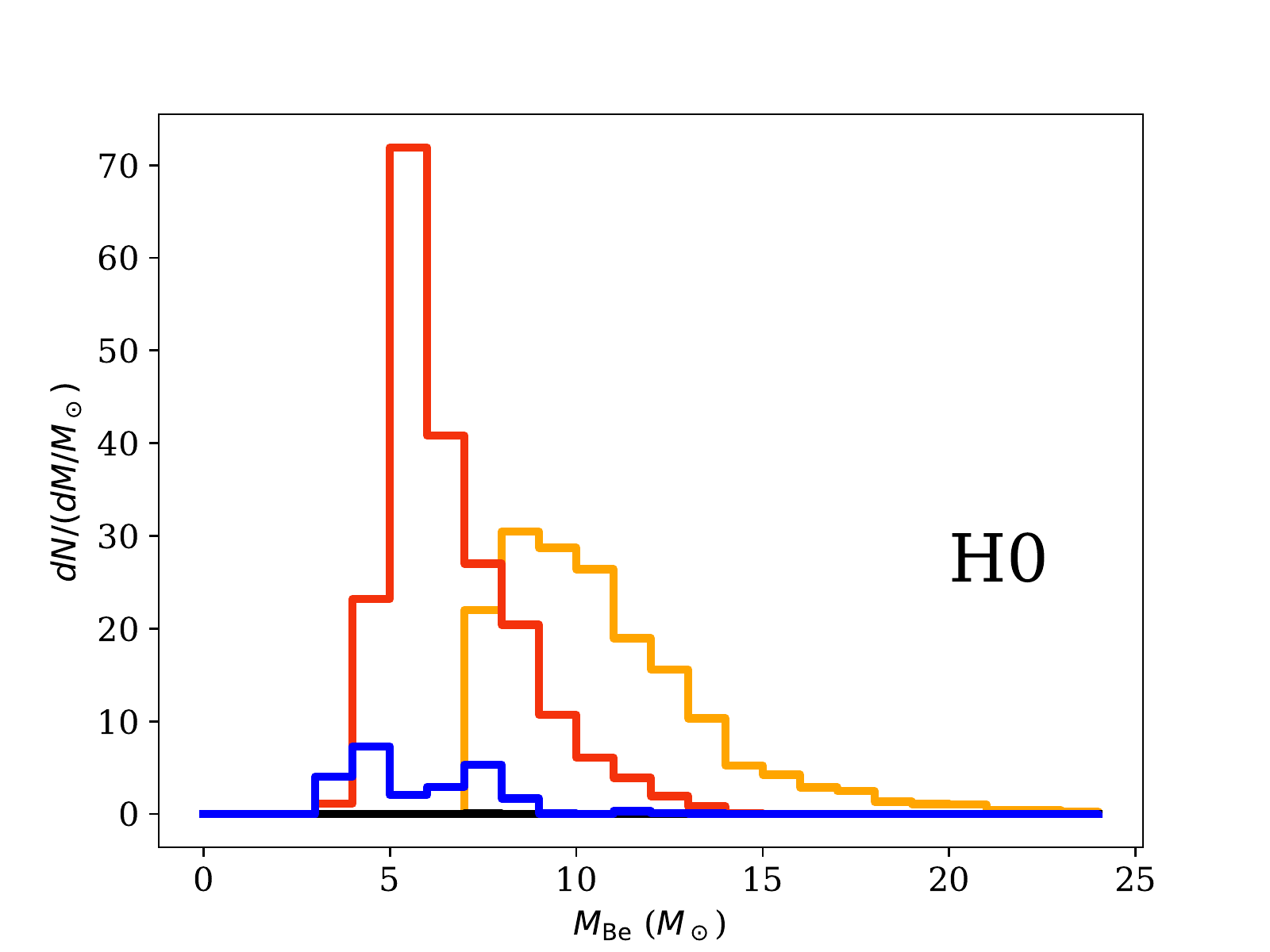}
    \end{minipage}
    \begin{minipage}{0.40\linewidth}
    \includegraphics[width=0.99\columnwidth]{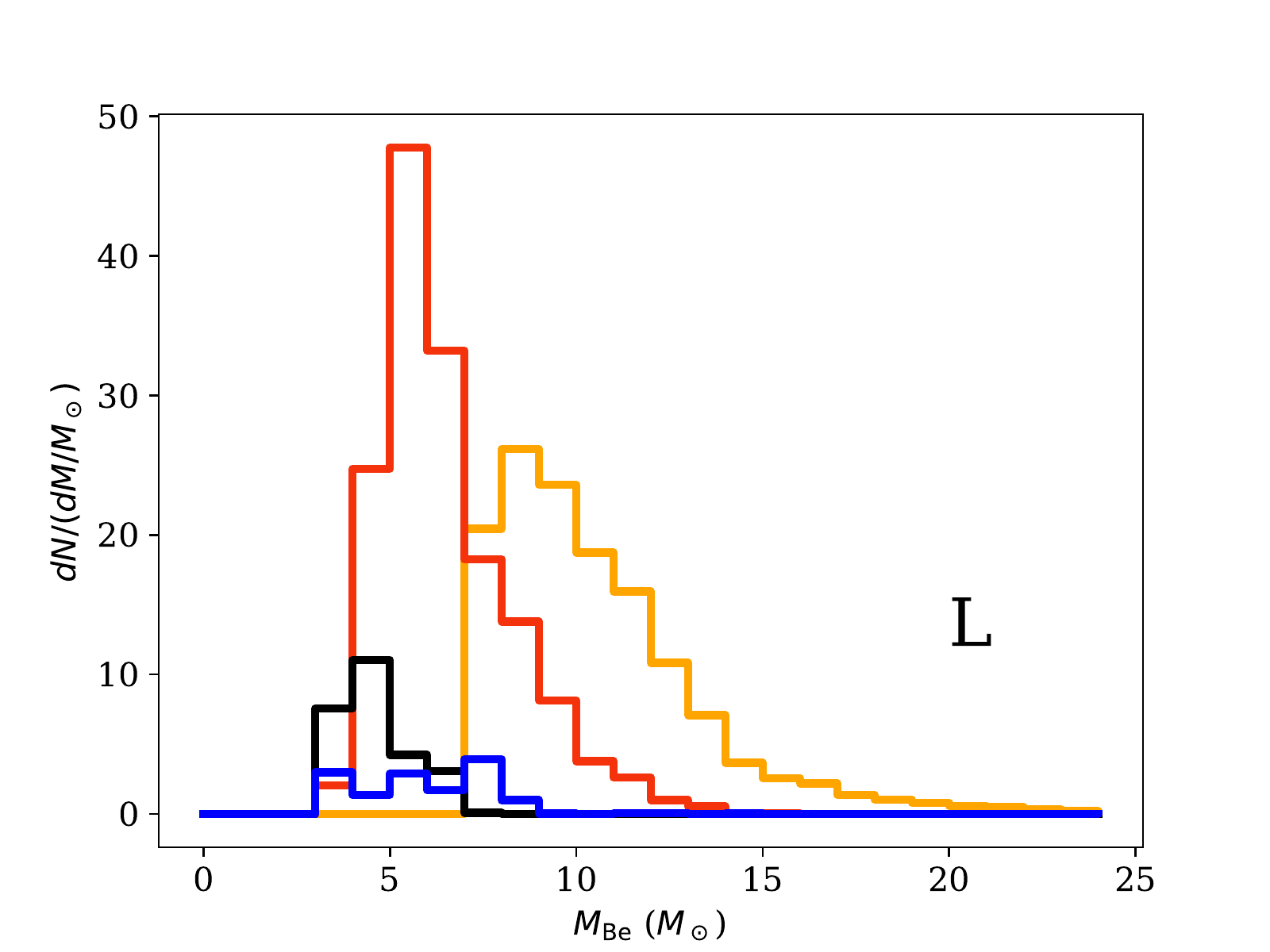}
    \end{minipage}
    \caption{Histogram for mass of the secondary stars for simulated systems for different models. We perform the simulation using the SMC star-formation history.
    Models A,B,C assume default SeBa mass transfer accretion efficiency prescription. Models D,E,F,H0 and L assume semi-conservative mass transfer accretion efficiency and provide a better match to the mass distribution of Be stars in Be X-ray binaries. Plots for the remaining models are available in the online supplementary materials.
    }
    \label{fig:be_mass_all_models}
\end{figure*}

\begin{figure*}
    \centering
    \begin{minipage}{0.44\linewidth}
    \includegraphics[width=0.99\columnwidth]{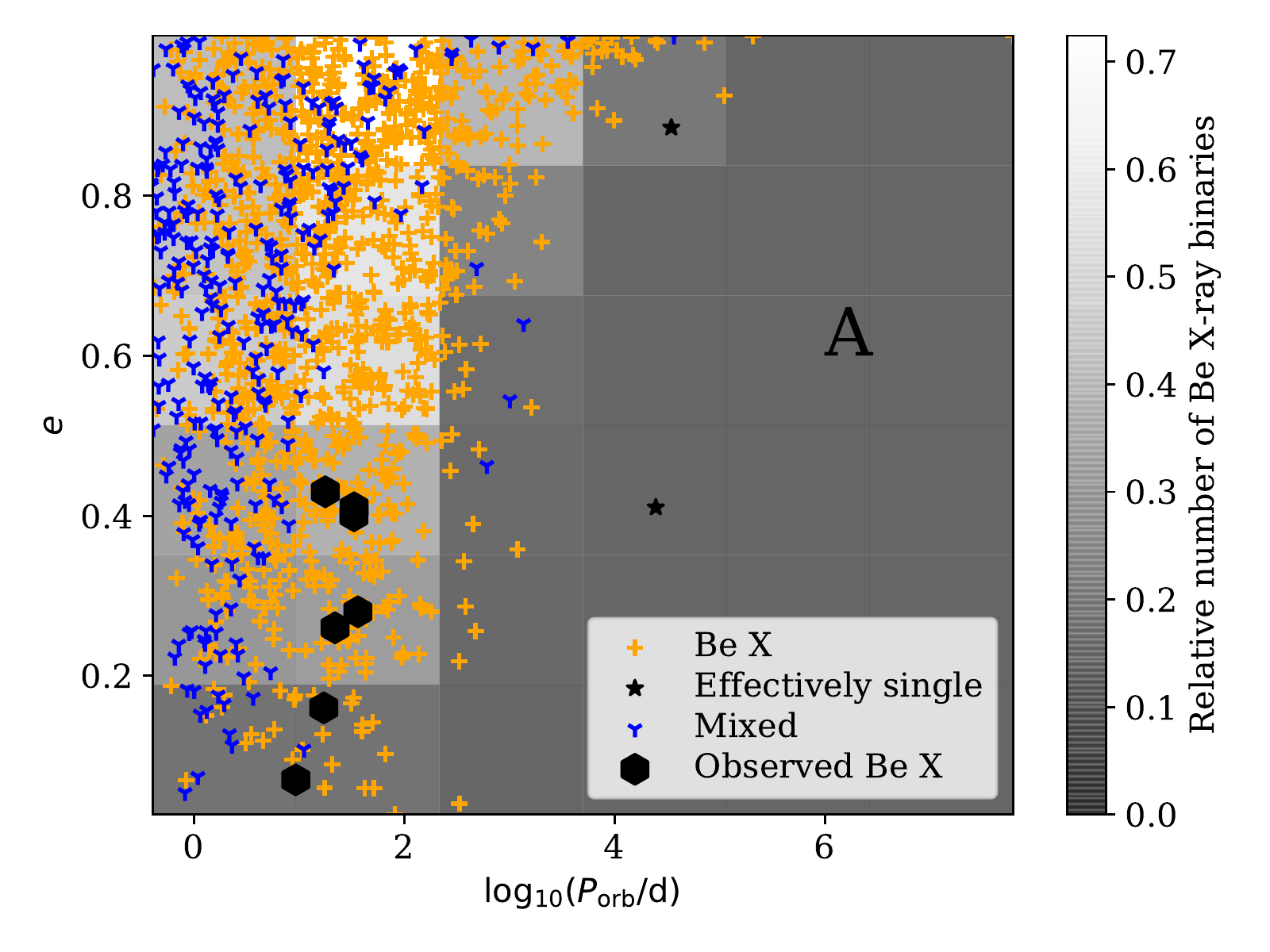}
    \end{minipage}
    \begin{minipage}{0.44\linewidth}
    \includegraphics[width=0.99\columnwidth]{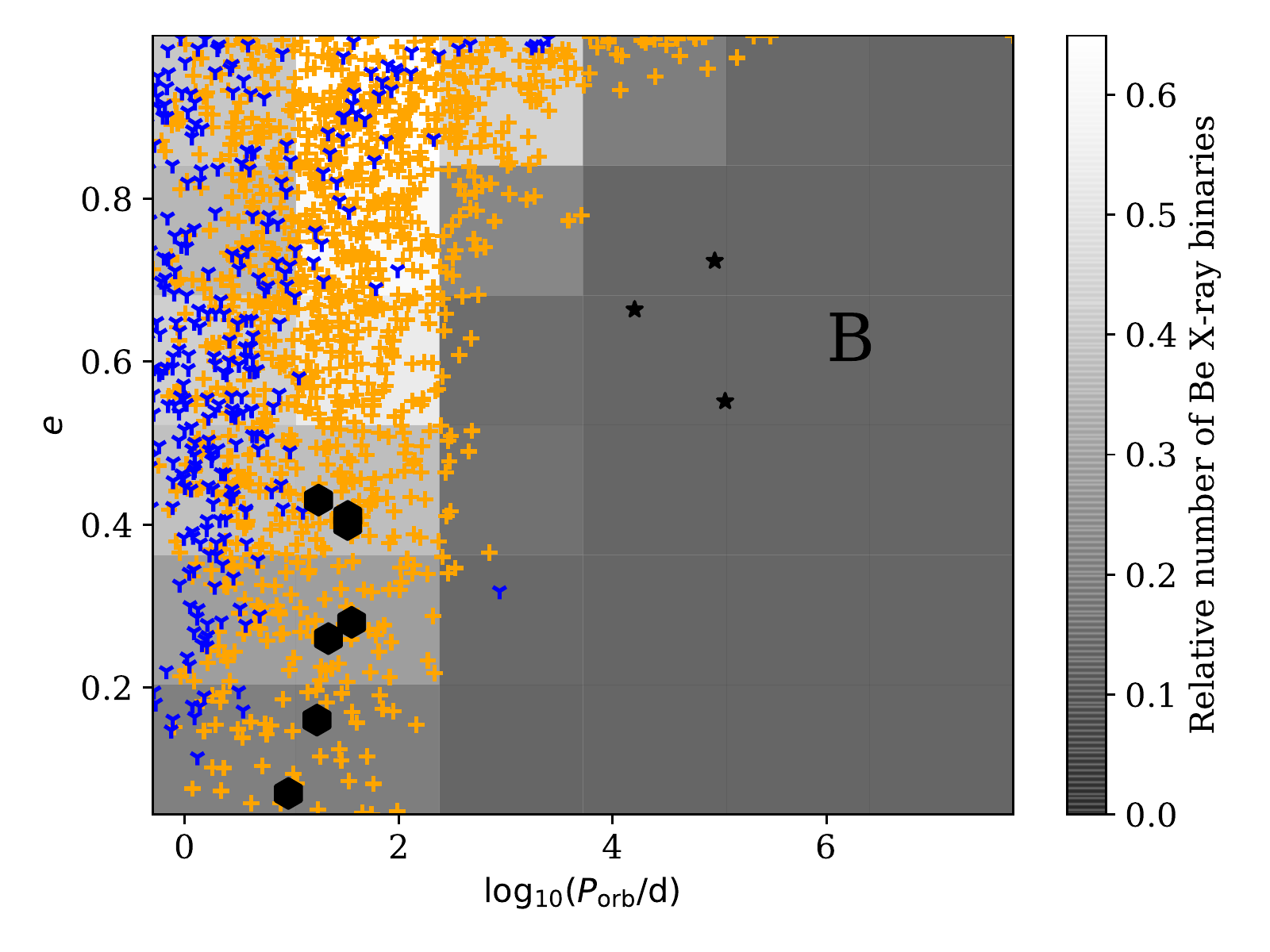}
    \end{minipage}
    \begin{minipage}{0.44\linewidth}
    \includegraphics[width=0.99\columnwidth]{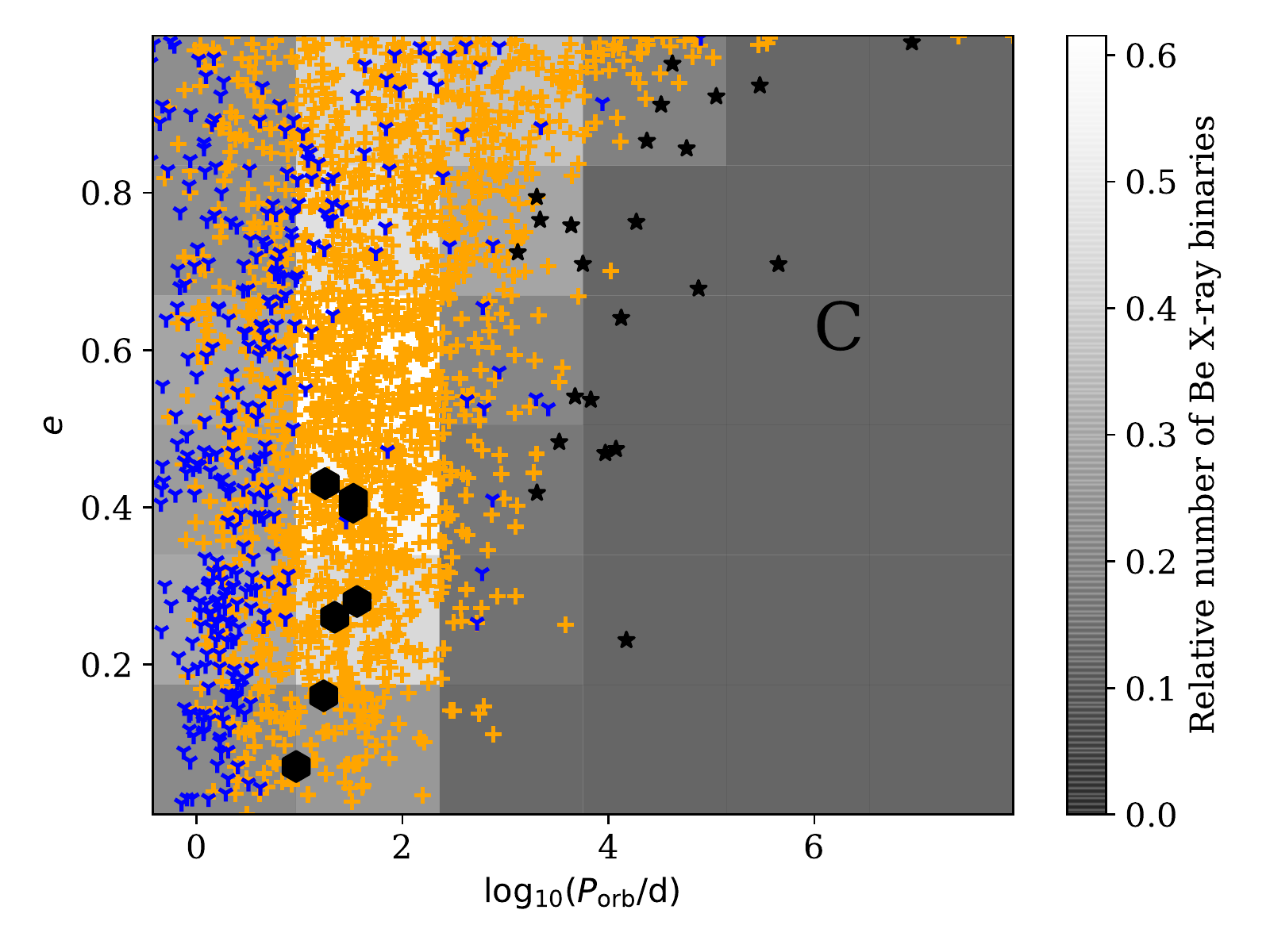}
    \end{minipage}
    \begin{minipage}{0.44\linewidth}
    \includegraphics[width=0.99\columnwidth]{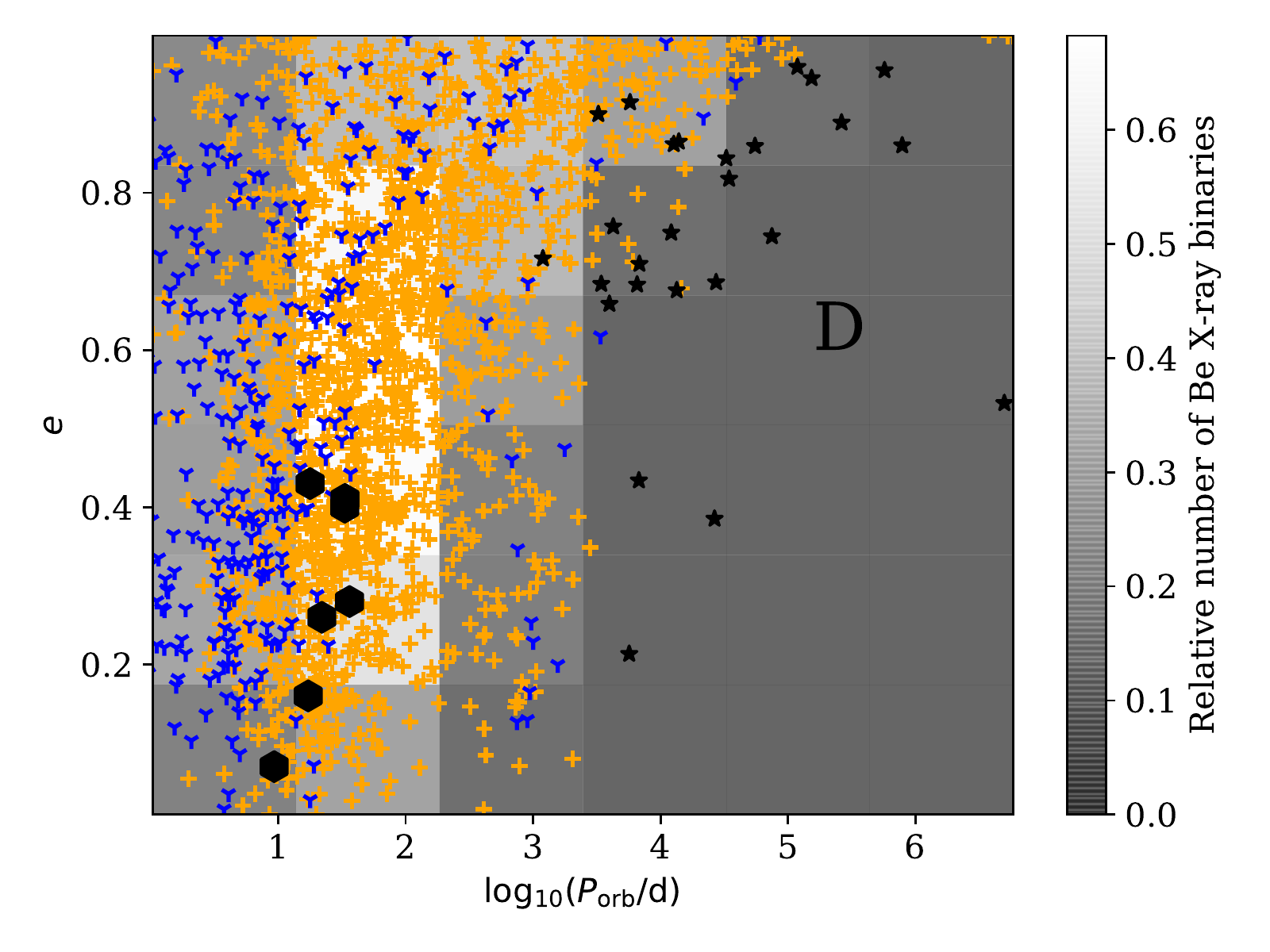}
    \end{minipage}
    \begin{minipage}{0.44\linewidth}
    \includegraphics[width=0.99\columnwidth]{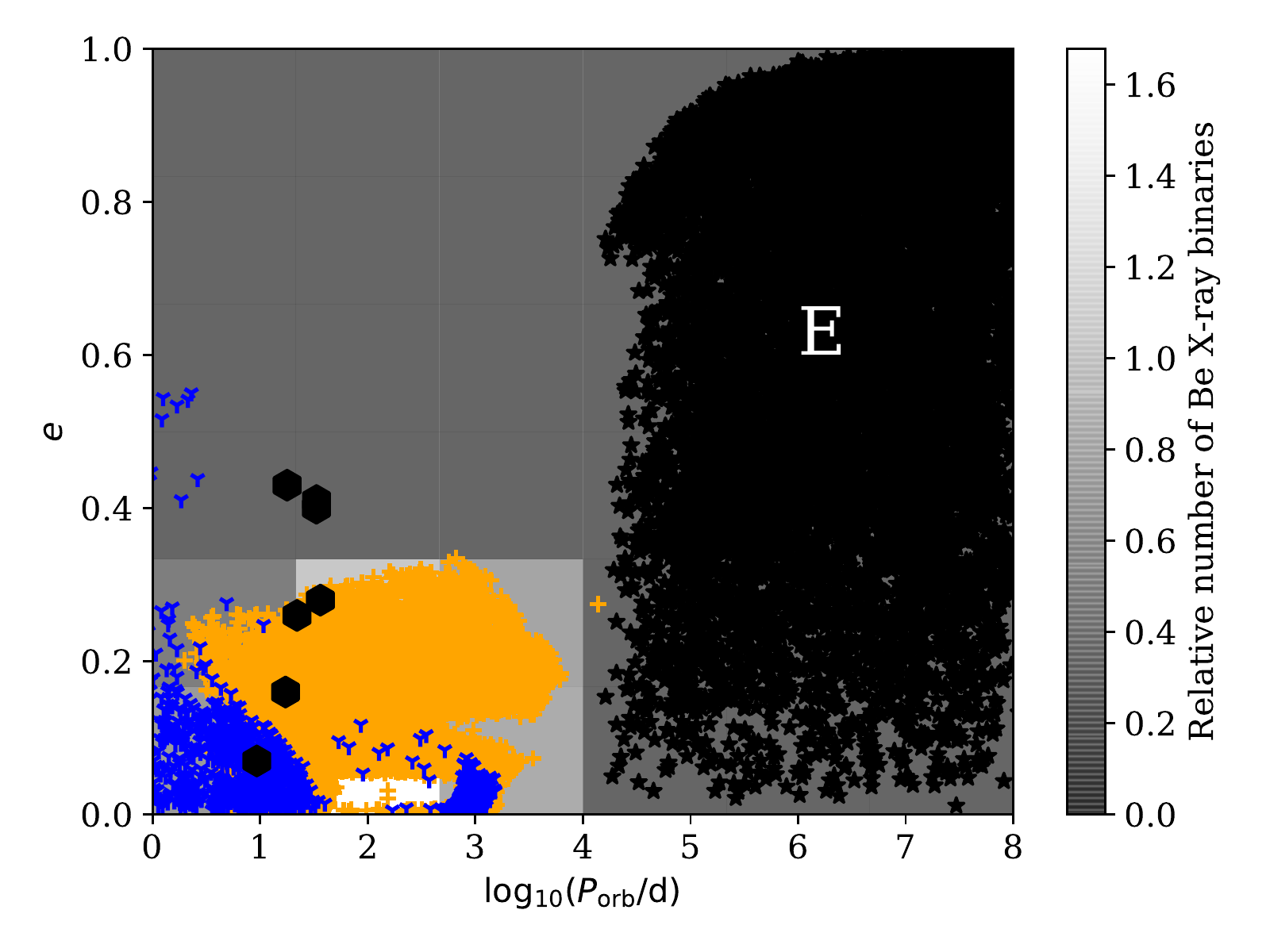}
    \end{minipage}
    \begin{minipage}{0.44\linewidth}
    \includegraphics[width=0.99\columnwidth]{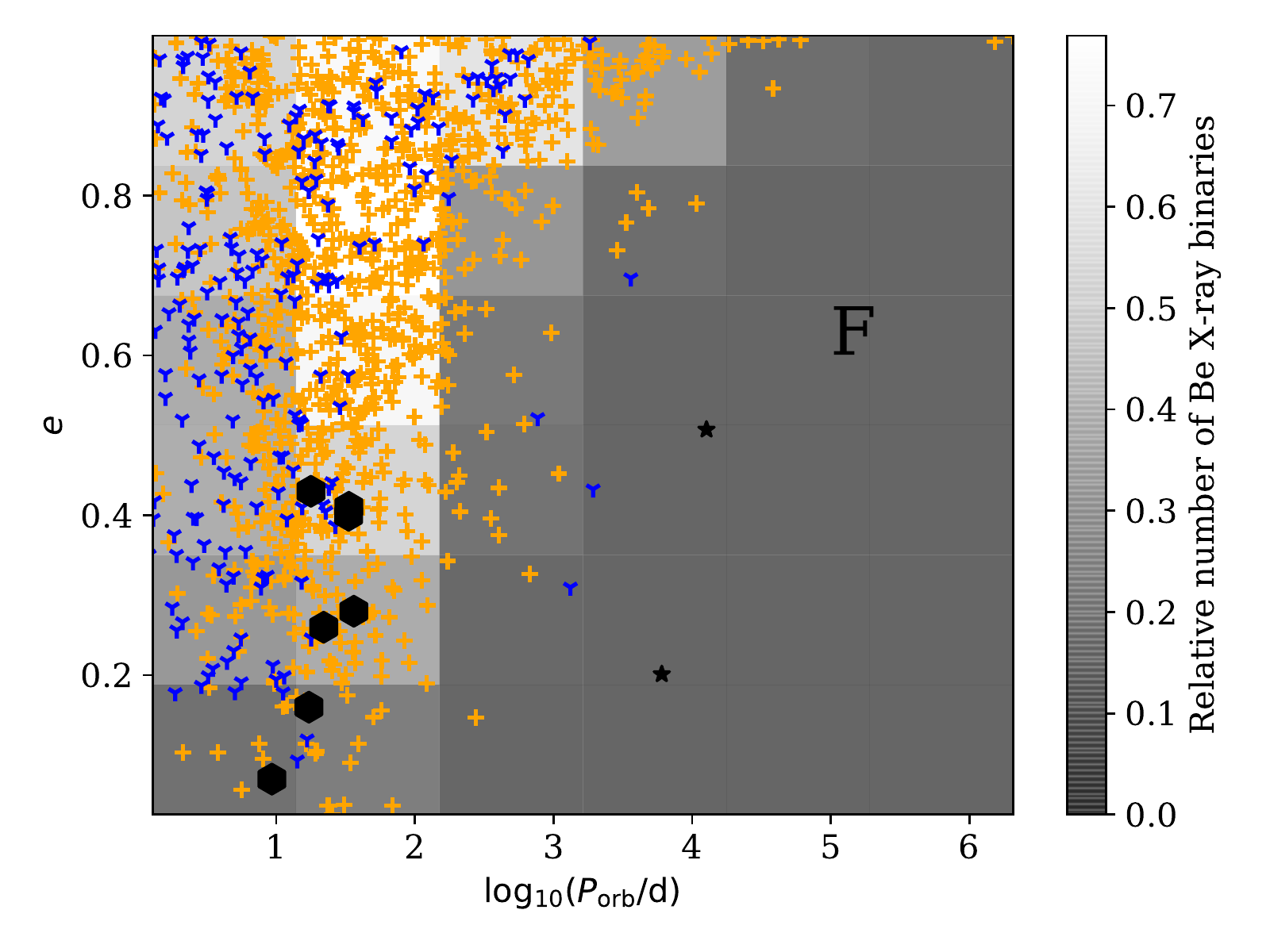}
    \end{minipage}
    \begin{minipage}{0.44\linewidth}
    \includegraphics[width=0.99\columnwidth]{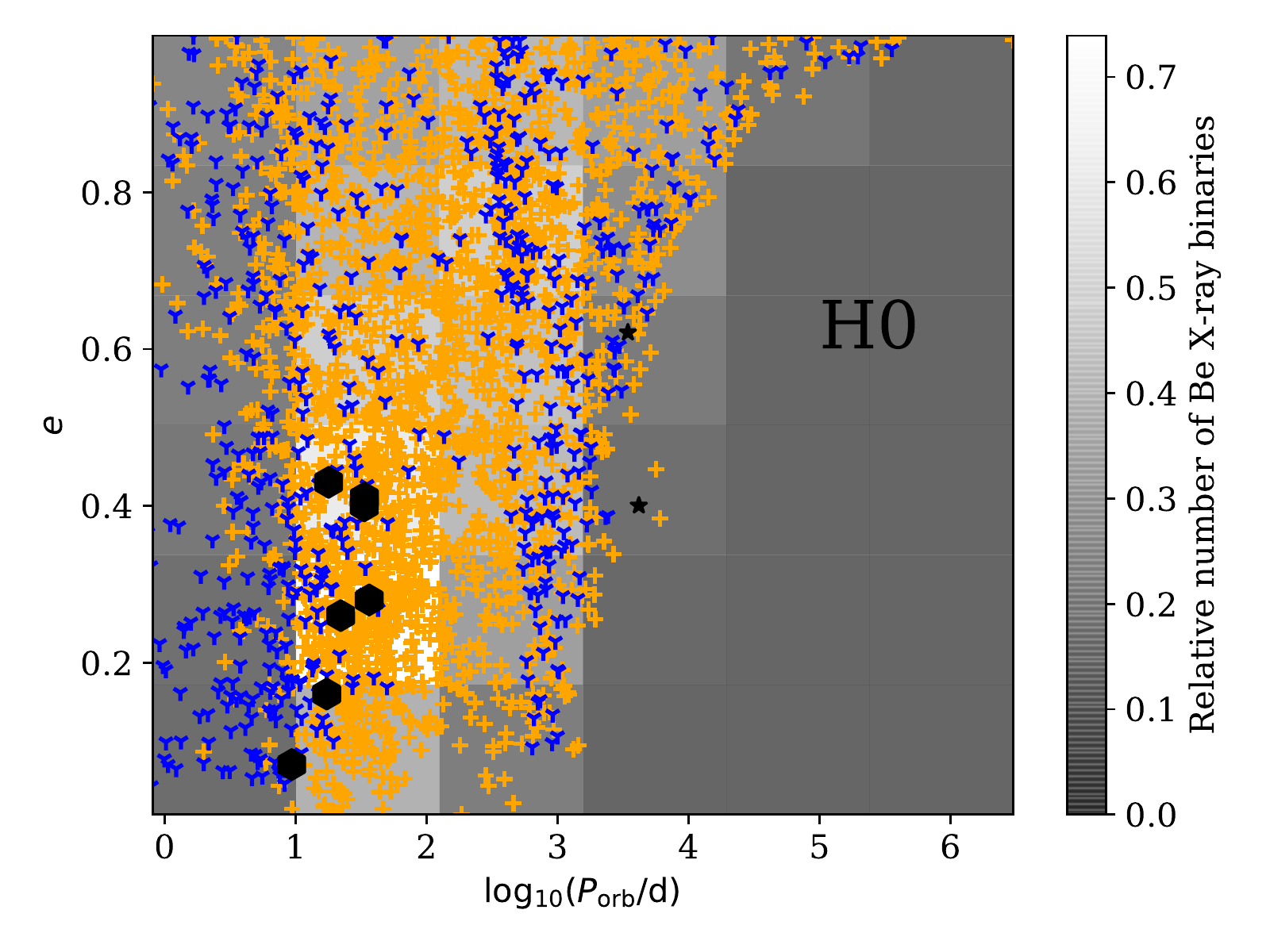}
    \end{minipage}
    \begin{minipage}{0.44\linewidth}
    \includegraphics[width=0.99\columnwidth]{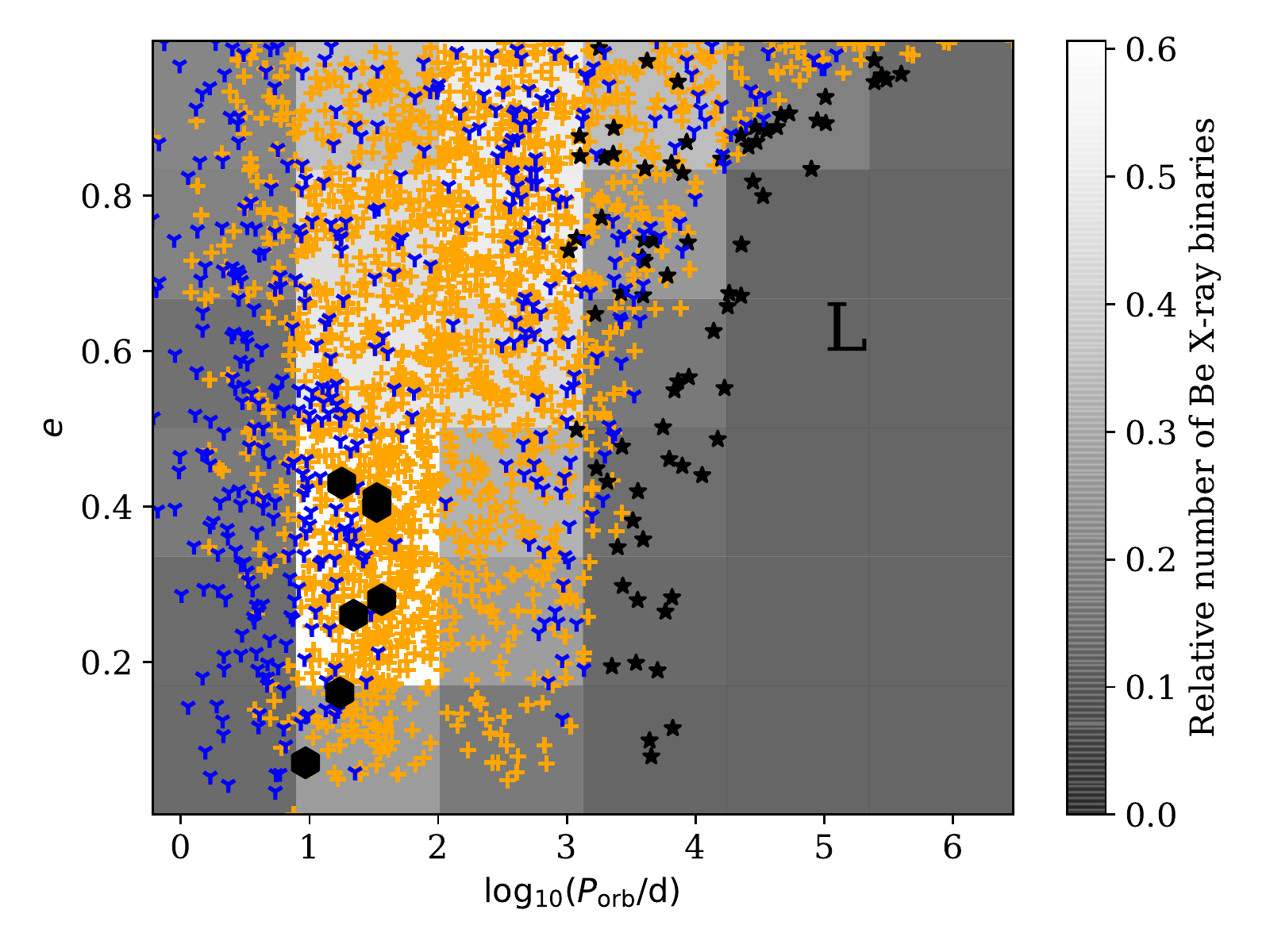}
    \end{minipage}
    \caption{Orbital periods and eccentricities for simulated binary systems
     for different models. Black hexagons are locations of observed Be X-ray binaries in SMC by \protect\cite{Coe2015}. Histogram shows density of Be X-ray systems only. Plots for the remaining models are available in the online supplementary materials.
    }
    \label{fig:pe_mass_all_models}
\end{figure*}

\begin{figure*}
    \begin{minipage}{0.49\linewidth}
	\includegraphics[width=0.99\columnwidth]{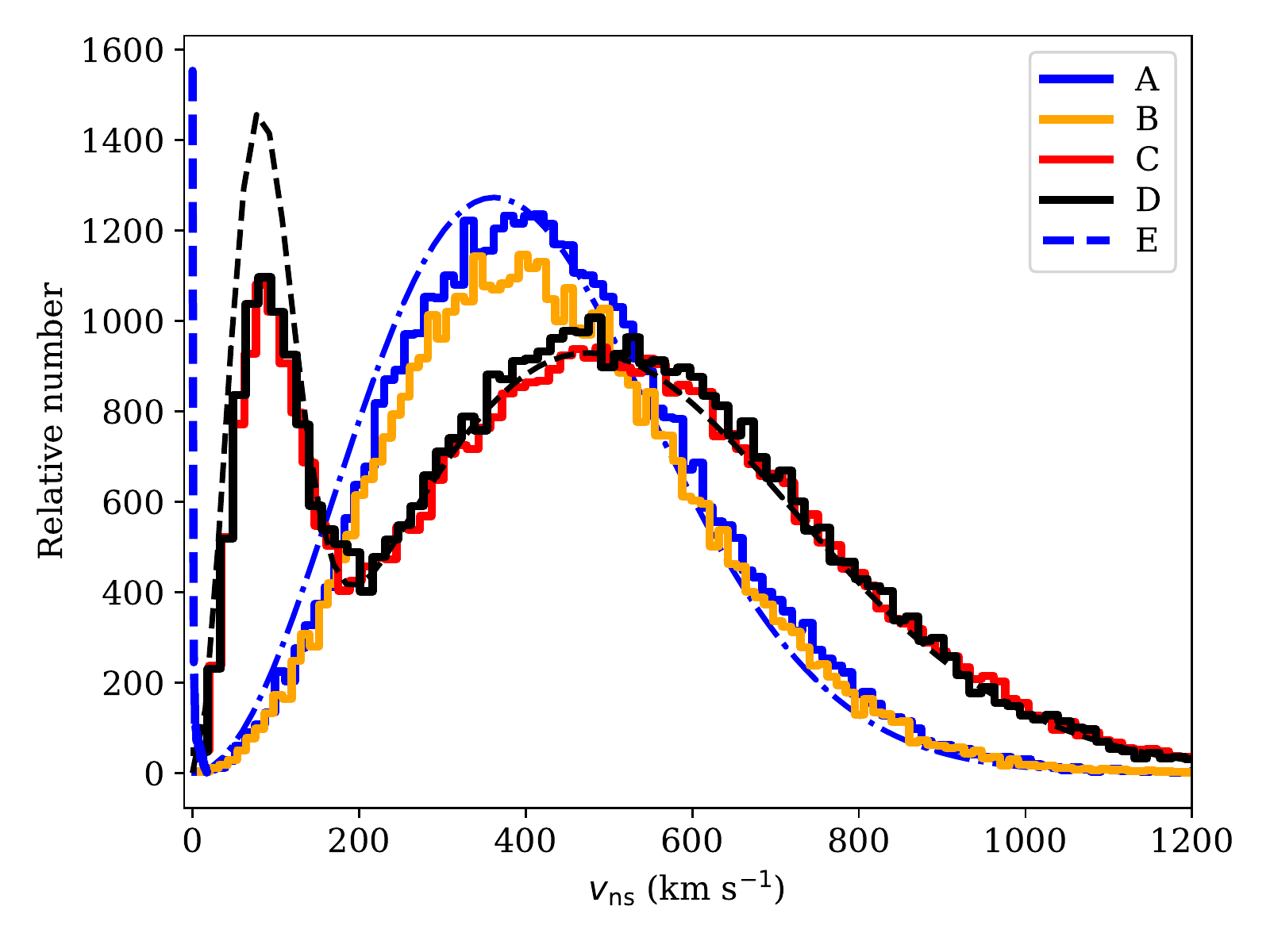}
	\end{minipage}
    \begin{minipage}{0.49\linewidth}
	\includegraphics[width=0.99\columnwidth]{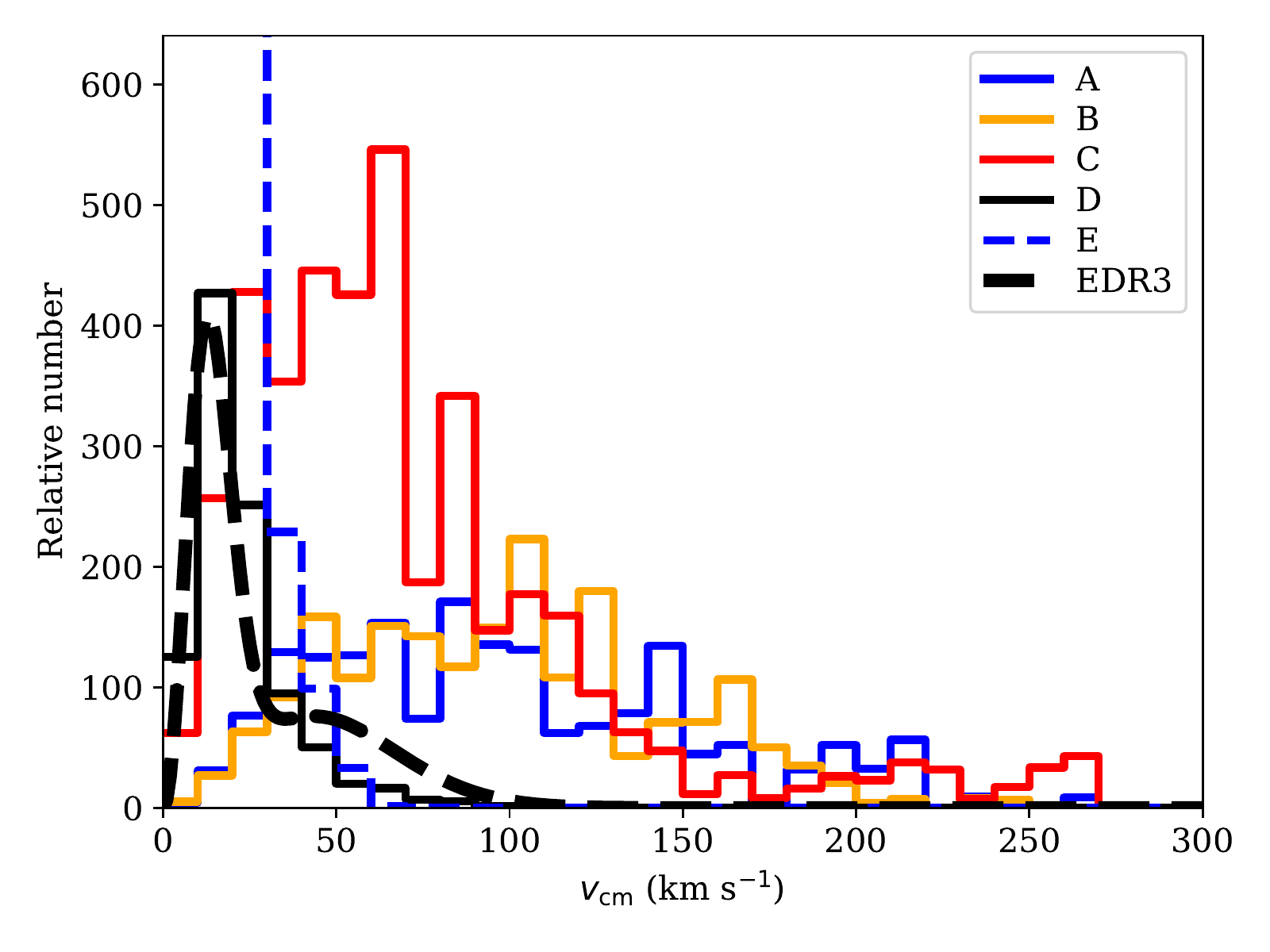}
	\end{minipage}	
	\begin{minipage}{0.49\linewidth}
	\includegraphics[width=0.99\columnwidth]{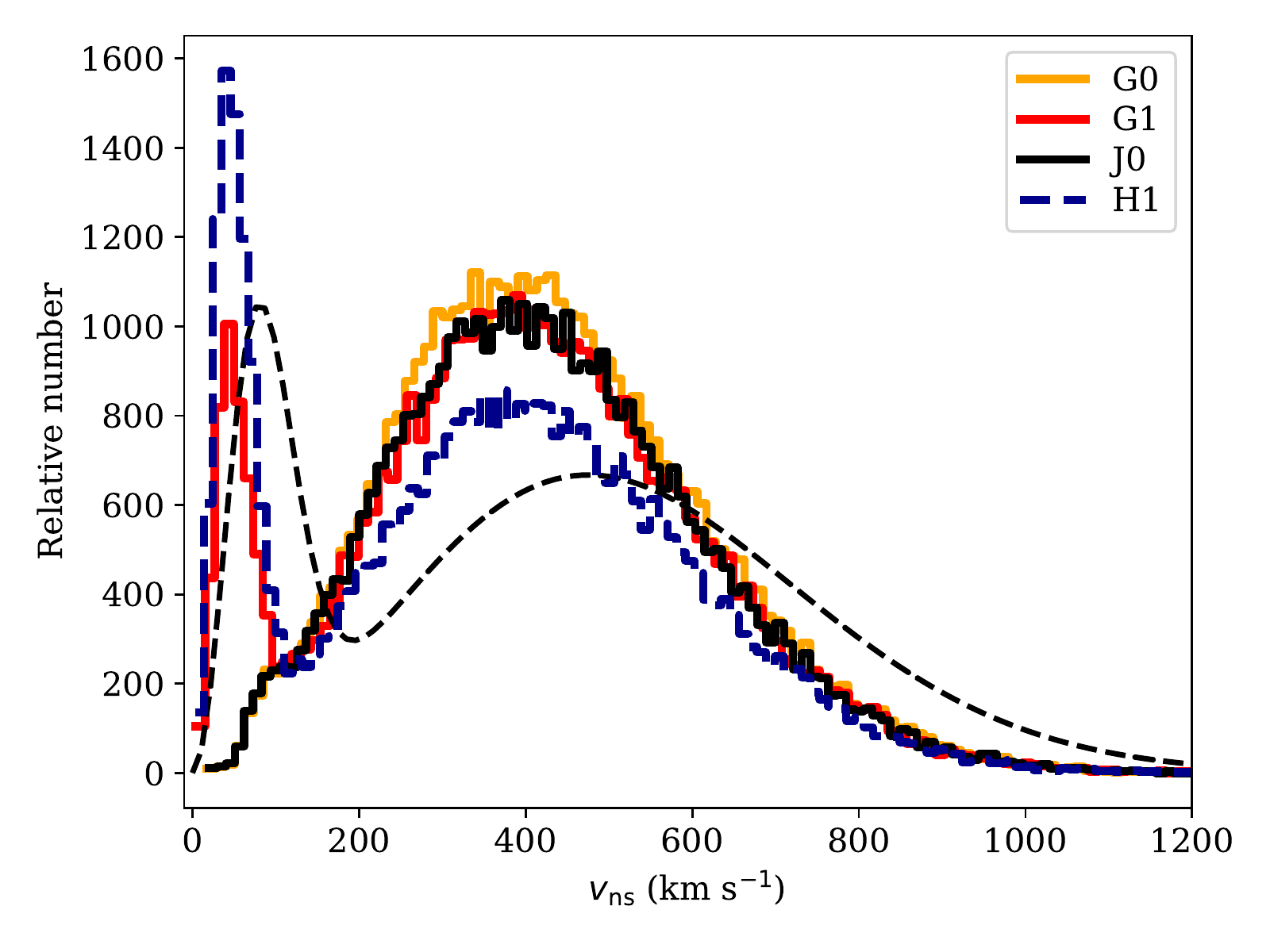}
	\end{minipage}
	\begin{minipage}{0.49\linewidth}
	\includegraphics[width=0.99\columnwidth]{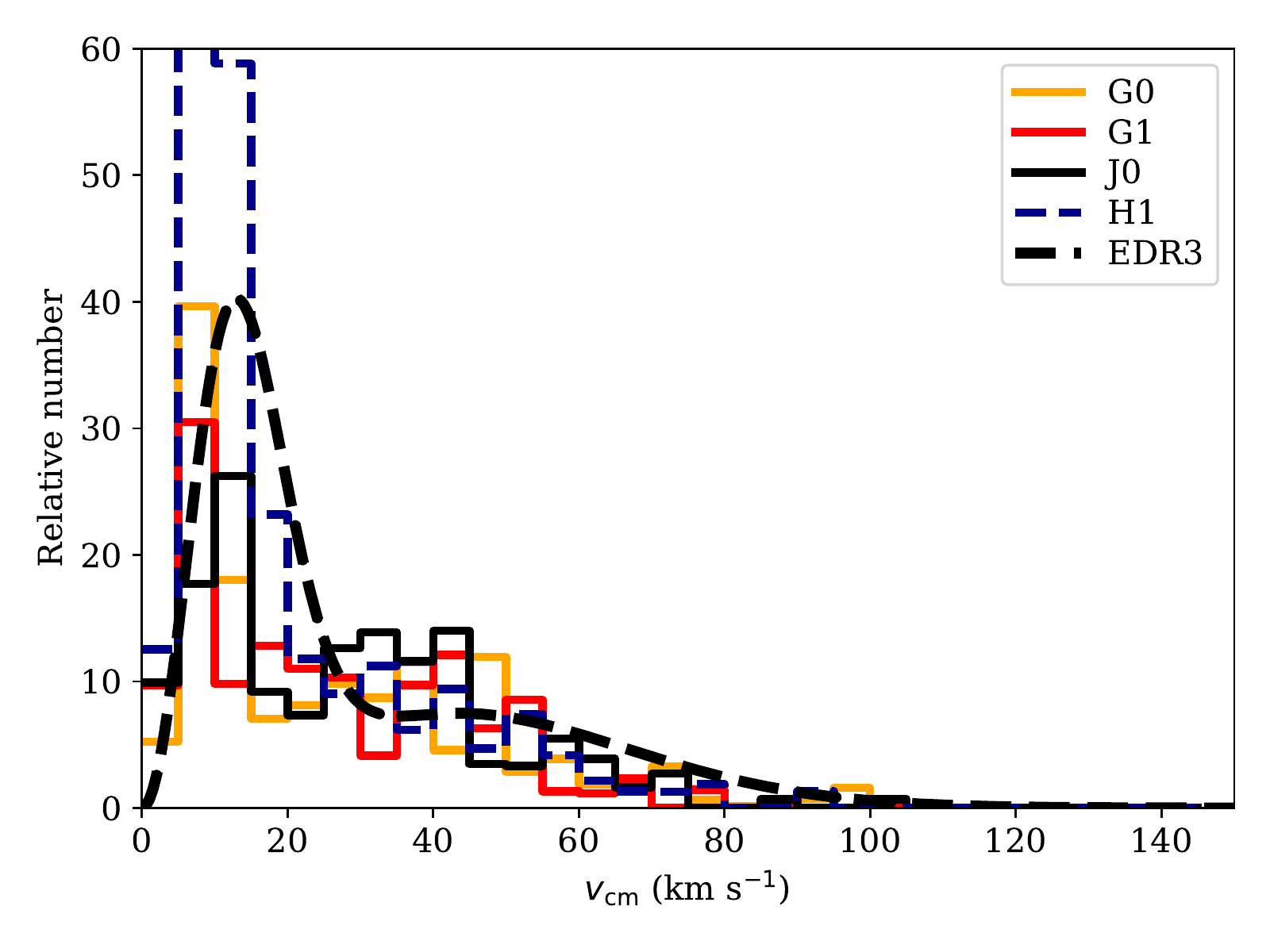}
	\end{minipage}	
    \caption{
    Left panels: the velocity distribution of synthetic isolated NSs (histograms) compared with natal kick distributions by \protect\cite{hobbs2005} (dot and dashed blue line) and \protect\cite{Igoshev2020} (dashed black line). The low-velocity peak of the distribution \protect\cite{Igoshev2020} is noticeably higher than the same peak of the histogram for model D because a large fraction of these NS stay bound in binaries. The SFH corresponds to Milky Way. Only a fraction of 5-7~per~cent of these systems could be ever discovered in radio surveys because of the beaming. Right panels: the peculiar velocity distribution of Be X-ray binaries obtained in simulations and restored based on Gaia EDR3 observations (black dashed line; normalisation is chosen arbitrarily). 
    Top panels: models without ecSN. Bottom panels: models with ecSN.
    It is worth noticing that model G1 (ecSN occurs also in effectively single stars) produces significantly more low-velocity isolated NSs in the velocity range of 10-100~km~s$^{-1}$ in comparison to model G0 (ecSN occurs only in stripped stars).  
    Plots for the remaining models are available in the online supplementary materials.
    }
    \label{fig:kick_disrupted_AD}
\end{figure*}

\section{Results: exploratory models}
\label{s:all_results}

With our exploratory model we investigate the influence of four factors on formation of Be X-ray binaries in SMC (see Table~\ref{tab:models}). These factors are as following: (1) natal kick distribution, (2) dependence between remnant mass and progenitor mass, (3) mass accretion efficiency during the mass transfer from primary to secondary, and (4) different natal kick for electron capture supernova explosion.

\subsection{Natal kicks}

Our models A and C are identical, except for the natal kick prescription, which are described following \cite{hobbs2005} or \cite{Igoshev2020} respectively. 
Figure~\ref{fig:be_mass_all_models} shows that the natal kick has little effect on the Be star mass distribution.
However, it significantly affects the distribution of eccentricities (see Figure~\ref{fig:pe_mass_all_models}). 
In case of model A, the peak of the distribution is shifted towards higher eccentricities ($\sim$0.9 compared to $\sim$0.5 in model C) and the low eccentricity systems - prevailing in the observed sample, are rare.
However, the limited size of the observed Be X-ray binary sample with measured eccentricities and potential biases (see Sec. \ref{s:selection effects}) do not allow for any strong conclusions from this comparison.
The natal kick prescription affects the total number of Be X-ray binaries. In particular, in SMC, the prescription by \citealt{Igoshev2020} produces 1.6 time more Be X-ray binaries than with the \cite{hobbs2005} prescription. The natal kick prescription affects the distribution of measured velocities for both Galactic isolated radio pulsars and Be X-ray binaries (see Figure~\ref{fig:kick_disrupted_AD}). The natal kick in form of \cite{hobbs2005} produces multiple Be X-ray binaries with systemic velocities above 100~km/s independently of assumptions about mass transfer.

We therefore restrict our further analysis to the discussion of the distributions of velocities of young radio pulsars and systemic velocities of Be X-ray binaries in the Galaxy.
Those distributions are very sensitive to the adopted natal kick prescription.

Comparing models D (natal kick distribution following \citealt{Igoshev2020}) and E (no natal kick), one can notice that in the no-kick scenario the size of surviving Be X-ray binary population is significantly increased (SMC could have up to 820 Be X-ray binaries)
and that Be X-ray binaries with NS that formed with no kick have eccentricities below 0.3. However, if all NS are formed with zero natal kick, the velocity distribution of radio pulsars is indistinguishable from the velocity distribution of massive stars (peaking at $1-2$~km/s and showing a sharp decay at 5~km/s), which is clearly incompatible with observations.


\subsection{Dependence between remnant mass and progenitor mass}

Our models A and B are identical, except for how we assign the NS mass depending on the mass of the progenitor. Figures~\ref{fig:be_mass_all_models} and \ref{fig:pe_mass_all_models} show that the distributions of Be masses and orbital elements are barely affected by the choice of prescription for the NS mass. The total number of Be X-ray binaries in SMC is also comparable for both prescriptions. The exact choice of this dependence does not affect the distribution of isolated pulsar velocities and weakly affects the distribution of Be X-ray velocities, see Figure~\ref{fig:kick_disrupted_AD}.
Therefore, we do not discuss the implications of alternative choice for relation between mass of the remnant and progenitor. 

\subsection{Mass accretion efficiency}\label{s:accretion efficiency}

Our models C and D are identical, except for the mass accretion efficiency.
We confirm conclusions of \cite{Vinciguerra2020}, showing that the mass accretion efficiency is a key variable influencing the Be star mass distribution of the simulated Be X-ray binaries. 
Using the default SeBa mass accretion efficiency prescription (model C), we find that the Be star mass distribution peaks at $4-5$~$M_\odot$ - which is incompatible with observations.  
In model D, with (fixed) semi-conservative mass accretion efficiency
the distribution of Be star masses peaks at $10-11$~M$_\odot$, more in line with the observed distribution (see Figure~\ref{fig:be_mass_all_models}). 
Those conclusions are insensitive to the adopted natal kick prescription.

The mass accretion efficiency also controls the total number of Be X-ray binaries in the SMC. For example, model F, which differs from model A only by prescription of this parameter, predicts 60 Be X-ray binaries while model A predicts 200. 
Figure \ref{fig:be_mass_all_models} shows that fixed semi-conservative mass accretion efficiency transforms systems, which could form Be X-ray binary, to systems passing through a CE stage, especially in the Be mass range of 3-7~M$_\odot$. That is why the number of Be X-ray binaries in model F is significantly less than in model A.

The mass accretion efficiency weakly affects the distribution of synthetic binaries on $P$~--~$e$ plot. This parameter is not important when comparing velocities of isolated radio pulsars. But the mass accretion efficiency is an important parameter for velocities of Be X-ray binaries. The distribution of synthetic velocities for Be X-ray binaries is very different for model C in comparison to model D. The model D is much more similar to the observed distribution. More quantitative comparison is available in Section~\ref{s:combined}.

\subsection{Electron capture supernova explosions}

The model of electron capture explosions has multiple hidden dimensions. Even with our simplified prescription used in population synthesis to completely specify the model, we have to make the following assumptions: (1) choose a mass range for helium core needed to get ecSN, (2) choose if effectively isolated stars also produce ecSN and (3) choose the magnitude of the natal kick, which is received by NS born in ecSN. It is not viable to restrict all these parameters in our study especially given the lack of observations for ecSN events. But it is worth to mention that current ecSN models are underconstrained and have to be further restricted by detailed simulations of supernova explosions and observations. In this section we compare typical outcomes of assumptions (1) and (2) mentioned above. 
 
We show the result of exploratory calculations for model H0 with electron capture supernova explosions in Figures~\ref{fig:be_mass_all_models} and \ref{fig:pe_mass_all_models}. A more detailed analysis is provided in Section~\ref{s:ecsn_long}. The model F has exactly the same assumptions as model H0 but no ecSN. Comparing these models, we see that allowing for electron capture supernova increases the total number of Be X-ray binaries. It also increases a fraction of binaries with lower eccentricities ($e\approx 0.4$) which makes it more similar to observations. 
This comparison does not depend much on observational selection. The model F produces $\approx 60$ Be X-ray binaries which is not enough to explain all systems observed in the SMC. This model produces even fewer synthetic binaries in the region where orbital period and eccentricities are easiest to measure (orbital periods in range 10-100 days and $e<0.5$).  
For model H0, the Be mass distribution peaks close to the value found by \cite{Vinciguerra2020} for their preferred model computed with the COMPAS population synthesis code but with similar assumptions. 
The total expected number of Be X-ray binaries in SMC in this model is $\approx 170$.
In comparison to all other models which do not include any ecSN, the peak of the period distribution is shifted towards values of $\sim100$ to $\sim 1000$~d.

Models where ecSN occurs only in stripped stars\footnote{Plots for all models are available in the online supplementary materials.} (G0, J0, H0) produce synthetic radio pulsars with velocities very similar to the \cite{hobbs2005} distribution, see Figure~\ref{fig:kick_disrupted_AD}. All models with ecSN produce Galactic Be X-ray velocities which reasonably match observations. Models where ecSN occur also in effectively isolated stars (G1, J1, H1) produce many more isolated NS with velocities $v<100$~km/s. The fraction of these NSs depend on the mass range assumed for ecSN progenitors: a wider mass range (H1) produces more slowly-moving isolated radio pulsars and nearly two times more Be X-ray binaries than a narrower range (G1).

\section{Combined analysis}
\label{s:combined}

In this Section, we introduce our combined analysis which allows for a comparison between the observed velocities of young isolated radio pulsars, peculiar velocities of Be X-ray binaries and the binary population synthesis models. 
This kind of analysis allows us to
examine multiple formation channels (with different natal kick prescriptions) simultaneously.

\subsection{Structure of the combined analysis}
The best representation of velocity distribution for pulsars and Be X-ray binaries, which one can obtain using the binary population synthesis code is the histogram of three-dimensional velocities (see example in Figure~\ref{fig:kick_disrupted_AD}). 
Therefore, it is important to develop a technique that allows to quantitatively compare these histograms with observations taking into account observational uncertainties. In our view, such a comparison can be achieved by using binned velocity distribution.
We define the binned velocity distribution as a function of parameters $\vec m$:
\begin{equation}
f_v(v, \vec m) = \sum_{i=1}^N m_i v^2 v_h (v),    
\end{equation}
where function $v_h(v)$ is a rectangular function (piece-wise constant) and it is equal to 1 in the range of velocities $[i\Delta v , (i+1) \Delta v )$ and is equal to 0 otherwise, $\Delta v$ is the size of velocity bin. The factor of $v^2$ is present in $f_v(\vec m)$  because it is a part of Jacobian for the spherical coordinate system. The vector of parameters $\vec m$ is normalised in such a way that:
\begin{equation}
\int_0^\infty f_v (\vec m) dv = \sum_{i=1}^N m_i \Delta v^3 \left(\frac{(i+1)^3 - i^3}{3}\right) = 1.    
\end{equation}
This velocity distribution is also part of the joint probability for parallaxes and proper motion measurements, see Appendix~\ref{s:binned}.
Therefore, given measured parallaxes and proper motions for Be X-ray binaries and isolated radio pulsars we can compute a likelihood for a distribution of peculiar velocities obtained from binary population synthesis simulations.

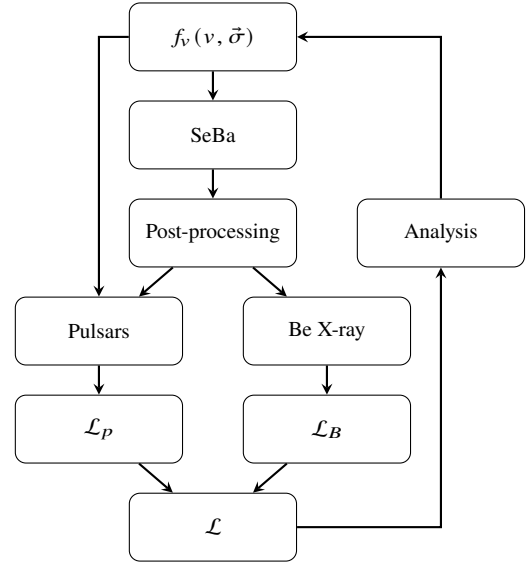
\begin{figure}
    \centering
    \begin{tikzpicture}
    \node (start) [startstop] {$f_v (v, \vec \sigma)$};
    \node (sb) [startstop, below of=start, yshift=-0.3cm] {SeBa};
    \draw [arrow] (start) -- (sb);
    \node (sb1) [startstop, below of=sb, yshift=-0.3cm] {Post-processing};
    \draw [arrow] (sb) -- (sb1);
    \node (analysis) [startstop, below of=sb, yshift=-0.3cm,  xshift=3cm] {Analysis};
    \node (l01) [startstop, below of=sb1, yshift=-0.3cm, xshift=-1.5cm] {Pulsars};
    \node (l1) [startstop, below of=l01, yshift=-0.3cm] {$\mathcal L_p$};
    \node (l02) [startstop, below of=sb1, yshift=-0.3cm, xshift=1.5cm] {Be X-ray};
    \node (l2) [startstop, below of=l02, yshift=-0.3cm] {$\mathcal L_B$};
    \draw [arrow] (sb1) -- (l01);
    \draw [arrow] (sb1) -- (l02);
    \draw [arrow] (l01) -- (l1);
    \draw [arrow] (l02) -- (l2);
    \node (L) [startstop, below of=l1, yshift=-0.3cm, xshift=1.5cm]
    {$\mathcal L$};
    \draw [arrow] (l1) -- (L);
    \draw [arrow] (l2) -- (L);
    \draw [arrow] (start) -| (l01);
    \draw [arrow] (L) -| (analysis);
    \draw [arrow] (analysis) |- (start);
    \end{tikzpicture}
    \caption{
    Schematic representation of the algorithm used to test various model natal kick distributions using the observational constraints based on the parallaxes and proper motions of the observed isolated radio pulsars and Be X-ray binaries. $f_v(v, \vec \sigma)$ is the natal kick distribution, SeBa is the binary population synthesis code used in this work, $\mathcal L_p$ and $\mathcal L_B$ correspond to a stage when the likelihood is computed separately for isolated radio pulsars and Be X-ray binaries.}
    \label{fig:algorithm}
\end{figure}

We present the algorithm for our combined analysis schematically in Figure~\ref{fig:algorithm}. This algorithm is the following: (1) we choose a natal kick distribution $f_v(v, \vec \sigma)$ and fix its parameters $\vec \sigma$, (2) we run SeBa binary population synthesis code for 20000 ZAMS binaries using the natal kick distribution chosen in the previous step and (3) analyse its output selecting isolated radio pulsars and Be X-ray binaries (see details in Section~\ref{s:model_exploration}). We prepare a histogram for velocities and convert number of systems in individual velocity bins $h_k$ to parameters of the binned velocity distribution $m_k$ using eq. \ref{e:mk}. At the next step (4), we compute likelihood that the observed parallaxes and proper motions for isolated radio pulsars and Be X-ray binaries are drawn from this model, see eq. \ref{e:final_ll}. Then (5), we combine these likelihoods. Finally, we select a next model or change the parameters and repeat the algorithm.

We can perform model selection using the Akaike information criterion (AIC). The AIC becomes the likelihood ratio test if model has no estimated parameters as in our case (besides model K):
\begin{equation}
\Delta \mathrm{AIC} = 2\Delta \mathcal L = 2 \log\left( \frac{L_1}{L_2}\right),    
\end{equation}
where $L_1$ and $L_2$ are the likelihoods of two compared models. We provide $2\Delta \mathcal L$ in our Table~\ref{tab:models} comparing all models with model K. It is possible to quantify how probable model 1 is in comparison to model 2 as:
\begin{equation}
p \sim \exp\left(-\mathcal L\right).    
\end{equation}
For example, the model E (no natal kick) is the most improbable model in our comparison. While model D is very similar to optimal model ($p=1.3$)\footnote{Strictly speaking, when we compare models with K we should subtract 4 from the difference because model K has two estimated parameters. But it is possible to compare models with D subtracting only 1.3 from the two log-likelihood difference because this model has no estimated parameters.}. The AIC$=2\mathcal{L}$ value of more than 9.2 means that model is hardly compatible with measurements $p < 0.01$.

Our algorithm has four significant advantages over natal kick studies that were performed in the past e.g. \cite{hobbs2005,Verbunt2017,Igoshev2020}: (1) it allows to include effects related to binary evolution, (2) it allows to rigorously test any velocity distribution that can be implemented in the population synthesis code
using exactly the same form of the likelihood function, (3) it can be easily extended to include more observational information for any given binary formation channel and (4) it is straightforward to extend this analysis and include Markov Chain Monte Carlo sampler to study correlations between parameters, if necessary.

\subsection{Results of the combined analysis}

Our exploratory model D (see Figure~\ref{fig:kick_disrupted_AD} and \ref{fig:pe_mass_all_models}) seems to fit well all available observational constrains: (1) a peak of Be masses is located around 11~M$_\odot$, (2) it produces $\approx 180$~Be X-ray binaries in SMC (compatible with estimates of 120 HMXBs by \citealt{Maravelias2019}, \citealt{HaberlSturm2016}), (3) Be X-ray binaries with measured periods and eccentricities coincide with maximum density of simulated systems (could be affected by observational selection), (4) it leads to natal kick distribution of isolated radio pulsars that is the most compatible with the one found in the recent work by \cite{Igoshev2020} (left panel Figure~\ref{fig:kick_disrupted_AD}) and (5) the velocity distribution of Be X-ray binaries provides a reasonable match to the observation-based distribution of the systemic velocities of Galactic Be X-ray binaries estimated in Section
\ref{s:data}
(right panel Figure~\ref{fig:kick_disrupted_AD}). 

Further in this Section, we use model D as our base model.
We only vary the parameters that describe the low velocity part of the natal kick distribution ($w$ and $\sigma_1$).
We do not vary parameter $\sigma_2$ because: (1) only the isolated pulsar population,whose velocity distribution is relatively well constrained \citep{Igoshev2020}, is  sensitive to large velocities and (2) it is numerically expensive to study three parameters instead of two.
We assume $\sigma_2 = 336$~km/s, as found by \cite{Igoshev2020}.

\begin{figure*}
    \begin{minipage}{0.49\linewidth}
    \includegraphics[width=\columnwidth]{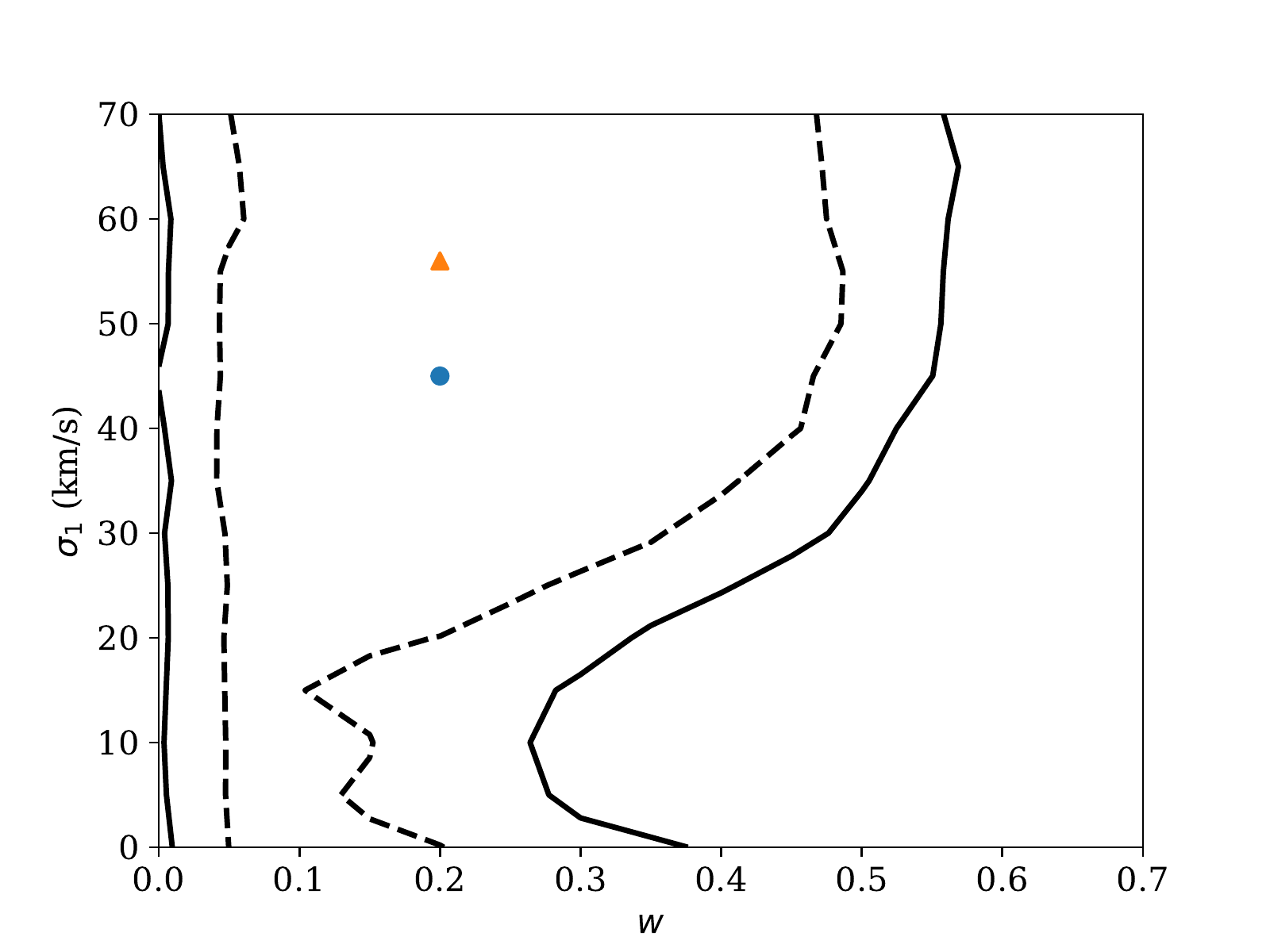}    
    \end{minipage}
	\begin{minipage}{0.49\linewidth}
    \includegraphics[width=\columnwidth]{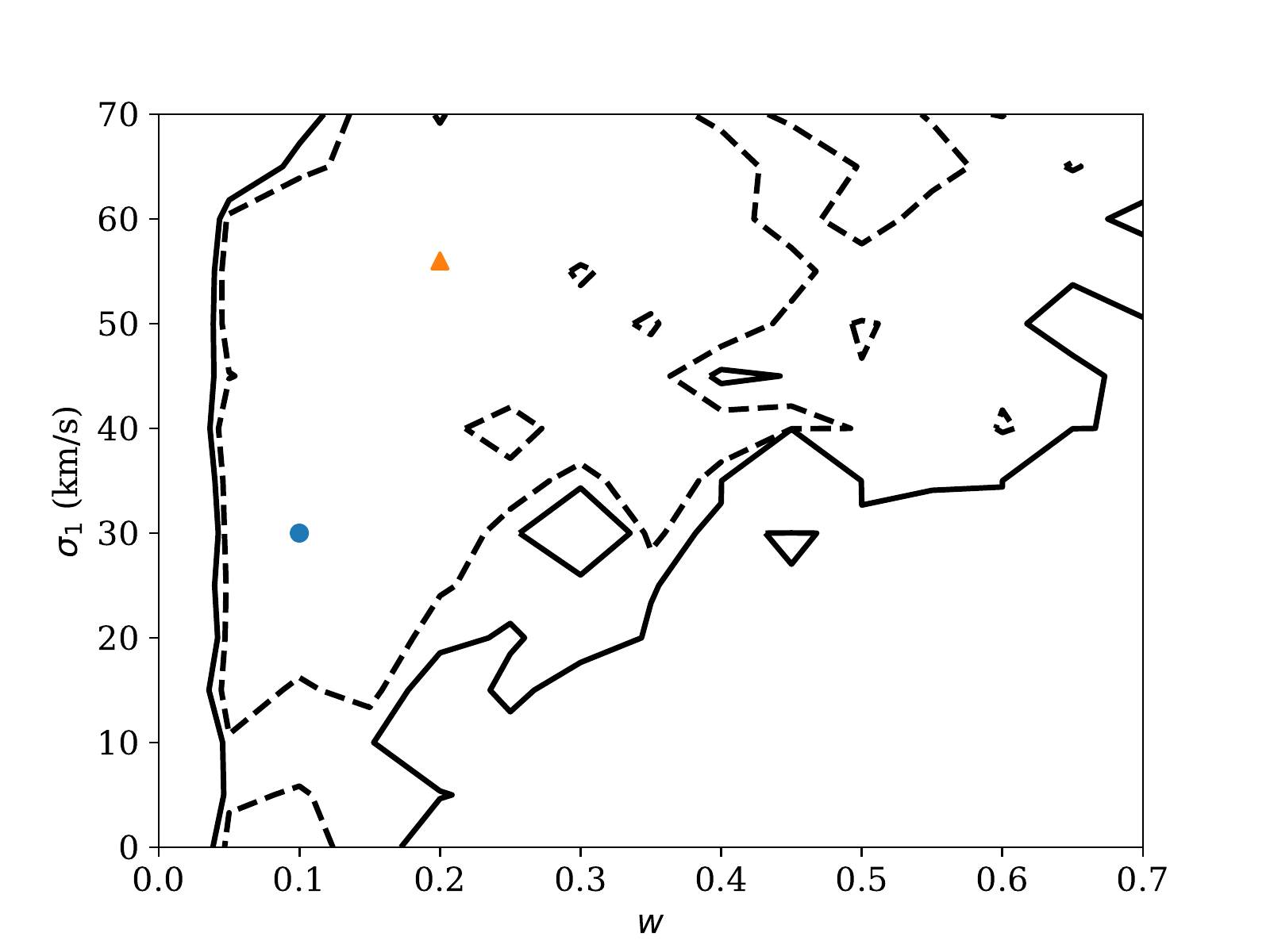}    
    \end{minipage}
    \caption{Likelihood profiles obtained for parallaxes and proper motions of isolated radio pulsars (left panel) and Be X-ray binaries (right panel). Blue dot corresponds to the maximum likelihood $w=0.2$, $\sigma_{1}=45$ (left panel), $w=0.1$, $\sigma_{1}=30$ (right panel); orange triangle corresponds to parameters of model D. Dashed line corresponds to 95~per~cent confidence interval, solid lines for 99~per cent confidence interval.  }
    \label{fig:contour_separate}
\end{figure*}

\begin{figure}
	\includegraphics[width=\columnwidth]{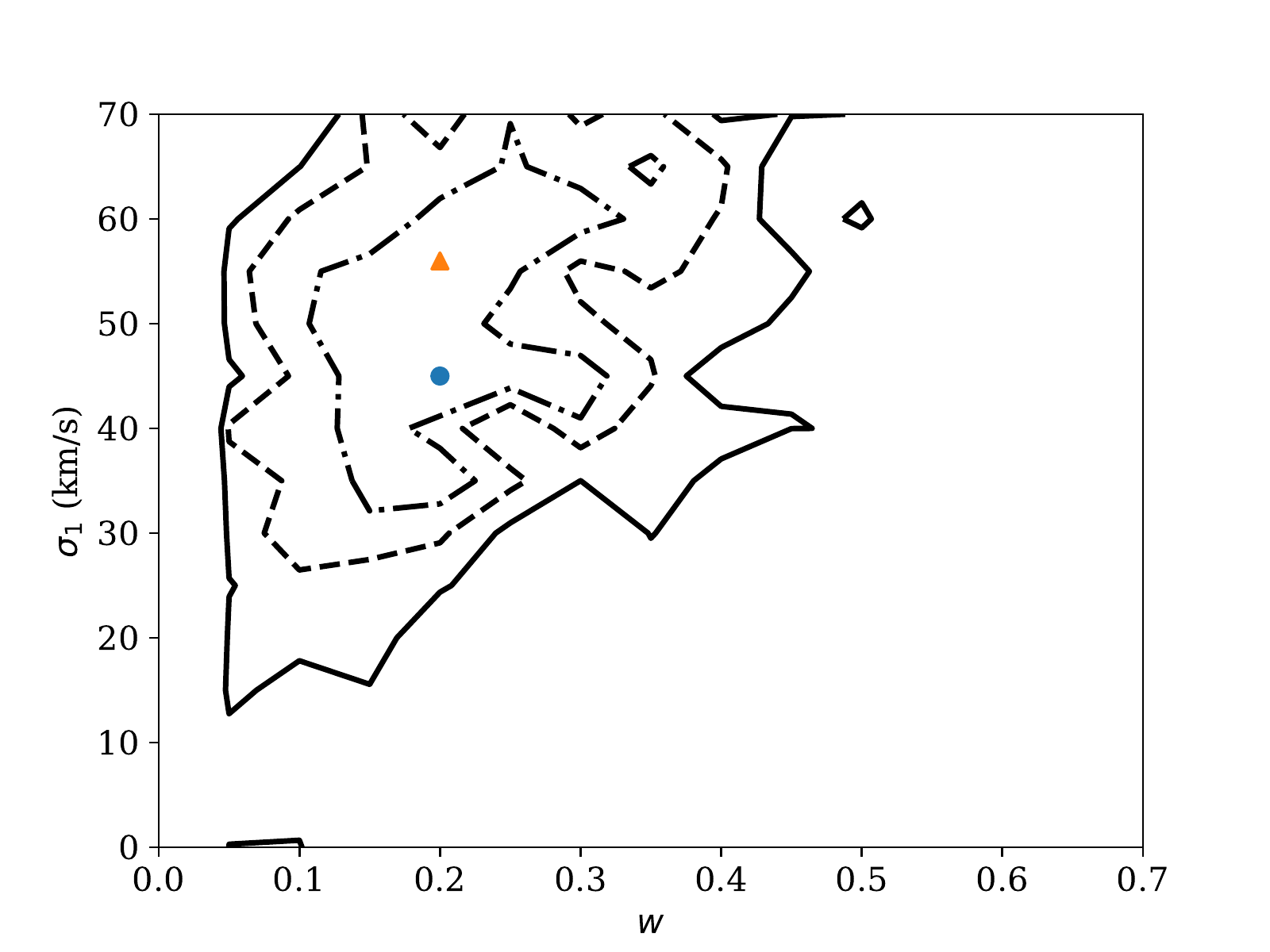}
    \caption{Likelihood profiles obtained in combined analysis of parallaxes and proper motions of isolated radio pulsars and Be X-ray binaries. Blue dot corresponds to the maximum likelihood $w=0.2$, $\sigma_{1}=45$ 
    orange triangle corresponds to parameters of model D. Dashed-and-dotted line corresponds to 68~per cent confidence interval, dashed line corresponds to 95~per~cent confidence interval, solid line for 99~per cent confidence interval.}  
    \label{fig:contour}
\end{figure}

We show the results of the optimisation procedure in Figure~\ref{fig:contour_separate} and \ref{fig:contour}. The acceptable values are concentrated in the top left corners of the corresponding figures. The confidence regions are somewhat noisy due to the stochastic nature of our models. Nevertheless, we can determine the maximum likelihood value from the combined analysis as $w = 0.2\pm 0.1$ and $\sigma_1 = 45_{-15}^{+25}$~km/s and fixed high-velocity component at $\sigma_2 = 336$~km/s. Based on our combined analysis we can completely exclude $w=0$, i.e. the situation in which only the high natal kick velocity component of the distribution is present. We also rule out cases with $w > 0.5$, i.e. when more than half of the NS population is born with weak natal kicks. The default parameters of model D are compatible with the optimal model, but are slightly less preferred.

\subsection{Electron capture supernova explosions}
\label{s:ecsn_long}

In this Section, we test a hypothesis that slowly moving radio pulsars ($|v| < 100$~km/s) are predominantly produced via ecSN. It is clear that describing the NS formation in the ecSN based on the fixed (either ZAMS or $M_{\rm C; BAGB}$) mass limits is a severe simplification and the uncertainties in the models are considerable.
Nonetheless, given the fact that such prescription is widely used in population synthesis, it is instructive to discuss whether our analysis can rule out such a scenario.

In order to rigorously test the origin of weak natal kicks via ecSN, we run six binary stellar evolution synthesis models (H0, H1, G0, G1, J0 and J1 models in Table~\ref{tab:models}). In these models  
core collapse supernova explosions lead to natal kicks described by the Maxwellian velocity distribution with $\sigma=265$~km/s and ecSN lead to natal kicks with $\sigma=30$~km/s. Our basic mass range is [1.83, 2.25]~M$_\odot$ (models G0, G1).
\cite{Chruslinska2018} consider an ad-hoc, 'optimistic' model variation that allows to form a larger fraction of NS with low natal kicks in the ecSN scenario, i.e. assuming that stars with $M_{\rm C; BAGB}$ in a wider range of [1.63, 2.45]~M$_\odot$ can lead to ecSN (models H0 and H1). 
We also consider such model variation in this study to speculate whether the highly efficient ecSN scenario could explain the low velocity natal kick component of the radio pulsar velocity distribution. Additionally, we try the mass range [1.6, 2.25] used by \cite{Willcox2021} (models J0, J1).


In all cases we find that our statistical optimal model (K) describes the observations better. All models where only stripped stars produce ecSN (G0, H0, J0) have $2\Delta\mathcal L > 9.2$, therefore their probability is below 1~per cent. It happens because these models do not produce enough low-velocity isolated radio pulsars (Figure~\ref{fig:kick_disrupted_AD}). On the other hand, models G1, H1 and J1 could be accepted at some level ($p=0.012$, $p=0.02$ and $p=0.04$ respectively). These models clearly produce a significant fraction of low-velocity isolated radio pulsars, see Figure~\ref{fig:kick_disrupted_AD} and the online supplementary materials.

It is possible to improve these models and produce a model which explains observations and includes ecSN. It is our model L. In this new ecSN model core-collapse SN impart Maxwellian natal kicks with $\sigma=336$~km/s and ecSN (mass range 1.63-2.45~M$_\odot$; ecSN occurs also in effectively single stars) impart Maxwellian natal kicks with $\sigma = 45$~km/s. This model has $2\Delta \mathcal L = 5.18$ which means $p = 0.075$. This model produces 140 Be X-ray binaries in the SMC. Therefore, this model describes both observations of isolated radio pulsars and Be X-ray binaries well. It is worth mentioning that it is unclear if ecSN could physically occur in effectively isolated stars (i.e. stars that have not passed through the stage of stable mass transfer).

In comparison to \cite{Willcox2021}, we do require ecSN to occur in effectively isolated stars. The main reason for this is that \cite{Willcox2021} used a non-parametric description for natal kick of NSs. This prescription already describes the natal kicks of isolated radio pulsars (this is how it was obtained in first place). Therefore, it is necessary for \cite{Willcox2021} to suppress the production of slowly moving isolated NS from binaries, which is done by prohibiting ecSN events in effectively isolated stars. In our case, the core collapse NS have natal kicks drawn from the Maxwellian velocity distribution with $\sigma = 336$~km/s instead of drawing natal kicks from observed parallaxes and proper motions. It means that a very small fraction of NSs have velocities below $50$~--~$100$~km/s. To add NS in this velocity range we do require ecSN in effectively isolated stars.

\section{Discussion}

\subsection{Consequence of binary fraction 85~percent}
\label{s:lower_binary_fraction}
In our basic simulations we assume 100~per cent binary origin for progenitors of isolated radio pulsars. In reality, a fraction of NS could be formed from truly isolated progenitors. 
To check how sensitive our simulations are to this assumption we perform additional simulations modifying model K by assuming that 15~per~cent of isolated radio pulsars originate from truly isolated progenitors. These systems receive exactly the same natal kick as other NS born in binaries\footnote{Note that the observed velocity of NS born in binaries is the sum of natal kick and orbital velocity, while observed velocities of NSs born from truly isolated stars consists of the natal kick only.}.

\begin{figure}
	\includegraphics[width=\columnwidth]{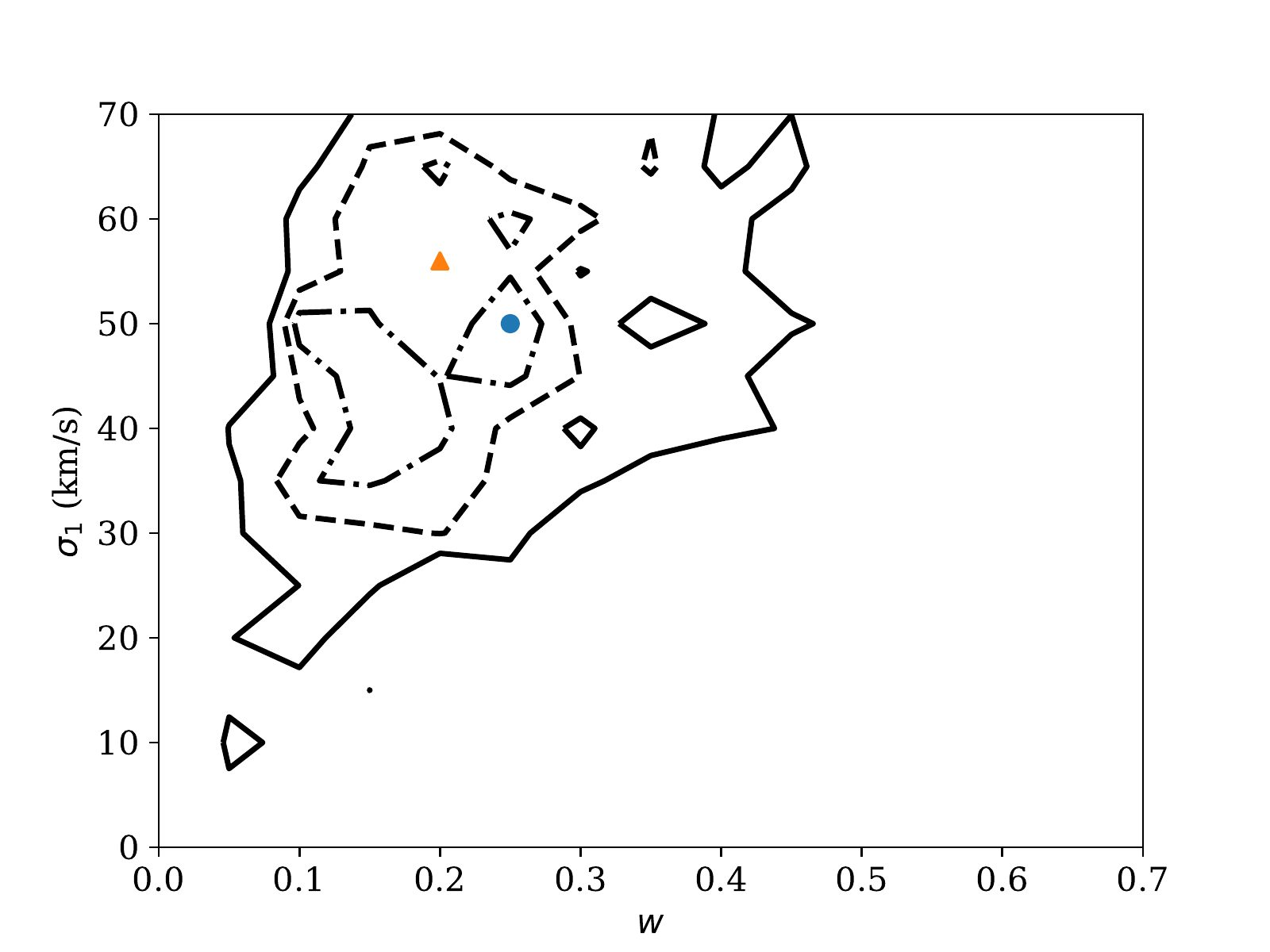}
    \caption{Likelihood profiles obtained in combined analysis of parallaxes and proper motions of isolated radio pulsars and Be X-ray binaries in the case when 15~per~cent of isolated radio pulsars are formed from isolated progenitors. Blue dot corresponds to the maximum likelihood $w=0.25$, $\sigma_{1}=50$; orange triangle corresponds to parameters of model D. Dashed-and-dotted line corresponds to 68~per cent confidence interval, dashed line corresponds to 95~per~cent confidence interval, solid line for 99~per cent confidence interval.  }
    \label{fig:contour_isol}
\end{figure}

We show the likelihood profiles obtained from our new combined analysis in Figures~\ref{fig:contour_isol}.
The resulting profiles resemble the ones obtained for our original model (see Fig. \ref{fig:contour_separate} and \ref{fig:contour}).
The overall maximum likelihood resulting from the combined analysis is shifted towards larger fraction of slower objects $w=0.25$ with slightly increased $\sigma_{1}=50$~km/s.
Nevertheless, this value is within the confidence interval of our original model. Thus, we conclude that the contribution of pulsars with truly isolated origin to the low kick velocity component of the distribution is minimal.

\subsection{Initial velocity dispersion of SN progenitors}

It is important to check how reliably Be X-ray binary velocities trace binary peculiar motion. 
ZAMS binaries might have an initial velocity dispersion of $\approx 10$~km/s \citep{deBruijne1999,Kiminki2018}. 
On top of that, their observed velocity may contain a contribution form the velocity dispersion of their parent association of the order of $\approx 5$~km/s \citep{reid2014}. In our analysis of binaries with B stars we find $\sigma = 11$~km~s$^{-1}$, see Appendix~\ref{s:bbinary}.

\begin{figure*}
 \begin{minipage}{0.49\linewidth}
 \includegraphics[width=\columnwidth]{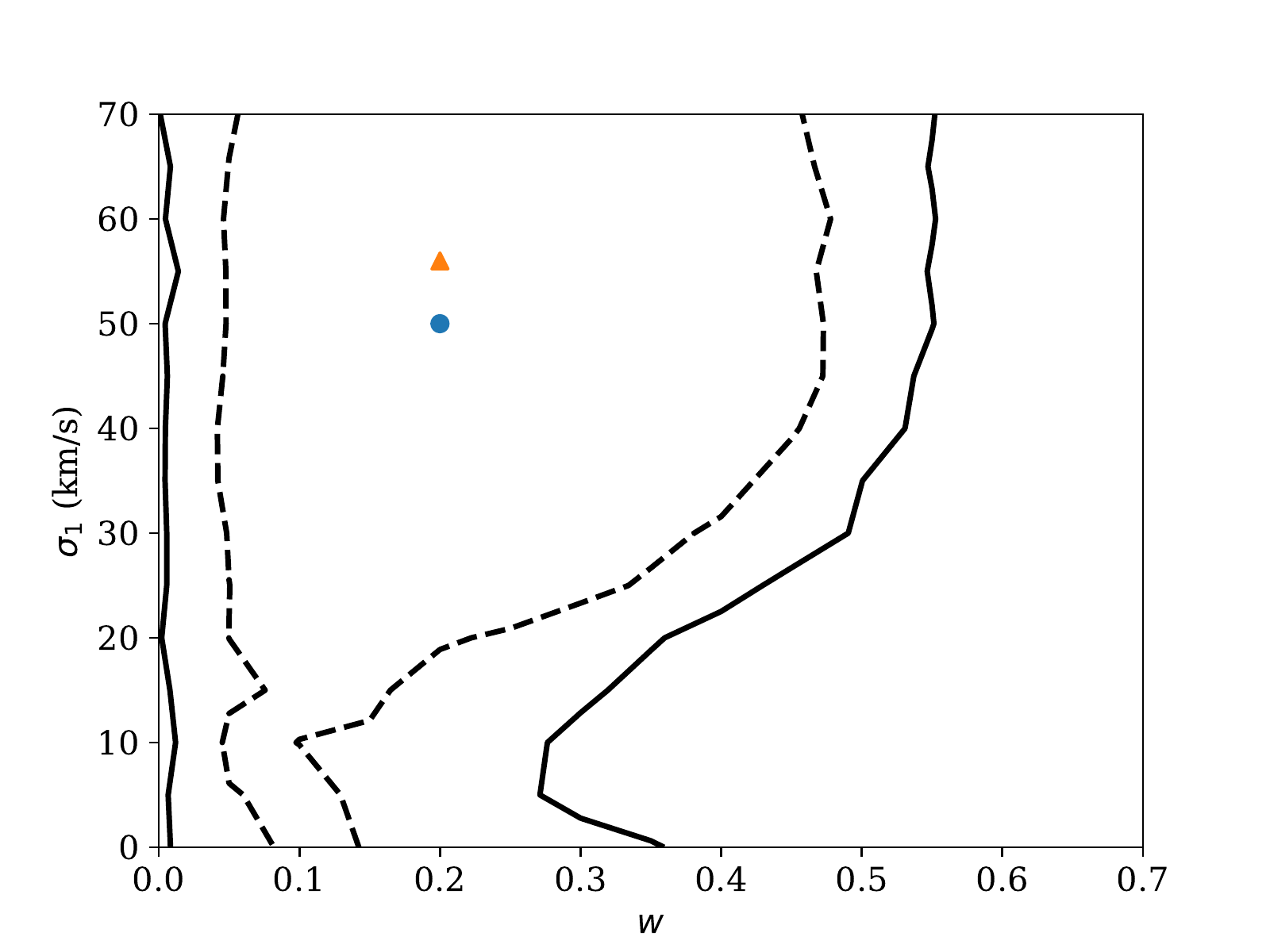} 
 \end{minipage}
	\begin{minipage}{0.49\linewidth}
 \includegraphics[width=\columnwidth]{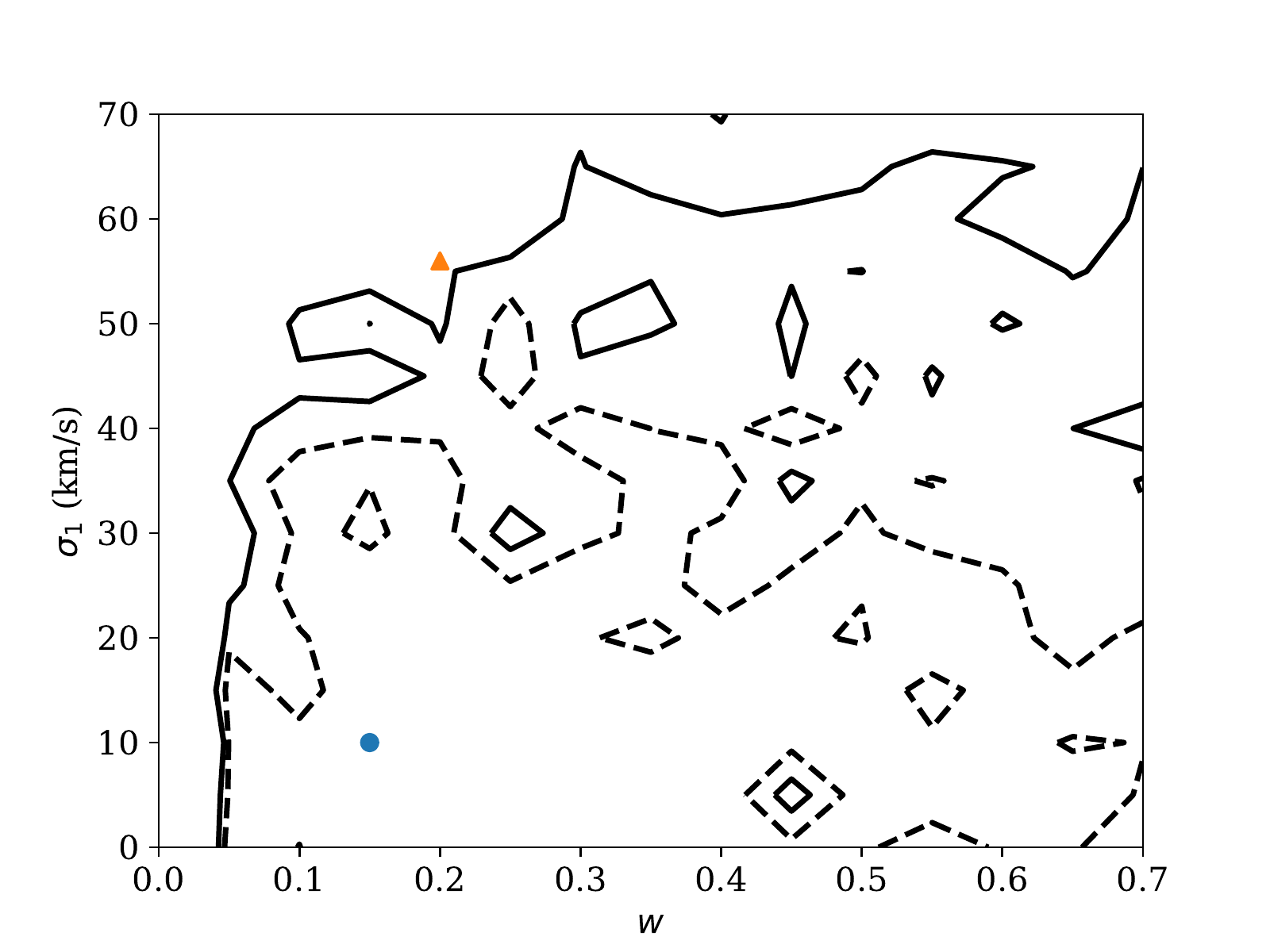} 
 \end{minipage}
 \caption{Likelihood profiles obtained for parallaxes and proper motions of isolated radio pulsars (left panel) and Be X-ray binaries (right panel) in the case when initial velocity dispersion is added to simulations. Blue dot corresponds to the maximum likelihood: $w=0.2$ and $\sigma_1=50$~km/s (left panel), and $w=0.15$ and $\sigma_1=10$~km/s (right panel). Orange triangle corresponds to parameters of model D. Dashed line corresponds to 95~per~cent confidence interval, solid lines for 99~per cent confidence interval.}
\label{fig:contour_separate_11}
\end{figure*}

To quantify the effect of the velocity dispersion on our results, we perform an additional combined analysis based on model D where we add to each simulated pulsar and Be X-ray binary a random velocity drawn from a Maxwellian distribution with $\sigma=11$~km/s.
We show the results of our simulations in Figure~\ref{fig:contour_separate_11}.

\begin{figure}
 \includegraphics[width=\columnwidth]{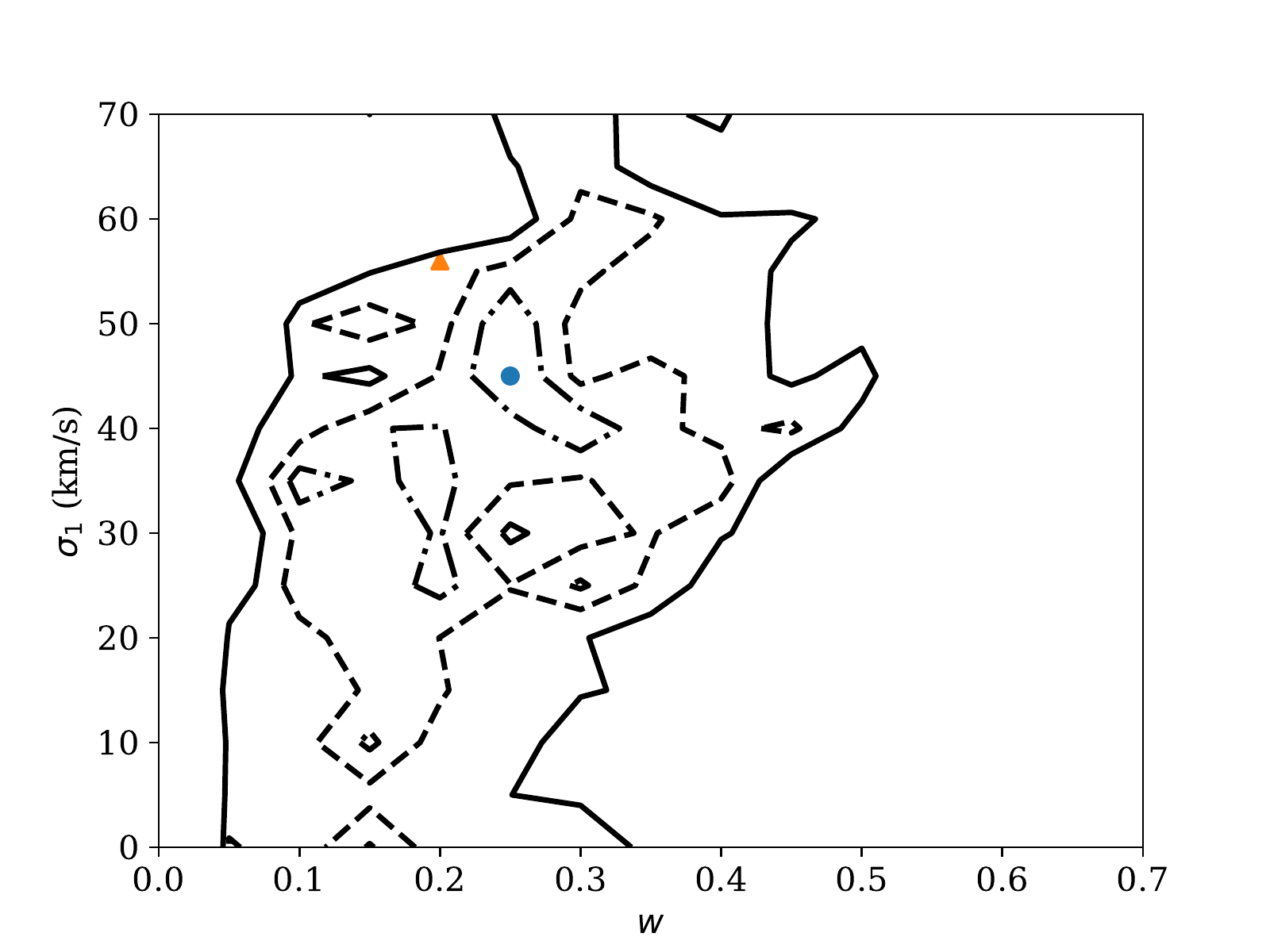} 
 \caption{Likelihood profiles obtained for parallaxes and proper motions of isolated radio pulsars and Be X-ray binaries in the case when initial velocity dispersion is added to simulations. Blue dot corresponds to the maximum likelihood $w=0.25$ and $\sigma_1=45$~km/s. Orange triangle corresponds to parameters of model D. Dashed line corresponds to 95~per~cent confidence interval, solid lines for 99~per cent confidence interval.}
\label{fig:contour_total_11}
\end{figure}

In the case of isolated radio pulsars, $\sigma_1$ is lowered by $10$~km/s. On the other hand, the response of Be X-ray binaries is not that trivial. The solution with $w = 0.2$ and $\sigma_1 = 45$~km/s is nearly excluded and the preferable solution is $w = 0.15$ $\sigma_1 = 10$~km/s. A range of solutions exists that are acceptable both by isolated radio pulsars and Be X-ray binaries. Optimisation of the combined model suggests that the minimum is located around $w = 0.25$ and $\sigma_1 = 45$~km/s (see Figure~\ref{fig:contour_total_11}), which is within the 68~per~cent confidence interval of our original optimisation.

\subsection{Effects of observational selection}\label{s:selection effects}

Although our model D seems to satisfy a large number of observational constraints, it is worth mentioning that some of these constrains are a matter of strong observational selection. For example, there are only seven measurements of orbital periods and eccentricities for Be X-ray binaries in the SMC, see e.g. in Figure~\ref{fig:pe_mass_all_models}. That is because eccentricity and orbital periods are difficult to obtain from X-ray observations. Many Be X-ray binaries are only seen close to their periastron passages for a short period of time. 
These observations are sufficient to identify presence of the Be X-ray binary and measure NS spin period (if it is less than $\approx 500$~s; \citealt{Laycock2010}), but do not allow to constrain the orbital period and especially eccentricity.
In the case of SMC, many Be X-ray binaries were discovered in RXTE scans. This technique is biased against the long-period binaries \citep{Laycock2010,Vinciguerra2020}. 

\begin{figure}
	\includegraphics[width=0.98\columnwidth]{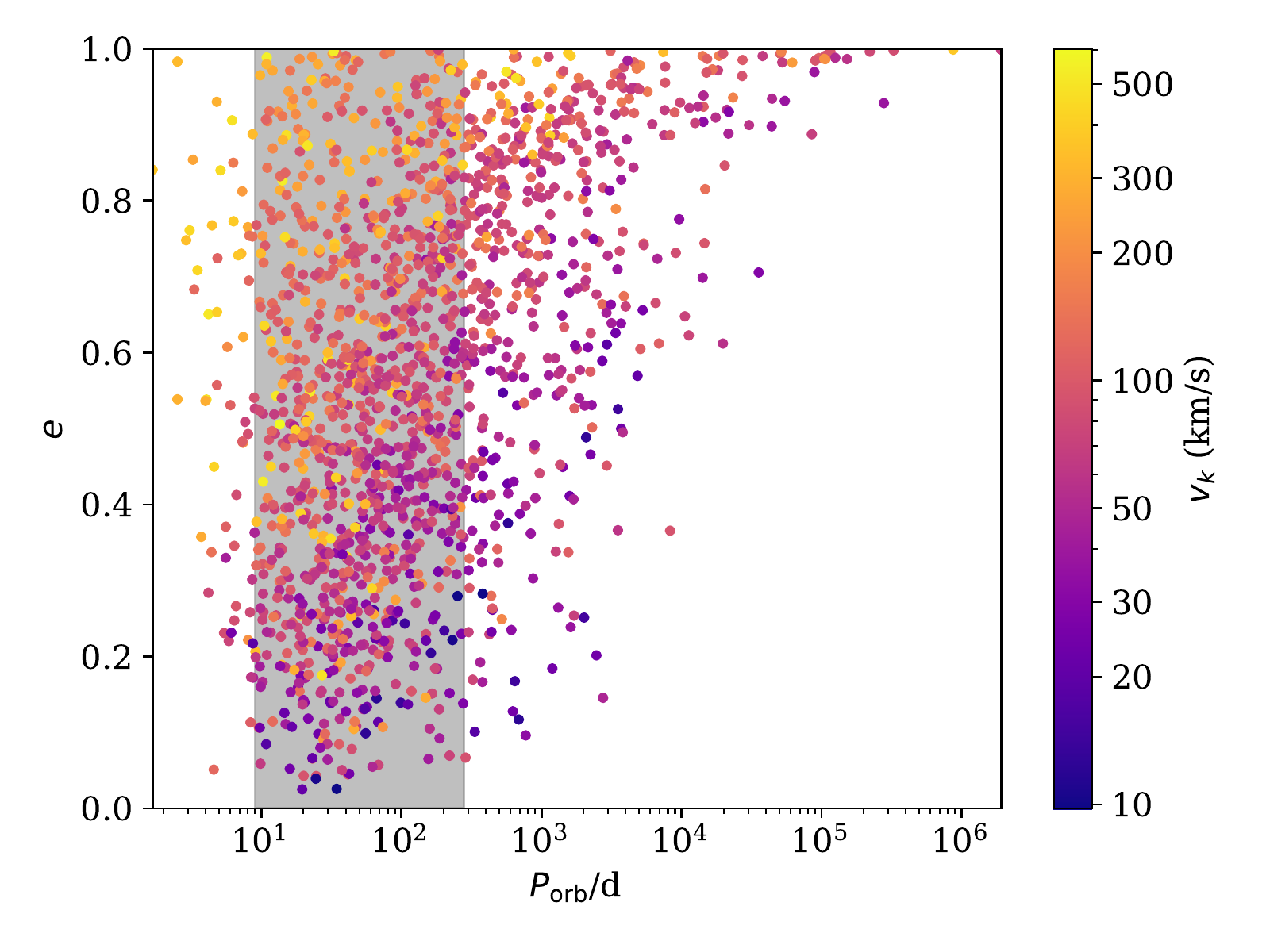}
    \caption{Distribution of natal kicks for synthetic Be X-ray binaries (colour) as a function of their orbital period and eccentricity. We show here the results of simulations D for Milky Way metallicity. Grey region shows a range of orbital periods measured for Galactic Be X-ray binaries based on \protect\cite{Raguzova2005}.  }
    \label{fig:pe_v_kick}
\end{figure}

Only two Be X-ray binaries with spin period longer than 1000~s are known within the Galactic sample, while our simplified simulations contain systems with much longer periods. 
Such systems might not be seen as Be X-ray pulsars because NS does not come close enough to interact with the decretion disk. 
Thus, some of those systems might be seen as single Be stars. 
The effect of this selection is the following: systemic velocity correlates with orbital and spin period.
In Figure~\ref{fig:pe_v_kick} we show the natal kick of NS as the function of the final orbital period and eccentricity for Be X-ray binaries. The Galactic Be X-ray binaries have measured orbital periods ranging $10$~--~$250$~days, see e.g. \cite{Raguzova2005}. 
In this range of orbital periods we see binaries which receive a natal kick ranging from extremely small values $\approx 10-15$~km/s to large values $>500$~km/s. Therefore, our analysis should be able to probe the full range of kick values.  It seems that due to these selection effects we may preferentially miss the low kick systems in the observed sample (there are hardly any high kick binaries in the long orbital period part of the population). But if we limit the model sample to the range of orbital periods probed observationally, their velocity distribution would be different. We check how important it is for the result of the optimisation procedure below.

\begin{figure}
	\includegraphics[width=0.98\columnwidth]{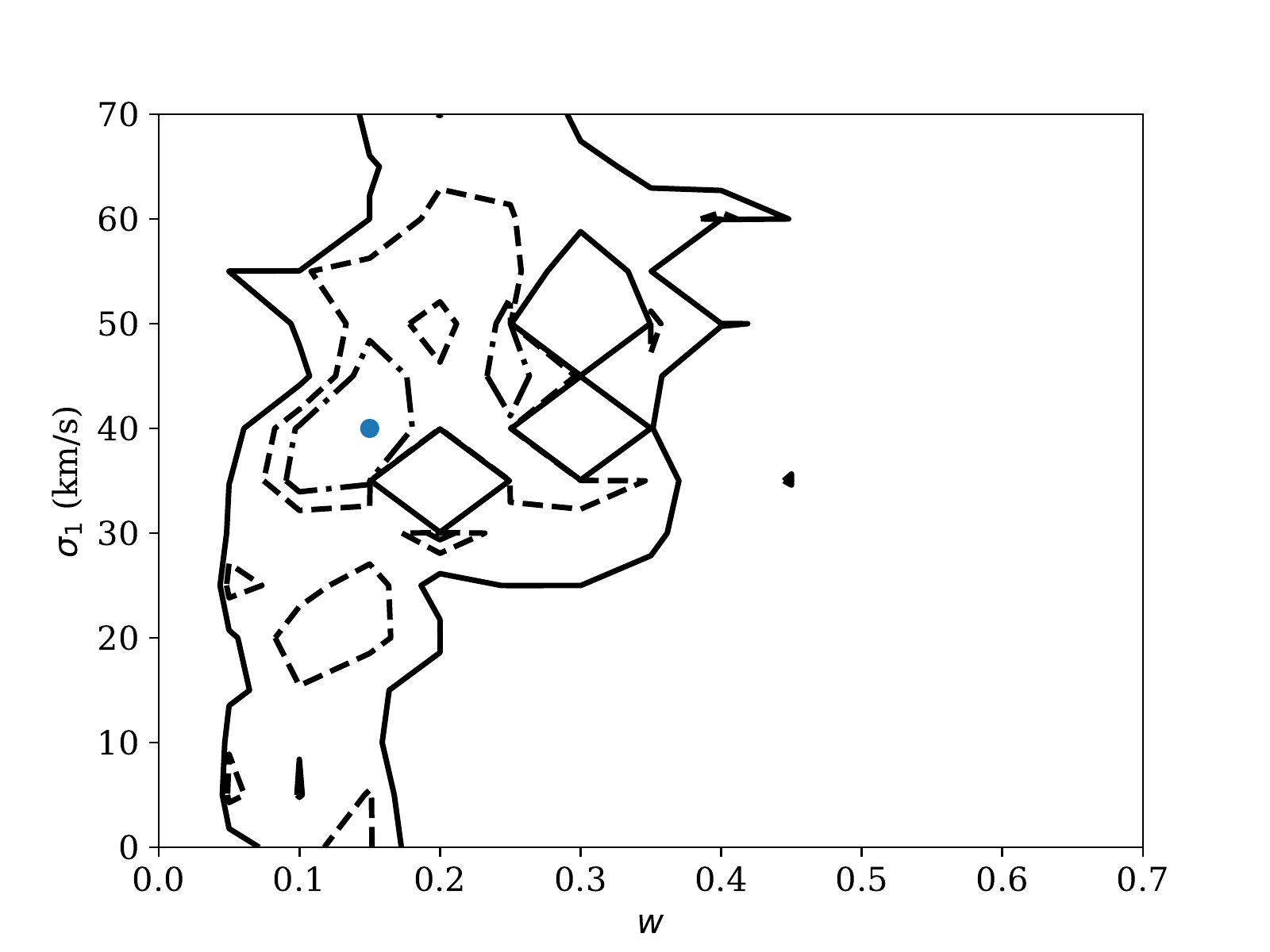}
    \caption{Likelihood profiles obtained in combined analysis of parallaxes and proper motions of isolated radio pulsars and Be X-ray binaries in the case when only Be X-ray binaries with orbital periods in range 10-250 days are included. Blue dot corresponds to the maximum likelihood $w=0.15$ and $\sigma_1=40$~km/s. Dashed-and-dotted line corresponds to 68~per cent confidence interval, dashed line corresponds to 95~per~cent confidence interval, solid line for 99~per cent confidence interval. }
    \label{fig:total_per}
\end{figure}

We perform our combined analysis once again, including only synthetic Be X-ray binaries with orbital periods in range 10-250~days to make it more comparable with observations. We show results in Figure~\ref{fig:total_per}. This result is more noisy because fewer Be X-ray binaries are included in the analysis. The result of optimisation is $w=0.15$ and $\sigma_1 = 40$~km/s. These values are well within the 1-$\sigma$ confidence interval of our original optimal model. Therefore, simple exclusion of binaries with very short and very long orbital periods does not lead to a significantly different result.

\begin{figure*}
    \begin{minipage}{0.49\linewidth}
	\includegraphics[width=0.99\columnwidth]{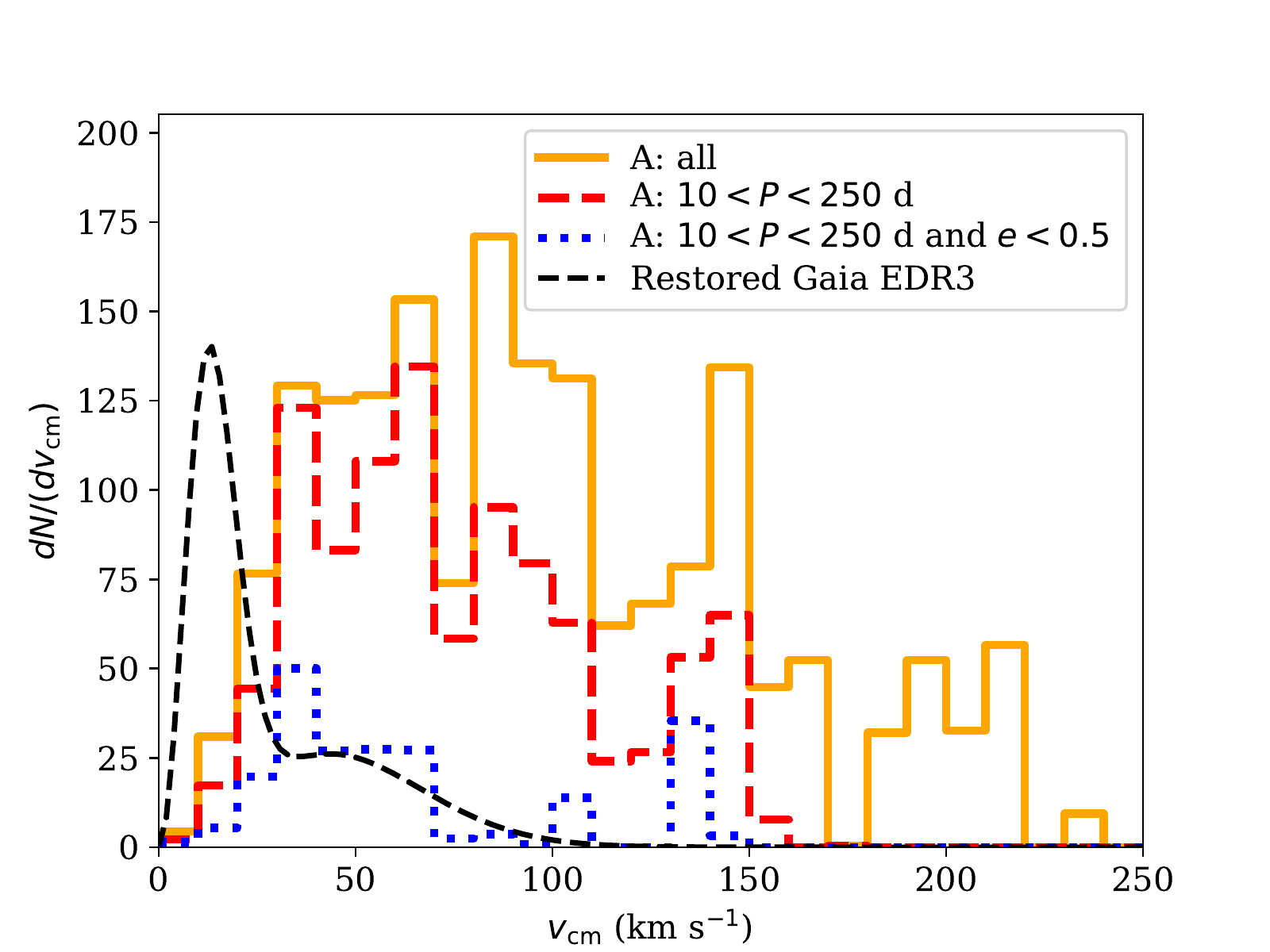}
	\end{minipage}
    \begin{minipage}{0.49\linewidth}
	\includegraphics[width=0.99\columnwidth]{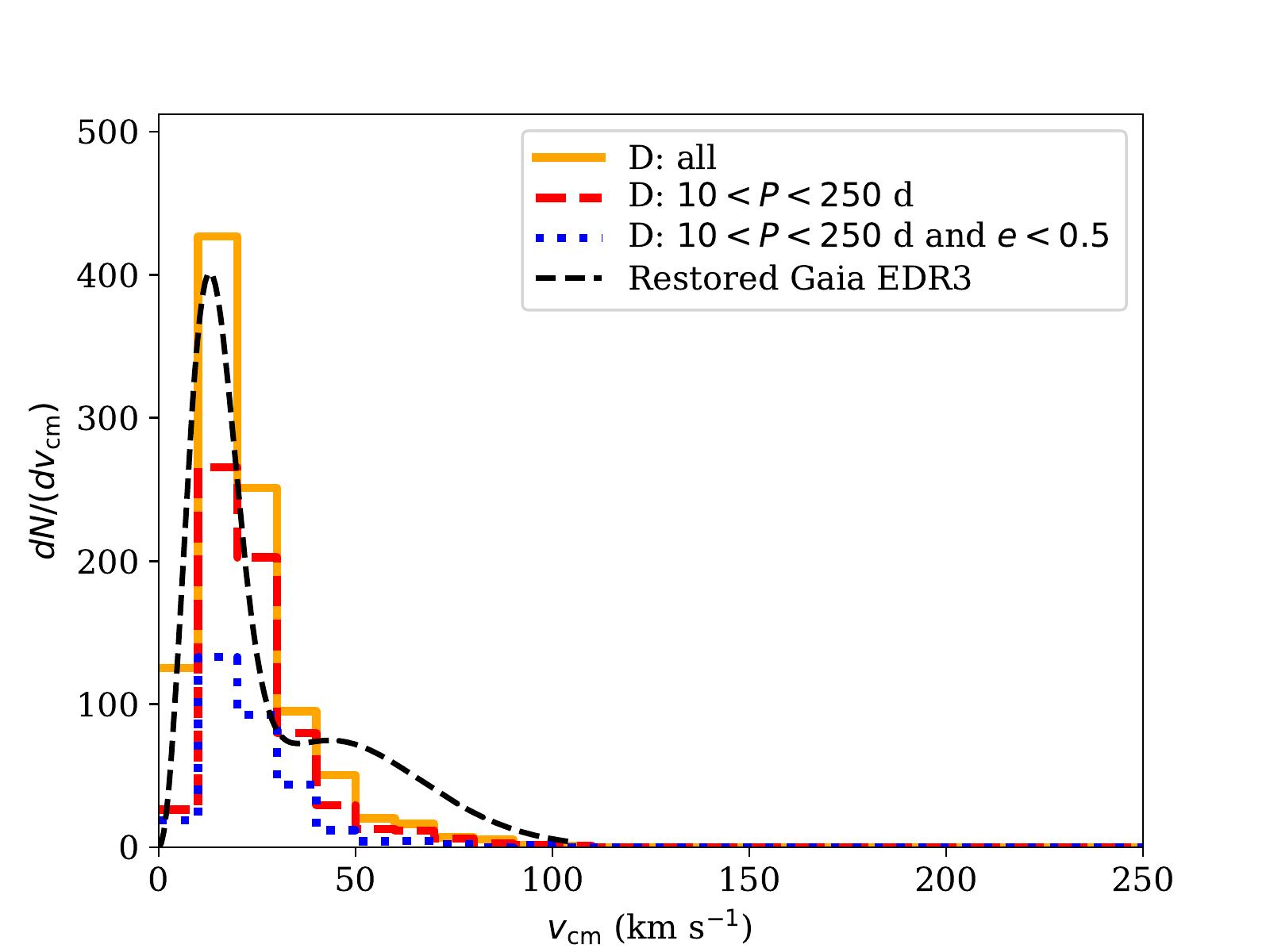}
	\end{minipage}	
    \caption{
    Possible influence of the observational selection on the peculiar velocity distribution of the model Be X-ray binary population (different panels and colours) compared with the distribution based on the Gaia EDR3 observations (black dashed line; height is chosen arbitrary). Left panel: model A. Right panel: model D. 
    Different types of lines correspond to different possible selection effects discussed in the text. The solid orange line shows all synthetic binaries (no observational selection). The red dashed line corresponds to the case when we can observe only Be X-ray binaries with orbital periods in range 10-250~days. The blue dotted line correspond to the case when we can only observe binaries with orbital periods in range 10-250~days and eccentricities smaller than 0.5.
    }
    \label{fig:kick_disrupted_AD_selection}
\end{figure*}

It is possible to argue that Be X-ray binaries with large eccentricities are absent in the observed sample based on Figure~\ref{fig:pe_mass_all_models}. Thus, the observed sample of Be X-ray binaries (see Figure~\ref{fig:pe_v_kick}) is biased towards smaller natal kicks ($<100$~km/s).  This is not entirely true, because we include in our sample multiple Be X-ray binaries with unknown eccentricity. In practice, it is much easier to discover Be X-ray binary with large eccentricity than to measure its eccentricity. In principle, for system with eccentricity $e>0.5$ periastron is located at much closer distance to Be star, thus such NS could cause more disruption to the decretion disk and accrete more as a result. In the Milky Way we know system J1845-024 with $e=0.88$ and $P_\mathrm{orb} = 242.18$~days \citep{hmxb_catalogue,Raguzova2005}. We simulate the extreme selection by removing all systems with $e>0.5$ and plot the result in Figure~\ref{fig:kick_disrupted_AD_selection}. The main effect of such selection is that the number of Be X-ray binaries with systemic velocities above 150~km/s is decreased in the case of the \cite{hobbs2005} natal kick distribution. In the case of natal kick distribution by \cite{Igoshev2020}, the shape of systemic velocity distribution is conserved. Overall, even after applying this severe (and quite unreasonable) observational selection, our models A and D predict very different velocities for Be X-ray binaries. Therefore, the effect of real observational selection is probably is much less significant.

\subsection{Applicability of our results}

Our result can be applied to model binary stellar evolution especially concerning the first supernova explosion. If the readers are concerned with modelling of isolated pulsar population, it is much better to use results by \cite{Igoshev2020} because the velocity distribution of isolated radio pulsars prefers slightly larger velocities than what results from the full natal kick distribution (NS with weaker kicks are preferentially bound in binaries).

One should be cautious when using our natal kick distribution to model the formation of double NS (DNS) systems.
Formation of such systems contains additional phases of interaction
that may affect the formation of the second NS (and the related natal kick)
and that cannot be constrained with Be X ray binary observations.
In particular, the secondary star in the progenitor of a merging DNS is thought to be almost completely stripped of its envelope before the SN (i.e. it undergoes the so-called ultra-stripped SN).
Such stars might receive a very different natal kick.
 It is not feasible to test this hypothesis using the isolated radio pulsars and Be X-ray binary populations, because the required extreme stripping is thought to occur during the late phases of mass transfer in compact binaries, where the secondary star is stripped by the first-born compact object.
 The natal kicks of DNSs have to be tested using the observational information on millisecond radio pulsar and DNS populations.

Note that our best natal kick velocity distribution resembles those found by \cite{Fryer1998}. 
The authors find that a fraction of $\approx 30$~per cent of NS should receive almost no kick while the high velocity pulsars form with natal kicks of the order of 600-700~km/s. \cite{Fryer1998} analyse multiple channels simultaneously including isolated radio pulsars, high- and low-mass X-ray binaries and double NSs. They only concentrate on the production rates but do not consider orbital parameters and peculiar velocities of Be X-ray binaries.
So, it might be the case that our natal kick model is acceptable for double NSs, but future studies are required.

\subsection{Caveat}
\label{s:caveats}

In our analysis we treat merged stars as effectively isolated. Merger occurs in $\approx 22_{-9}^{+26}$~per~cent of binaries according to simulations by \cite{Renzo2019}.  Some of these stars could explode as supernova and form isolated radio pulsars which will effectively increase a number of radio pulsars with isolated progenitors. We assume that natal kicks of these NSs will be drawn from the same distribution as natal kicks of NS born from initially isolated progenitors. This assumption has a caveat that these merger products might have a different structure of the core and might result in a different supernova explosion with somewhat different natal kick distribution for produced NSs. This hypothesis is worth of future investigation.

\section{Conclusions}

In this work, we revisit the natal kick velocity distribution of neutron stars
in light of the available constraints of the observed 
populations of isolated radio pulsars and Be X-ray binaries.

We introduce a framework which allows us to test any 
natal kick velocity distribution by confronting the velocity distributions
of the mock isolated radio pulsars and Be X-ray binary populations with the distributions obtained from observations. We modelled Be X-ray binaries  self-consistently to account for the effects of binary interactions.

Observational distributions of velocities are obtained using the parallaxes and proper motion measurements from radio interferometry for isolated radio pulsars (similar to the earlier study of isolated pulsars velocities by \citealt{Verbunt2017} and \citealt{Igoshev2020}), and peculiar velocities of Galactic Be X-ray binaries from Gaia EDR3.

As a starting point for our analysis, we simulate the population of Be X-ray binaries in the SMC and identify evolutionary models that can reasonably reproduce the demographics of the population:
the size of the population and the mass distribution of Be stars.
We also discuss the resulting distributions of orbital parameters (periods and eccentricities),
however, in that case the potential severe observational selection effects 
limit the meaningful conclusions from the comparison with observations.
Similarly as \cite{Vinciguerra2020}, we find that the properties of the population are
strongly affected by the assumptions about the mass transfer phase prior to NS formation.
We find that the properties of the SMC Be X-ray binaries are best reproduced if we assume that their progenitors evolve through a phase of stable, semi-conservative mass transfer, consistent with the results of \cite{Vinciguerra2020}.
Therefore, in our default model for the formation of (Galactic and SMC) Be X-ray binaries we require an episode of the stable, semi-conservative mass transfer. 


We identify the following comparison metrics: (1) total number of SMC Be X-ray binaries, (2) dependence of eccentricity on orbital period  (3) mass of Be stars and (4) velocities of Galactic radio pulsars and Be X-ray binaries. 
We find that the natal kick distributions significantly affect the total number of produced Be X-ray binaries in the SMC, the eccentricities of Be X-ray binaries and the velocities of both Galactic radio pulsars and Be X-ray binaries. The exact dependence between remnant and progenitor mass hardly affects any of our metrics. The mass accretion efficiency affects the total number of Be X-ray binaries, Be mass and distribution of velocities for Galactic Be X-ray binaries.   

We find that using the bimodal Maxwellian natal kick distribution from \cite{Igoshev2020} (based on the isolated radio pulsar data) we can obtain a reasonable match of the model with the observation-based distribution of Galactic systems.
We further apply our framework to find the optimal parameters of the low velocity component of this distribution in light of the additional constraints from Be X-ray binaries.
The optimisation procedure results in a fraction of NS forming with kicks in the low velocity component of w=0.2$\pm$0.1 and the Maxwellian parameter $\sigma_{1}$=45$^{+25}_{-15}$ km/s and we fix $\sigma_2 = 336$~km/s.

Finally, we show that the natal kick prescription commonly used in population synthesis 
studies: combining the distribution from \cite{hobbs2005} and allowing for low-velocity 
NS formation in ecSN with a Maxwellian distribution with $\sigma = 30$~km/s, is incompatible with observations of isolated radio pulsars if ecSN occur only in stripped stars. If ecSN occur also in effectively isolated stars, this distribution is marginally compatible with observations.
It is possible to improve this model and suggest that core collapse supernova explosions impart a natal kick with a Maxwellian distribution with $\sigma = 336$~km/s while ecSN occurs in both effectively isolated and stripped stars with helium core mass in a range $1.63$~--~$2.25$~M$_\odot$ at the base of asymptotic giant branch. These NS receive Maxwellian natal kick with $\sigma = 45$~km/s.
This model is slightly less preferable than our statistical model K, but it is still compatible with all available observations. This model can be improved further when a larger sample of low-velocity pulsars becomes available.
Whether it is physically possible to form a NS in ecSN explosions in this broad mass range is unclear at the moment. Further modelling is required to resolve this issue.

\section*{Data Availability Statement}

The data underlying this article will be shared on reasonable request to the corresponding author.

\section*{Acknowledgements}

A.I.P. is grateful to Prof. Gijs Nelemans, Eva Laplace, Dr Serena Vinciguerra, Anastasia Frantsuzova and Prof. Sergei Popov as well as all participants of SeBa and binary stellar evolution meeting for multiple fruitful discussions. AI was supported by STFC grant no.\ ST/S000275/1. ST acknowledge support from the Netherlands Research Council NWO (VENI 639.041.645 grants).
This work was undertaken on ARC4, part of the High Performance Computing facilities at the University of Leeds, UK.
This work has made use of data from the European Space Agency (ESA) mission
{\it Gaia} (\url{https://www.cosmos.esa.int/gaia}), processed by the {\it Gaia}
Data Processing and Analysis Consortium (DPAC,
\url{https://www.cosmos.esa.int/web/gaia/dpac/consortium}). Funding for the DPAC
has been provided by national institutions, in particular the institutions
participating in the {\it Gaia} Multilateral Agreement.
MC acknowledges support  from the  Netherlands  Organisation  for  Scientific  Research  (NWO).




\bibliographystyle{mnras}
\bibliography{bibl} 




\appendix

\section{Catalogue of Galactic Be X-ray binaries}
\label{s:systems_BeX}

We found the following information about individual HMXBs in the literature.

$[$KRL2007b$]$ 335 also known as IGR J18450–0435 is a supergiant system  \citep{systIGRJ18450}. This system is excluded from analysis.

2MASS J16193220-4944305 also known as IGR J16195-4945 is a supergiant system \citep{systIGRJ16195}. This system is excluded from analysis.

2MASS J21342037+4738002 also known as IGR J21343+4738 is B1IVe star \citep{systIGRJ21343}. This system is included in the analysis. 

UCAC2 4813819 also known as IGR J11435-6109 is B0Ve or B2Ve \citep{systIGRJ11435,systIGRJ11435a}. This system is included in the analysis. 

AX J1739.1-3020  is probably supergiant system \citep{systAXJ1739}.  This system is excluded from analysis.

V* GP Vel also known as Vela X-1 is a supergiant system, see e.g. \cite{systVelaX1}. This system is excluded from analysis.

4U 1907+09 is a supergiant system \citep{syst4U1907}. This system is excluded from analysis.

SS 188 also known as IGR J08262-3736 is a supergiant system \citep{systSS188}. This system is excluded from analysis.

V* V479 Sct is a possible counterpart of  LS 5039 which seems to be a black hole \citep{systV479Sct}. This system is excluded from analysis.

BD+60 73 also known as IGR J00370+6122 is a supergiant system \citep{systBD6073}. This system is excluded from analysis.

EM* AS 14 also known as TYC 3681-695-1 is B1-2 III-Ve star \citep{systEMAS14}. This system is included in the analysis.

HD 74194 also known as LM Vel is supergiant system \citep{systHD74194}. This system is excluded from analysis.

HD 63666 is a dwarf star, so it cannot be HMXB \citep{systHD63666}. This system is excluded from analysis.

HD 49798 is a subdwarf system \citep{systHD49798}. This system is excluded from analysis.

IGR J17544-2619 is a supergiant system \citep{systIGRJ17544}. This system is excluded from analysis.

SS 433 is a microquasar with accretion disk around the compact object \citep{systss433}. The spectral type might be A and accretion does not proceed through the decretion disk. This system is excluded from analysis.

HD 226868 also known as Cyg X-1 is probably HMXB with a black hole \citep{systCygX1}. This system is excluded from analysis.

\begin{table*}
   \centering
   \begin{tabular}{lrccccccccc}
    \hline  
    HMXB  & Sp. type & Inc &  Gaia counterpart & V     & g     &  $\varpi\pm \sigma_\varpi$ & $\mu_\alpha \pm \sigma_\alpha$ & $\mu_\delta \pm \sigma_\delta$\\
          & & &  Gaia EDR3 ID      & (mag) & (mag) & (mag)  &  (mas / year) & (mas/year)\\
    \hline
$[$KRL2007b$]$ 335        &  O9Ia  &         N & 4258160560148155648  &  14.06  &  12.76  & $ 0.164  \pm  0.024 $ & $ -1.366  \pm  0.024 $ & $ -5.595  \pm  0.022 $  \\
GRO J2058+42              &  O9.5-B0IV-Ve  & Y & 2065653598916388352  &  14.74  &  14.13  & $ 0.078  \pm  0.015 $ & $ -2.21  \pm  0.015 $ & $ -3.351  \pm  0.017 $  \\
2MASS J16193220-4944305   &  B1sg  &         N & 5935509395659726592  &  16.8  &  16.37  & $ 0.359  \pm  0.051 $ & $ -0.184  \pm  0.062 $ & $ -0.545  \pm  0.044 $  \\
2MASS J21342037+4738002   &  BII Ve  &       Y & 1978365123143522176  &  14.16  &  14.0  & $ 0.084  \pm  0.014 $ & $ -2.212  \pm  0.015 $ & $ -2.558  \pm  0.015 $  \\
LS 1698                   &  B0III/V:e  &    Y & 5352018121173519488  &  11.48  &  11.24  & $ 0.171  \pm  0.016 $ & $ -6.305  \pm  0.021 $ & $ 3.01  \pm  0.018 $  \\
UCAC2  39636510           &  B0.5Ve  &       Y & 3423526544838563328  &  12.21  &  12.17  & $ 0.138  \pm  0.018 $ & $ 0.573  \pm  0.02 $ & $ -0.608  \pm  0.014 $  \\
LS   I +61  303           &  B0Ve  &         Y & 465645515129855872  &  10.75  &  10.4  & $ 0.378  \pm  0.013 $ & $ -0.423  \pm  0.011 $ & $ -0.256  \pm  0.012 $  \\
UCAC2   4813819           &  B2III/B0V  &    Y & 5335021599905643264  &  13.4  &  14.73  & $ 0.242  \pm  0.021 $ & $ -3.448  \pm  0.021 $ & $ 1.165  \pm  0.021 $  \\
Ginga 0834-430            &  B0-2III-Ve  &   Y & 5523448274762133632  &  20.4  &  19.17  & $ 1.105  \pm  0.217 $ & $ -3.235  \pm  0.215 $ & $ 3.755  \pm  0.266 $  \\
$[$KRL2007b$]$ 84         &  B0e  &          Y & 5258414192353423360  &  15.27  &  13.88  & $ 0.243  \pm  0.013 $ & $ -4.702  \pm  0.016 $ & $ 3.559  \pm  0.014 $  \\
2E  1752                  &  O9.7Ve  &       Y & 3052677318793446016  &  10.9  &  11.99  & $ 0.154  \pm  0.015 $ & $ -0.638  \pm  0.015 $ & $ 1.256  \pm  0.014 $  \\
GSC 03588-00834           &  B0Ve  &         Y & 2162805896614571904  &  14.2  &  13.77  & $ 0.131  \pm  0.013 $ & $ -3.505  \pm  0.014 $ & $ -3.16  \pm  0.013 $  \\
V* V635 Cas               &  B0.2Ve  &       Y & 524677469790488960  &  15.19  &  14.3  & $ 0.136  \pm  0.016 $ & $ -1.684  \pm  0.013 $ & $ 0.504  \pm  0.017 $  \\
V* BQ Cam                 &  O8.5Ve  &       Y & 444752973131169664  &  15.42  &  14.2  & $ 0.134  \pm  0.02 $ & $ -0.268  \pm  0.02 $ & $ 0.44  \pm  0.02 $  \\
HD  34921                 &  B0IVpe  &       Y & 184497471323752064  &  7.48  &  7.23  & $ 0.721  \pm  0.03 $ & $ 1.305  \pm  0.041 $ & $ -3.999  \pm  0.028 $  \\
2MASS J17002524-4219003   &  B2e  &          Y & 5966213219190201856  &  9.15  &  8.71  & $ 0.641  \pm  0.023 $ & $ 1.181  \pm  0.03 $ & $ -1.47  \pm  0.023 $  \\
V* V441 Pup               &  O5Ve  &         Y & 5613494119551805184  &  11.83  &  11.6  & $ 0.096  \pm  0.017 $ & $ -0.881  \pm  0.012 $ & $ 1.785  \pm  0.018 $  \\
AX J1739.1-3020           &  O8.5Iab(f)  &   N & 4056922100878037120  &  14.8  &  16.22  & $ 0.68  \pm  0.053 $ & $ 2.954  \pm  0.062 $ & $ 1.821  \pm  0.041 $  \\
BD+53  2790               &  O9.5Vep  &      Y & 2005653524280214400  &  9.84  &  9.74  & $ 0.305  \pm  0.014 $ & $ -4.173  \pm  0.015 $ & $ -3.317  \pm  0.014 $  \\
V* V572 Pup               &  B0.2IVe  &      Y & 5548261400354128768  &  12.74  &  12.42  & $ 0.118  \pm  0.012 $ & $ -1.455  \pm  0.011 $ & $ 2.146  \pm  0.016 $  \\
V* GP Vel                 &  B0.5Ia  &       N & 5620657678322625920  &  6.87  &  6.74  & $ 0.496  \pm  0.015 $ & $ -4.822  \pm  0.015 $ & $ 9.282  \pm  0.016 $  \\
4U 1907+09                &  O8.5Iab  &      N & 4309225217336733824  &  16.35  &  16.82  & $ 0.233  \pm  0.073 $ & $ -2.749  \pm  0.074 $ & $ -2.455  \pm  0.076 $  \\
LS   V +44   17           &  B0.2Ve  &       Y & 252878401557369088  &  10.73  &  10.4  & $ 0.379  \pm  0.015 $ & $ 0.101  \pm  0.016 $ & $ -1.186  \pm  0.014 $  \\
SS 188                    &  OB  &           N & 5541793213959987968  &  12.33  &  12.16  & $ 0.178  \pm  0.01 $ & $ -2.367  \pm  0.009 $ & $ 3.177  \pm  0.013 $  \\
V* V479 Sct               &  ON6V((f))z  &   N & 4104196427943626624  &  11.27  &  10.8  & $ 0.49  \pm  0.015 $ & $ 7.425  \pm  0.014 $ & $ -8.151  \pm  0.012 $  \\
BD+60    73               &  B1Ib  &         N & 427234969757165952  &  9.66  &  9.45  & $ 0.272  \pm  0.012 $ & $ -1.796  \pm  0.011 $ & $ -0.525  \pm  0.014 $  \\
EM* AS   14               &  B1-2 III-Ve  &  N & 414196617287885312  &  11.36  &  11.41  & $ 0.339  \pm  0.018 $ & $ -2.463  \pm  0.015 $ & $ -0.546  \pm  0.017 $  \\
4U 2238+60                &  Be  &           Y & 2201091578667140352  &  14.8  &  14.1  & $ 0.104  \pm  0.014 $ & $ -2.344  \pm  0.015 $ & $ -1.015  \pm  0.014 $  \\
EM* GGA  104              &  B1IIIe  &       Y & 511220031584305536  &  11.42  &  11.22  & $ 0.328  \pm  0.022 $ & $ -1.029  \pm  0.016 $ & $ -0.082  \pm  0.017 $  \\
V* V662 Cas               &  B1Iae  &        Y & 524924310153249920  &  11.14  &  10.52  & $ 0.196  \pm  0.011 $ & $ -1.243  \pm  0.009 $ & $ 0.761  \pm  0.012 $  \\
2MASS J01581848+6713234   &  B2IVe+  &       Y & 518990967445248256  &  14.43  &  13.69  & $ 0.133  \pm  0.013 $ & $ -1.198  \pm  0.011 $ & $ 0.3  \pm  0.013 $  \\
HD  74194                 &  O8.5Ib-II(f)p  &N & 5522306019626566528  &  7.55  &  7.45  & $ 0.443  \pm  0.017 $ & $ -7.465  \pm  0.02 $ & $ 6.1  \pm  0.019 $  \\
HD  63666                 &  B7IV/V  &       N & 5489434710755238400  &  7.6  &  7.54  & $ 1.536  \pm  0.021 $ & $ -4.572  \pm  0.027 $ & $ 8.53  \pm  0.028 $  \\
WRAY 15-793               &  O9.5III/Ve  &   Y & 5336957010898124160  &  12.12  &  11.59  & $ 0.329  \pm  0.011 $ & $ -5.421  \pm  0.012 $ & $ 1.37  \pm  0.012 $  \\
V* QV Nor                 &  B0.2Ia:e  &     Y & 5886085557746480000  &  14.5  &  13.16  & $ 0.128  \pm  0.015 $ & $ -6.711  \pm  0.015 $ & $ -4.111  \pm  0.014 $  \\
2MASS J01354986+6612433   &  B1Ve  &         Y & 519352324516039680  &  13.31  &  12.46  & $ 0.167  \pm  0.011 $ & $ -1.626  \pm  0.009 $ & $ -0.027  \pm  0.011 $  \\
HD  49798                 &  sdO6  &         N & 5562023884304074240  &  8.287  &  8.22  & $ 1.92  \pm  0.05 $ & $ -4.162  \pm  0.066 $ & $ 5.926  \pm  0.058 $  \\
HD 259440                 &  B0pe  &         Y & 3131822364779745536  &  9.12  &  8.88  & $ 0.54  \pm  0.023 $ & $ -0.026  \pm  0.02 $ & $ -0.428  \pm  0.016 $  \\
V* CI Cam                 &  B0/2I[e]  &     Y & 276644757710014976  &  11.77  &  10.77  & $ 0.21  \pm  0.015 $ & $ -0.474  \pm  0.018 $ & $ -0.51  \pm  0.013 $  \\
V* BP Cru                 &  B1.5Iaeq  &     Y & 6054569565614460800  &  10.66  &  9.75  & $ 0.251  \pm  0.016 $ & $ -5.227  \pm  0.016 $ & $ -2.071  \pm  0.019 $  \\
HD 245770                 &  O9/B0III/Ve  &  Y & 3441207615229815040  &  9.39  &  8.6  & $ 0.525  \pm  0.023 $ & $ -0.59  \pm  0.031 $ & $ -2.88  \pm  0.016 $  \\
IGR J17544-2619           &  O9Ib  &         N & 4063908810076415872  &  12.94  &  11.66  & $ 0.396  \pm  0.027 $ & $ -0.506  \pm  0.029 $ & $ -0.668  \pm  0.018 $  \\
SS 433                    &  A7Ib:  &        N & 4293406612283985024  &  13.0  &  12.6  & $ 0.118  \pm  0.023 $ & $ -3.027  \pm  0.024 $ & $ -4.777  \pm  0.024 $  \\
HD 226868                 &  O9.7Iabpvar  &  N & 2059383668236814720  &  8.91  &  8.54  & $ 0.444  \pm  0.015 $ & $ -3.812  \pm  0.015 $ & $ -6.31  \pm  0.017 $  \\
CPD-63  2495              &  O9.5Ve  &       Y & 5862299960127967488  &  9.98  &  9.63  & $ 0.443  \pm  0.013 $ & $ -7.093  \pm  0.012 $ & $ -0.342  \pm  0.014 $  \\
    \hline
    \hline
    \end{tabular}
    \caption{List of HMXBs with optical counterpart in the Gaia EDR3 with well measured parallax ($\varpi'/\sigma_\varpi > 3$). Y in the third column indicates that object is included in the velocity analysis. }
    \label{t:xrb_gaia}
\end{table*}


\section{Maximum likelihood analysis of the velocity distribution}
\label{s:appendix_ml}
The maximum likelihood technique was described in complete details by \cite{Verbunt2017}. Here we briefly summarise the most important details.
The joint probability to measure parallax and proper motion for a star given velocity distribution $f_v (\vec v, \sigma)$ with parameters $\sigma$ can be written as:
\begin{align}
P (\varpi'_i, \mu_{\alpha * i}', \mu_{\delta i}', D, v_\alpha, v_\delta, v_r) \propto f_D (D) f_v (\vec v, \sigma) \nonumber \\
 \times  \exp \left[ - \frac{(1/D - \varpi_i')^2}{2\sigma_D^2} \right]  \nonumber \\ 
 \times  \exp \left[ - \frac{(\mu_{\alpha * i} (D) + cv_\alpha/D - \mu_{\alpha * i}')^2}{2\sigma_\alpha^2}\right] \nonumber \\
 \times  \exp \left[ - \frac{(\mu_{\delta i} (D) +  cv_\delta/D - \mu_{\delta * i}')^2}{2\sigma_\delta^2}\right]. 
 \label{e:joint}
\end{align}
Here $f_D (D)$ is the initial distribution for actual distance $D$ which depends on location in the sky. Values $\varpi_i'$, $\mu_{\alpha * i}'$ and $\mu_{\delta * i}'$ correspond to measured parallax and proper motion components for star with index $i$ in catalogue. Values of $\mu_{\alpha * i} (D)$ and $\mu_{\delta i} (D)$ stay for values of proper motions which appears due to the rotation of the Galaxy at distance $D$. They depend on location in the sky. Coefficients $A_\alpha$ and $A_\delta$ allows a transformation of velocity $v$ to component of proper motion along $\alpha$ and $\delta$ directions. Values $\sigma_D$, $\sigma_\alpha$ and $\sigma_\delta$ corresponds to measurements errors for parallax and proper motion components. Coefficient $c=4.74$ allows a transformation of measurements to units of km/s.

The Maxwellian velocity distribution for individual velocity components $v_\alpha, v_\delta$ and $v_r$ consists of normal distributions:
\begin{equation}
f^M_v (\vec v, \sigma) = \frac{1}{2\pi \sqrt{2\pi}\sigma^3} \exp\left(-\frac{(v_\alpha^2 + v_\delta^2 + v_r^2)}{2\sigma^2}\right).    
\label{e:maxw}
\end{equation}
Sum of two Maxwellians has the following form:
\begin{equation}
f^{2M}_v(\vec v, \vec \sigma) =  w f^M_v (\vec v, \sigma_1) + (1-w) f^M_v (\vec v, \sigma_2).     
\end{equation}
The actual velocity components $v_\alpha$, $v_\delta$, $v_r$ and distances $D$ are not known. Therefore, we integrate the joint probability eq.~(\ref{e:joint}) over all unknown variables:
\begin{equation}
P(\varpi'_i, \mu_{\alpha * i}', \mu_{\delta i}') = \iint P (\varpi'_i, \mu_{\alpha * i}', \mu_{\delta * i}', D, \vec v) d^3 \vec v dD.
\end{equation}
Some of these integrations are performed analytically and some numerically. Practical equations used for calculations of this joint probability can be found in section 4 by \cite{Verbunt2017}. As soon as exact model for velocity distribution is chosen and measured values are fixed this joint probability becomes likelihood for a model parameter:
\begin{equation}
L_i (\sigma) =  P(\varpi'_i, \mu_{\alpha * i}', \mu_{\delta i}' | \sigma).
\end{equation}
Working with multiple independent measurements we introduce log-likelihood:
\begin{equation}
\mathcal L (\sigma) = - 2 \sum_{i=1}^N \log L_i (\sigma).    
\end{equation}
This log-likelihood is maximised further to find the best parameter $\sigma$ or a vector of parameters $\vec \sigma$.

\section{Binned velocity distribution}
\label{s:binned}
First, we introduce a local polar coordinate system to describe the velocity vector:
\begin{flalign}
v_\alpha  = & v \sin \psi_1 \cos \psi_2 = A_\alpha v \\
v_\delta  = & v \sin \psi_1 \sin \psi_2 = A_\delta v \\
v_r       = & v \cos \psi_1.
\end{flalign}
Our binned velocity distribution is based on individual velocity components and defined as following:
\begin{equation}
f^B_v (\vec v, \vec m) = m_j v^2,
\end{equation}
if $\Delta v j < \sqrt{v_\alpha^2 + v_\delta^2 + v_r^2} \leq \Delta v (j+1)$.
We substitute the  binned velocity distribution is  in eq.(\ref{e:joint}).
It is possible to analytically compute the most inner integral over absolute value of velocity:
$$
I_{1,i,j} = m_i \int_{\Delta v i}^{\Delta v (i+1)} v^2 \exp \left[ -\frac{1}{2} \left\{ \frac{A_\alpha v + D (\mu_{\alpha * } (D) -  \mu_{\alpha * }' )^2}{D^2 \sigma_\alpha^2}  \right.\right.    
$$
\begin{equation}
\hspace{2.6cm} \left. \left. + \frac{A_\alpha v + D (\mu_{\alpha * } (D) -  \mu_{\alpha * }' )^2}{D^2 \sigma_\alpha^2} \right\} \right] dv.
\label{s:int}
\end{equation}
This integral has following form:
\begin{equation}
I_v = m_i e^{-A_3} \int_{\Delta v i}^{\Delta v (i+1)} v^2 \exp\left(-A_1 v^2 - A_2 v \right) dv,     
\end{equation}
with the following values:
\begin{equation}
A_1 = \frac{c^2 A_\alpha^2}{2\sigma_\alpha^2 D^2} + \frac{c^2 A_\delta^2}{2\sigma_\delta^2 D^2}    
\end{equation}
\begin{equation}
A_2 = \frac{cA_\alpha ( \mu_{\alpha * } (D) - \mu_{\alpha * }' )}{D\sigma_\alpha^2} + \frac{cA_\delta ( \mu_{\delta } (D) - \mu_{\delta }' )}{D\sigma_\delta^2}    
\end{equation}
\begin{equation}
A_3 = \frac{(\mu_{\alpha * } (D) - \mu_{\alpha * }')^2}{2\sigma_\alpha^2} + \frac{(\mu_{\delta } (D) - \mu_{\delta }')^2}{2\sigma_\delta^2}.    
\end{equation}
The result of integration for eq. (\ref{s:int}) is:
$$
I_v =  m_i \frac{\exp\left(E^2-A_3\right)}{A_1 \sqrt{A_1}} \left(\frac{\sqrt{\pi}}{4} \left[1 + 2E^2\right] \mathrm{erf}(\sqrt A_1 v + E)\right. \hspace{1.8cm}   
$$
\begin{equation}
\hspace{1.5cm}\left.\left. - \frac{1}{2}\left[2E-1\right] \exp(-(\sqrt{A_1}v+E)^2) \right) \right|_{\Delta v i}^{\Delta v (i+1)},   
\end{equation}
where $E = A_2 / (2\sqrt{A_1})$.
There are three more integrals over angles and distance:
$$
P(\varpi'_i, \mu_{\alpha * i}', \mu_{\delta i}' | m_i) = m_i \int_0^{D_\mathrm{max}} f_D (D)\hspace{2.5cm}.
$$
\begin{equation}
\hspace{3.5cm} \times \iint I_v (D, \psi_1, \psi_2, \sigma) d\Omega dD   
\end{equation}
These integrals are computed numerically following the same procedure as described by \cite{Verbunt2017}.

For each individual pulsar and for each velocity bin we can compute the likelihood as:
\begin{equation}
L_i^{v_k, v_{k+1}} = P(\varpi'_i, \mu_{\alpha * i}', \mu_{\delta i}' | 1),    
\end{equation}
because model is described using a constant parameter $m_k$, we can move this parameter out of all integrations. In this case the likelihood for a particular model but for a single pulsar can be computed as:
\begin{equation}
L_i = \sum_{k=1}^{N} m_k L_i^{v_k, v_{k+1}}.   
\end{equation}
And the complete log-likelihood is computed then as:
\begin{equation}
\mathcal L (\vec m) = -2 \sum_{i=1}^N \log \left( \sum_{k=1}^{N} m_k L_i^{v_k, v_{k+1}} \right).
\label{e:final_ll}
\end{equation}
This equation has an interesting property. Given the data it is possible to compute numerically expensive part $L_i^{v_k, v_{k+1}}$ just once and substitute this matrix into eq.(\ref{e:final_ll}) together with $m_k$ obtained from a model to get the log-likelihood. 

The most natural usage of the log-likelihood eq.(\ref{e:final_ll}) is simply to minimise it. This exercise is however quite useless at the moment because a number of young pulsars and Be X-ray binaries is quite small, therefore not enough pulsars is seen in each velocity bin to really constrain the velocity distribution.
However, if a binary evolution model provides a model for velocity distribution of isolated radio pulsars or Be X-ray binaries, the eq. (\ref{e:final_ll}) provides a natural way to test such a model against real measurements.
Given a fact that velocity bins have exactly the same extent in model and $L_i^{v_k, v_{k+1}}$, the relation between $m_k$ and relative number of objects in velocity bin $h_k$ is as following:
\begin{equation}
m_k = \frac{3 h_k }{\Delta v^3 ((i+1)^3 - i^3)}. 
\label{e:mk}
\end{equation}
Therefore, after we separate Be X-ray binaries and isolated NS from the results of the population synthesis we prepare a histogram with a step in velocity of 33.4~km/s for normal radio pulsars and $10$~km/s for Be X-ray binaries and compute $m_k$ which is further used in calculations of likelihood according to eq. (\ref{e:final_ll}).




\section{Calculation of distributions for secondary mass, orbital periods etc}
\label{s:orbitperiod}
Here we use same approach as \cite{Vinciguerra2020}. Namely we assume that orbital properties does not change during the Be X-ray binary stage. In this case each individual binary contributes a weight of:
\begin{equation}
w = \frac{1}{M_\mathrm{tot}}\int \mathrm{SFH} (-t) dt,
\end{equation}
where $M_\mathrm{tot}$ is the total stellar mass required to produce 100000 high-mass binaries. 
We use the SFH for SMC by \cite{rubele2015}. It is a piece-wise function, so we can replace integral with a sum:
\begin{equation}
w = \frac{1}{M_\mathrm{tot}} \sum_i \mathrm{SFH} (-t_i) \Delta t_i.    
\end{equation}
Further we plot a histogram for each individual orbital parameter using these weights. 

\section{Velocities of spectroscopic B binaries}
\label{s:bbinary}
It is important to check if the velocity dispersion of Be X-ray binary progenitors is indeed small and comparable to $\approx 10-15$~km~s$^{-1}$. We estimate the velocity dispersion using our code (see Appendix~\ref{s:appendix_ml}), parallaxes and proper motions measured by Gaia.
In order to perform this analysis we use the catalogue of spectroscopic binaries by \cite{Pourbaix2004}. Our choice for the catalogue can be explained as the following: isolated B stars could be formed as a result of binary stellar evolution (walk away and ran away stars) which changes their velocities significantly. As for the spectroscopic binaries, we select those where primary did not evolve of the main sequence yet,  thus these binaries probably did not interact yet.
Although a small fraction of these stars could move fast because of triple and multiple stars disruption or due to OB association dissolution, this sample should provide us some information about velocity dispersion of B stars before any interaction occurs.

From this catalogue, we select only binaries where the primary is main sequence B star without any chemical peculiarity. Chemical peculiarity could appear as a result of mass transfer or supernova explosion which could change the velocity of the system. Further we find optical counterpart for each star in the Gaia EDR3. After choosing only stars with good quality astrometry (relative error of parallax is less than 0.33), we have a list of 117 systems. 

At the next step we run our analysis for velocities. For model with single Maxwellian distribution we obtain $\sigma=11$~km/s. For model which is a sum of two Maxwellians, we obtain a fraction of low-velocity objects $w = 20\pm 8$ with $\sigma_1 = 3$~km/s, remaining binaries have parameter $\sigma_2 = 11$~km/s. The model with two Maxwellian is more probable. Overall, we confirm that velocity dispersion of progenitors for Be X-ray binaries is $\sigma = 11$~km/s.

\bsp	
\label{lastpage}
\end{document}